\documentclass[longauth]{aa}

\usepackage{graphicx}
\usepackage{txfonts}
\usepackage{amsmath}    
\usepackage{mathtools}
\usepackage{amssymb}
\usepackage{color}
\usepackage{multirow}
\usepackage{xcolor}
\usepackage{ulem}

\begin{document}

   \title{Retrieval of the physical parameters of galaxies from WEAVE-StePS-like data using machine learning}
    \titlerunning{Retrieval of physical parameters using machine learning}

\author{J. Angthopo\inst{1} \and B.R. Granett   \inst{1} \and
  F. La Barbera 
  \inst{2} \and 
  M. Longhetti 
  \inst{1} \and 
  A. Iovino 
  \inst{1} \and 
  M. Fossati 
  \inst{3} \and 
  F.R. Ditrani 
  \inst{1, 3}  
  \and 
  L. Costantin\inst{4} 
  \and 
  S. Zibetti\inst{5} 
  \and 
  A. Gallazzi\inst{5} 
  \and 
  P. S\'anchez-Blázquez\inst{6,7} 
  \and
  C. Tortora\inst{2}
  \and C. Spiniello\inst{8, 2}
  \and B. Poggianti\inst{9} 
  \and A. Vazdekis\inst{10, 11}
  \and M. Balcells\inst{10,11,12} 
  \and S. Bardelli\inst{13} 
  \and C. R. Benn\inst{12}
  \and M. Bianconi\inst{14}
  \and M. Bolzonella\inst{13}
  \and G. Busarello\inst{2} 
  \and L. P. Cassar\`a\inst{15}
  \and E. M. Corsini\inst{16,9} 
  \and O. Cucciati\inst{13}
  \and G. Dalton\inst{8,17}  
  \and A. Ferr\'e-Mateu\inst{10,11}
  \and R. Garc\'ia-Benito\inst{18} 
  \and R.M. Gonz\'alez Delgado\inst{18}
  \and E. Gafton\inst{12}
  \and M. Gullieuszik\inst{9} 
  \and C. P. Haines\inst{19,1} 
  \and  E. Iodice\inst{2} 
  \and A. Ikhsanova\inst{16}
  \and S. Jin\inst{8,20} 
  \and J. H. Knapen\inst{10,11}
  \and S. McGee\inst{14}
  \and A. Mercurio\inst{21,2}
  \and P. Merluzzi\inst{2} 
  \and L. Morelli\inst{19,1}
  \and A. Moretti\inst{9}
  \and D.N.A. Murphy\inst{22}
  \and A. Pizzella\inst{16,9}
  \and L. Pozzetti\inst{13}
  \and R. Ragusa\inst{2} 
  \and S. C. Trager\inst{20} 
  \and D. Vergani\inst{13}
  \and B. Vulcani\inst{9} 
  \and M. Talia\inst{23,13}  
  \and E. Zucca\inst{13} 
  }

\institute{INAF-Osservatorio Astronomico di Brera, via Brera 28, I-20121 Milano, Italy\\
      \email{james.angthopo@inaf.it}
      \and {INAF - Osservatorio Astronomico di Capodimonte, Via Moiariello 16, I-80131 Napoli, Italy}
      \and{Università degli studi di Milano-Bicocca, Piazza della scienza, 20125 Milano, Italy} 
      \and{Centro de Astrobiolog\'{\i}a (CAB), CSIC-INTA, Ctra. de Ajalvir km 4, Torrej\'on de Ardoz, E-28850, Madrid, Spain} 
      \and{INAF - Osservatorio Astrofisico di Arcetri, Largo Enrico Fermi 5, I-50125 Firenze, Italy} 
      \and{Departamento de F\'{i}sica Te\'{o}rica, Universidad Aut\'{o}noma de Madrid, E-28049 Madrid, Spain} 
      \and{Instituto de F\'{i}sica de Part\'{i}culas y del Cosmos (IPARCOS), Universidad Complutense de Madrid, E-28040 Madrid, Spain}
      \and{Dept. Physics, University of Oxford, Keble Road, Oxford OX1 3RH, U.K.}
      \and{INAF--Osservatorio Astronomico di Padova, vicolo dell'Osservatorio 5, I-35122 Padova, Italy}
      \and{Instituto de Astrof\'isica de Canarias, IAC, Vía L\'actea s/n, E-38205, La Laguna (S.C. Tenerife), Spain}
      \and{Departamento de Astrof\'isica, Universidad de La Laguna, E-38206, La Laguna (S.C. Tenerife), Spain}
      \and{Isaac Newton Group of Telescopes, ING, 38700 La Palma (S.C. Tenerife), Spain}
      \and{INAF – Osservatorio di Astrofisica e Scienza dello Spazio, Via P. Gobetti 93/3, I-40129 Bologna, Italy}
      \and{School of Physics and Astronomy, University of Birmingham, Birmingham, B15 2TT, UK}
      \and{INAF - IASF Milano, Via Alfonso Corti 12, 20133 Milano, Italy}
      \and{Dipartimento di Fisica e Astronomia ``G. Galilei'', Universit\`a
      di Padova, vicolo dell'Osservatorio 3, I-35122 Padova, Italy}
      \and{RAL, Space, Science and Technology Facilities Council, Harwell, Didcot OX11 0QX, U.K.}
      \and{Instituto de Astrofísica de Andaluc\'ia (CSIC), P.O. Box 3004, E-18080, Granada, Spain}
      \and{Instituto de Astronomía y Ciencias Planetarias de Atacama (INCT), Universidad de Atacama, Copayapu 485, Copiapó, Chile}
      \and{Kapteyn Astronomical Institute, Rijksuniversiteit Groningen, Landleven 12, 9747 AD Groningen, the Netherlands}
      \and{Dipartimento di Fisica “E.R. Caianiello”, Università degli studi di Salerno, Via Giovanni Paolo II 132, I-84084 Fisciano (SA)}
      \and{Institute of Astronomy, University of Cambridge, Madingley Road, Cambridge CB3 0HA, U.K.}
      \and{Universit\'a di Bologna – Department of Physics and Astronomy, Via Gobetti 93/2, 40129 Bologna, Italy}}

   \date{Last edited - \textbf{\today}}

  \abstract
   {The \textit{William Herschel} Telescope Enhanced Area Velocity Explorer (WEAVE) is a new, massively multiplexing spectrograph that allows us to collect about one thousand spectra over a 3 square degree field in one observation. The WEAVE Stellar Population Survey (WEAVE-StePS) in the next 5 years will exploit this new instrument to obtain high-S/N spectra for a magnitude-limited ($I_{AB} = 20.5$) sample of $\sim$ 25 000 galaxies at moderate redshifts ($ z \geq 0.3$), providing insights into galaxy evolution in this as yet unexplored redshift range. 
   }
   {We aim to test novel techniques for retrieving the key physical parameters of galaxies from WEAVE-StePS spectra using both photometric and spectroscopic (spectral indices) information for a range of noise levels and redshift values.}
   {We simulated $\sim\,$105 000 galaxy spectra assuming star formation histories with an exponentially declining star formation rate, covering a wide range of ages, stellar metallicities, specific star formation rates (sSFRs), and dust extinction values. We considered three redshifts (i.e. $z=0.3, 0.55,$ and $0.7$), covering the redshift range that WEAVE-StePS will observe. 
   We then evaluated the ability of the random forest and K-nearest neighbour algorithms to correctly predict the average age, metallicity, sSFR, dust attenuation, and time since the bulk of formation, assuming no measurement errors. We also checked how much the predictive ability deteriorates for different noise levels, with S/N$_{\rm I,obs}=10$, 20, and 30, and at different redshifts. Finally, the retrieved sSFR was used to classify galaxies as part of the blue cloud, green valley, or red sequence.
   }
   {We find that both the random forest and K-nearest neighbour algorithms accurately estimate the mass-weighted ages, u-band-weighted ages, and metallicities with low bias. The dispersion varies from 0.08--0.16$\,$dex for age and 0.11--0.25$\,$dex for metallicity, depending on the redshift and noise level. For dust attenuation, we find a similarly low bias and dispersion. For the sSFR, we find a very good constraining power for star-forming galaxies, $\log\,$sSFR$\gtrsim-11$, where the bias is $\sim 0.01\,$dex and the dispersion is $\sim 0.10\,$dex. However, for more quiescent galaxies, with $\log\,$sSFR$\lesssim-11$, we find a higher bias, ranging from 0.61 to 0.86$\,$dex, and a higher dispersion, $\sim 0.4\,$dex, depending on the noise level and redshift. In general, we find that the random forest algorithm outperforms the K-nearest neighbours. Finally, we find that the classification of galaxies as members of the green valley is successful across the different redshifts and S/Ns. 
   }
   {We demonstrate that machine learning algorithms can accurately estimate the physical parameters of simulated galaxies for a WEAVE-StePS-like dataset, even at relatively low S/N$_{\rm I,obs} = 10$ per \AA\ spectra with available ancillary photometric information. A more traditional approach, Bayesian inference, yields comparable results. The main advantage of using a machine learning algorithm is that, once trained, it requires considerably less time than other methods.}

   \maketitle
%

\section{Introduction}

Over the last two decades, several wide-area photometric and spectroscopic surveys have greatly improved our understanding of galaxy formation and evolution. 
Most notably, the combination of wide-area and pencil-beam spectroscopic surveys, including the Sloan Digital Sky Survey \citep[SDSS;][]{SDSS}, the Galaxy And Mass Assembly \citep[GAMA;][]{2013GAMA}, zCOSMOS \citep{Lilly:2009}, the VIMOS Public Extragalactic Redshift Survey \citep[VIPERS;][]{Guzzo:2014}, and 3D-HST \citep{Momcheva:2016}, have pushed the boundaries of galaxy formation studies to a few billion years after the Big Bang. These spectroscopic efforts usually target well-known fields, where multi-wavelength imaging campaigns provide deep complementary datasets, often covering the UV to near-infrared (NIR) parts of the electromagnetic spectrum. These surveys include the SDSS imaging survey, the Cosmic Assembly Near-infrared Deep Extragalactic Legacy Survey \citep[CANDELS;][]{Koekemoer:2011}, UltraVISTA \citep{McCrack:2012}, and the Hyper Suprime-Cam Subaru Strategic Prime \citep[HSC-SSP;][]{Aihara:2018}. Exploitation of these data in combination with numerical hydrodynamic simulations \citep{Crain:15, Schaye:15, SpringTNG:17} and semi-analytic models \citep[and references within]{Somerville:2015} has allowed us to greatly advance our understanding of galaxy formation and evolution mechanisms. 

One of the key discoveries of recent decades has been the existence of a bimodality in the star formation activity of galaxies \citep{Strateva:01, 2004Bald, Faber:07}, with two distinct populations of galaxies: one star-forming population characterised by blue optical colours and a separate redder population of quiescent galaxies.  
This bimodality holds locally and at high redshifts, up to $z\approx 3$--$4$ \citep{Fritz:2014, Graaff:2021}. Moreover, a broad separation of the galaxy population into two classes is found in different parameters, including colour--colour 
\citep{2009Will}, star formation rate (or colour)--stellar mass \citep{Salim:14, Trayford:2016, Phill:19, Wright:2019, Nelson:2019}, and D$_n$(4000)--velocity dispersion \citep{Angthopo:2019} planes.
Finally, an important sample of transitioning galaxies straddle the two main populations, often referred to as galaxies in the green valley (GV). The study of GV galaxies is essential to understanding the quenching mechanism of star formation. \citet{Faber:07} proposed multiple evolutionary paths across the GV depending on the quenching mechanism. This was further supported by \citet{Schawinski:2014}, who found that there are two distinct quenching timescales depending on the morphology of the galaxy: elliptical galaxies are thought to have a more rapid quenching than spiral galaxies. 

To gain insights into the physical mechanisms related to the quenching of star formation and the origin of the aforementioned bimodality, it is essential to obtain accurate estimates of galaxy physical parameters. These parameters can be obtained by fitting observed galaxies to synthetic spectral energy distribution templates \citep{Gavazzi:2002, Ilbert:2006} 
or by comparing spectral indices of observed galaxies to templates obtained with different star formation histories \citep[SFHs;][]{Gallazzi:2005, Costantin:2019, Angthopo:2020, Ditrani:2023}. Furthermore, with access to a large 
quantity of spectroscopic and photometric data, sophisticated spectral fitting 
codes such as \texttt{STARLIGHT} \citep{SLight}, \texttt{pPXF} \citep{2017pPXF, Cappellari:2023}, \texttt{BAGPIPES} 
\citep{Carnall:2018}, and \texttt{Prospector} \citep{Johnson:21} have been developed to 
estimate various physical parameters and to reconstruct the SFHs of galaxies from joint fits of the available spectroscopy and photometry. A large scatter, however, remains in the estimates of these parameters due to the different physical recipes assumed by different codes. 
\citet{Pacifici:2023} explore such differences for 15 different spectral fitting algorithms when used to estimate stellar masses, star formation rates (SFRs), and effective dust attenuation (A$_{\rm V}$) at $z\sim 1$ and 3. They find systematic differences in the estimations of physical parameters by different algorithms of 0.1--0.3 dex.

Despite these systematics, modern spectral fitting codes provide accurate estimates of the physical parameters of galaxies, often thanks to Bayesian inference methods and advanced algorithms used to explore the high-dimensional likelihood space. 
However, existing and upcoming large surveys that will measure the spectro-photometry of tens of thousands to millions of galaxies are bringing the field into the realm of big data, making the computational time of traditional fitting codes a variable of growing interest. 
As an alternative, machine learning (ML) algorithms have been applied to 
estimate galaxy parameters. These codes have been used to estimate photometric redshifts \citep{Ball:2008,LiR:2022}, 
SFRs or specific star formation rates \citep[sSFRs;][]{SS:2017,Davidzon:2019,Euclid:2023}, and 
metallicities \citep{Simet:2021}, with a good agreement found between true and predicted 
values. ML algorithms can be broadly divided into two categories -- supervised and unsupervised. Supervised ML algorithms require data with labels and are mostly used for the purposes of regression and classification. Unsupervised ML algorithms do not use any labels for the data and instead learn to characterise the distribution of the dataset on their own. Unsupervised methods are predominantly used to group data points with similar properties as a tool for dimensionality reduction and data compression or to identify outliers.

Our study uses the former method -- supervised ML for regression. 
We make use of random forest (RF) and 
K-nearest neighbour (KNN) algorithms to estimate physical parameters 
-- including the average age of the stellar population, both the mass-weighted age (mwa) and the $u$-band-weighted age (uwa), the metallicity, the sSFR, the A$_{\rm V}$, and the time from the bulk of the star 
formation -- using both spectral and photometric information.
Most of the analysis is performed on a simulated dataset that mimics the upcoming 
observations of the WHT Enhanced Area Velocity Explorer Stellar Population Survey \citep[WEAVE-StePS;][]{Iovino:2023} at the 4.2\,m \textit{William Herschel }Telescope (WHT) in La Palma.  
The WEAVE spectrograph has a large 
field of view, $\sim 3\,$square degrees, and huge multiplexing capabilities, with nearly 1000 spectra observed in a single pointing \citep{Jin:2024}. The spectral coverage spans from $\sim 3660-9590\,$\AA\ at a resolution of R$\sim$ 5000 \citep{Dalton:2012, Dalton:2016}. WEAVE-StePS aims to observe a magnitude-limited ($I_{AB} = 20.5$) sample  of $\sim 25,000$ galaxies, mostly 
between $z=0.3$ and $0.7$ with median S/N$_{\rm I,obs}=10$ per \AA. 
This survey is designed to bridge the redshift gap between SDSS and LEGA-C \citep{LEGAC:2016}. 
The combination of the three surveys will allow us to directly study galaxy evolution on a long and continuous span of cosmic time, over nearly 8 billion years.

The paper is structured as follows: In Sect.~\ref{sec:Method} we describe the two ML algorithms and the models used to create the synthetic galaxy templates. We also describe the procedure to simulate realistic spectra and photometry with measurement uncertainties. Section~\ref{sec:results} outlines our main findings, and we test how well we can retrieve the physical parameters, both in the absence of measurement errors and with realistic uncertainties on spectra and photometry. Additionally, we investigate the variations in retrieval capabilities for data with different S/N values and redshifts. 
In Sect.~\ref{sec:Class} we classify galaxies into three groups -- blue cloud (BC), GV, and red sequence (RS) -- and we discuss the completeness of the classification. Section~\ref{sec:Disc} discusses our results, outlining the methodology's caveats and potential limitations. Finally, in Sect.~\ref{sec:Sum} we summarise our main results. Throughout the paper, we assume $H_0=69.6\,$km$\mathrm{s^{-1}\,Mpc^{-1}}$, $\Omega_\mathrm{M}=0.286$, and $\Omega_{\Lambda} = 0.714$ \citep{Bennett:2014}. Magnitudes are given in the AB system \citep{Oke:1974} unless otherwise stated.

\section{Methodology}
\label{sec:Method}
This section describes the two ML algorithms applied in this paper, the library of templates adopted, and how we used them to obtain observed magnitudes and spectra that simulate WEAVE-StePS observations. 

\subsection{Machine learning methods}
\label{sec:mlmethods}
To retrieve the physical parameters, we made use of two ML
algorithms for regression, RF \citep{Breiman:2001} and KNN \citep{NSA:1992}, as they are two well-known algorithms that are simple to implement and yield accurate results \citep{SS:2017, Bonjean:2019}. These algorithms are implemented using the \texttt{scikit-learn v1.4.0} Python package \citep{scikit-learn}. The algorithms take as input the set of galaxy spectro-photometric observables and output the predicted physical parameters. They are first trained on a galaxy dataset with known physical parameters. Once trained, new observables can be input to determine estimates of the physical parameters. 
We used a simulated dataset of galaxy spectro-photometric measurements for which we have the underlying model physical parameters.
We used 90$\%$ of the dataset for training the algorithms, and the remaining 
10$\%$ of the sample for testing and validation. We randomised the data selection so that both the training 
and testing samples are fully representative of each other. Generally, ML algorithms perform better when the input values are normally distributed and centred on 0 with a standard deviation of 1. Thus, before running the algorithms for training or testing, we pre-processed the galaxy parameters by subtracting the mean values and dividing by the standard deviation, computed from the training set. For parameters that range over many orders of magnitude, such as the sSFR, we used the base ten logarithm of the value and then standardise to mean 0 and standard deviation 1. 

The first algorithm we considered, the RF, uses an ensemble of decision trees.
A decision tree consists of a sequence of rules applied to determine the output value. It is constructed by iteratively splitting the training sample according to the values of the observables (features). A different feature is used to make the partition at each tree level. The process can be stopped after a certain number of splits -- this stopping point is known as the maximum depth. The final nodes, known as leaf nodes, have physical parameter values assigned to them, which are returned to give the parameter estimate. The final estimate from the RF is determined by averaging the estimates from each decision tree. We tested the algorithm with a number of trees ranging from 1 to 100 (see Sect. \ref{sec:results}). We find that a larger number of trees in the forest gave more robust results, although the performance is not sensitive to the precise number (see Appendix \ref{sec:OP_ML} for details). We thus opted for 100 trees. The trees were expanded until each leaf node contained a single galaxy.

The KNN algorithm is the second algorithm that we used for the 
purposes of estimating the physical parameters of our templates. The algorithm stores the training sample, then when given observables to evaluate, it finds the nearest neighbours in the multi-dimensional parameter space of the training sample. A weighted average of the estimates from each neighbour determines the returned parameter estimate. We weighted the neighbours according to the inverse of the Euclidean distance. Similarly to the RF, we tested the performance with the number of neighbours ranging from 1 to 100. The optimal value depends on the dataset's size and distribution; in our case, we opted to use 100 neighbours for our runs. 

\subsection{Template library}
\label{sec:TemplLib}
We tested the ML algorithms described in the previous section using a library of spectral templates. 
The library is based on the 
\citet{BC03} models (2016 revised version, hereafter CB16) assuming a Chabrier initial mass function \citep[][]{Chab:03}. 
The CB16 models cover 
a wide spectral range using different stellar spectral libraries. 
The optical wavelength is 
based on the MILES stellar library \citep{MILES:2006}, which covers the range
$3525 <\lambda <7500\,$\AA\ at the resolution of 2.5\,\AA. The UV part of 
the spectra, $\lambda \leq 3525\,$\AA, is purely theoretical, based on \citet{Martins:05} 
models with a resolution of 1\,\AA. Finally, the NIR part of the spectra is 
based on the BaSeL semi-empirical library \citep{Westera:2002} at a resolution of 3\,\AA. 

\begin{figure*}
    \centering
    \includegraphics[width=\linewidth]{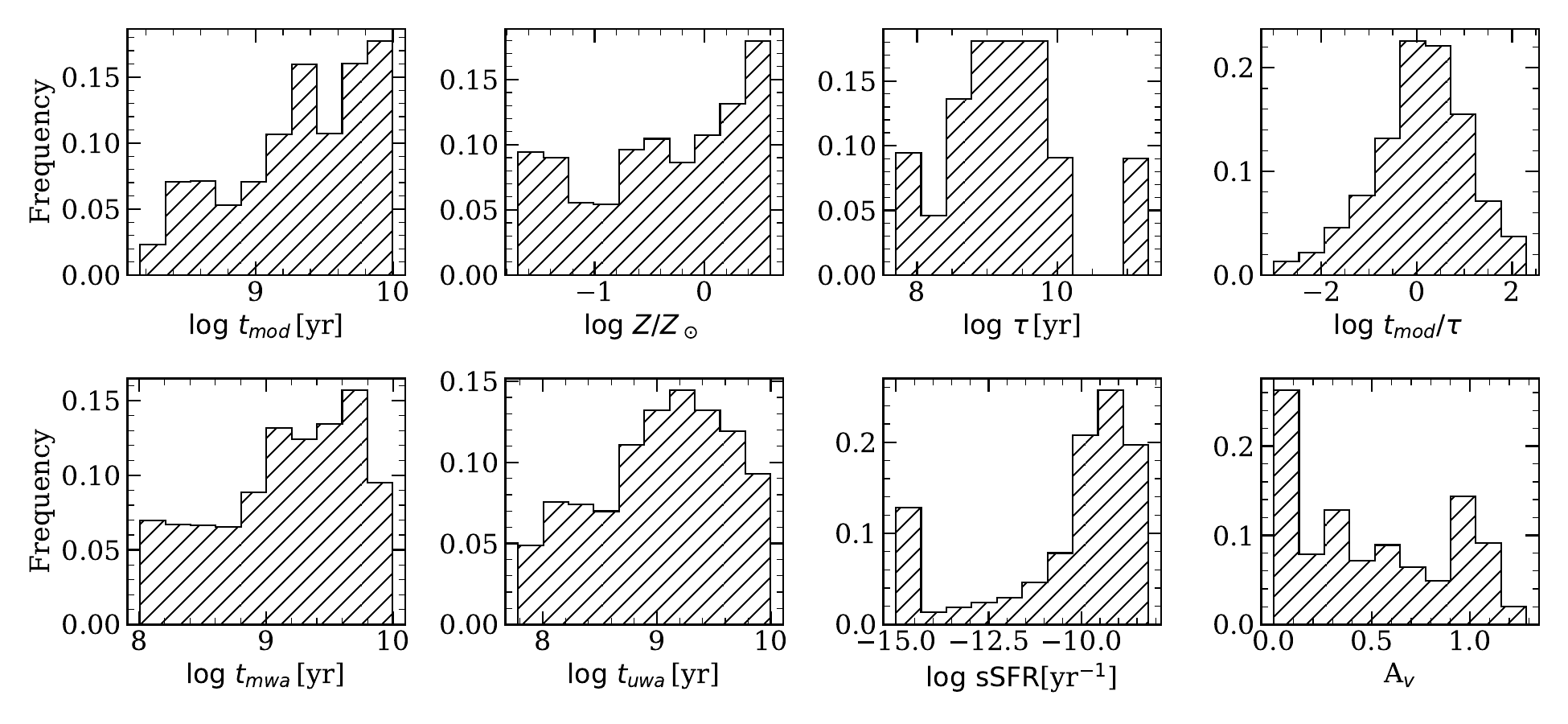}
    \caption{Histograms of the physical parameters of the templates created using the BC16 models.
    \textit{Top panels:} Distribution of, from left to right, 
    model age, i.e. age since the birth of the simulated 
    galaxy, metallicity, rate of decay of star formation ($\tau$), and time since the bulk of the star formation ($t_{\rm mod}/\tau$).
    \textit{Bottom panels:} Histograms of mwa and uwa, sSFR, and effective dust attenuation (A$_{\rm V}$). See the main text for more details on the distribution of these parameters.
    }
    \label{fig:ML_Fig1}
\end{figure*}
We built our template library assuming a simple SFH represented by an exponentially decaying SFR:
\begin{equation}
    \mathrm{SFR}(t) \propto e^{-t/\tau}
    \label{eq:SFR_eq}
,\end{equation}
where $t$ is the time elapsed since the onset of star formation in Gyrs, and $\tau$ is the 
rate of decay of the SFR. The model ages, $t$, vary from $0.1$ to $10.04\,$ Gyr,  where we have a total of 59 different ages. 
\footnote{The changes in the observables, such as the spectral indices, do not scale linearly with the changes in the age of the stellar population. Therefore, the time steps, t$_{\rm mod}$, are chosen ad hoc so that we have a more uniform distribution in the average ages of our templates while ensuring that we do not have a strong over-density region in the observable plane, such as the H$\beta$ vs [MgFe]$^\prime$. Note that since we tried to ensure uniformity in multiple dimensions, we do not have complete uniformity in any single given dimension.} For $\tau$ we considered 20 different values ranging from 0.05 Gyr to 10 Gyr. Between 0.05 and 1 Gyr, the values of $\tau$ were set with a linear step size of $0.1\,$Gyr to represent quenched old galaxies, while for $\tau \geq 1\,$Gyr,  we used a logarithmic step of $\sim0.09\,$dex. We also added two more sets of templates with $\tau \sim\,$100 and 200$\,$Gyr to mimic a constant SFH. The metallicity varies from $\log\, Z/Z_\odot = -1.69$ to $+0.6$. Since BC16 models are distributed at discrete values of metallicity $\log\, Z/Z_\odot = -1.69, -0.40, +0.0, +0.40$, we created templates covering a continuous distribution of this parameter by interpolating between the spectra on random values between those obtained on the fixed grid. We reached a maximum value of $+$0.6 by linear extrapolation of the template spectra obtained with $\log\, Z/Z_\odot = +0.00$ and $+0.40$. 
Finally, we adopted the \citet{Charlot2000} dust prescription, which applies dust attenuation depending on the galaxy's SFH. We chose the total optical depth in the V band, $\hat{\tau}_{\rm V}$, to range from 0 to 3, where for each $\tau$ and metallicity we have a model for $\hat{\tau}_{\rm V} = 0$ and $3$. In addition to these two optical depths, we randomly generated two additional values from a uniform distribution between 0 and 3.\ We note that the prescription treats dust in younger stellar populations, $\leq 10^{7}\,$yr, differently to that in older stellar populations. Less dust affects the older population, owing to the diffuse interstellar medium, and is formulated as $\mu \hat{\tau}_{\rm V}$. We selected  $\mu=0.3$.
In total, we have $\sim 105\, 000$ galaxy templates, which do not contain any emission lines, as we assumed that their contribution has already been removed from the simulated spectra. We note that correcting for nebular emissions from stellar contributions is challenging and may introduce systematic errors in the estimation of certain absorption lines. As a worst-case scenario, we assumed that the systematics associated with emission line corrections are so high that the absorption features potentially affected 
by emission line residuals (namely Mgb, Balmer lines, and Fe5015) are no longer usable. 
We therefore performed a test run excluding all these indices from the spectral analysis and verified that this extreme choice induces only minimal variations in the age and metallicity estimates. The major effect is present for the age estimates, where removing all the Balmer lines (H$\beta$, H$\gamma$, and H$\delta$) in the absence of photometric information causes an increase in the uncertainty in all age estimates by roughly 0.06\,dex.

Figure~\ref{fig:ML_Fig1} shows the distribution of the physical parameters of our templates. 
We used the uwa and mwa to retrieve ages that are sensitive mostly to young stellar populations and to all stars regardless of age, respectively.  We defined the sSFR as
\begin{equation}
    \mathrm{sSFR} = \frac{\mathrm{SFR}}{\mathrm{M_\star^{live} + M_\star^{rem.}}}
    \label{eq:defsSFR}
,\end{equation}
where SFR is the instantaneous SFR in M$_\odot$/yr, and $\mathrm{M_\star^{live}}$ and $\mathrm{M_\star^{rem.}}$ are the stellar mass in M$_\odot$ stored in living stars and remnants, respectively. For galaxies with  $\log$(sSFR)$\leq -15$, we set it to $-15$ as it has been found that measurements of SFRs, and thus sSFRs, are highly uncertain below a certain threshold \citep{Brinchmann:2004, Donnari:2019} since there is a high degree of degeneracy between observables and sSFRs. Finally, we note that the \citet{Charlot2000} prescription applies to dust depending on the SFH; thus, for a given $\hat{\tau}_{\rm V}$, the A$_{\rm V}$ will differ for different $\tau$. This leads to the non-uniform distribution shown above.

\subsubsection{Observables}
\label{sec:observables}
\begin{figure*}
    \centering
    \includegraphics[width=\linewidth]{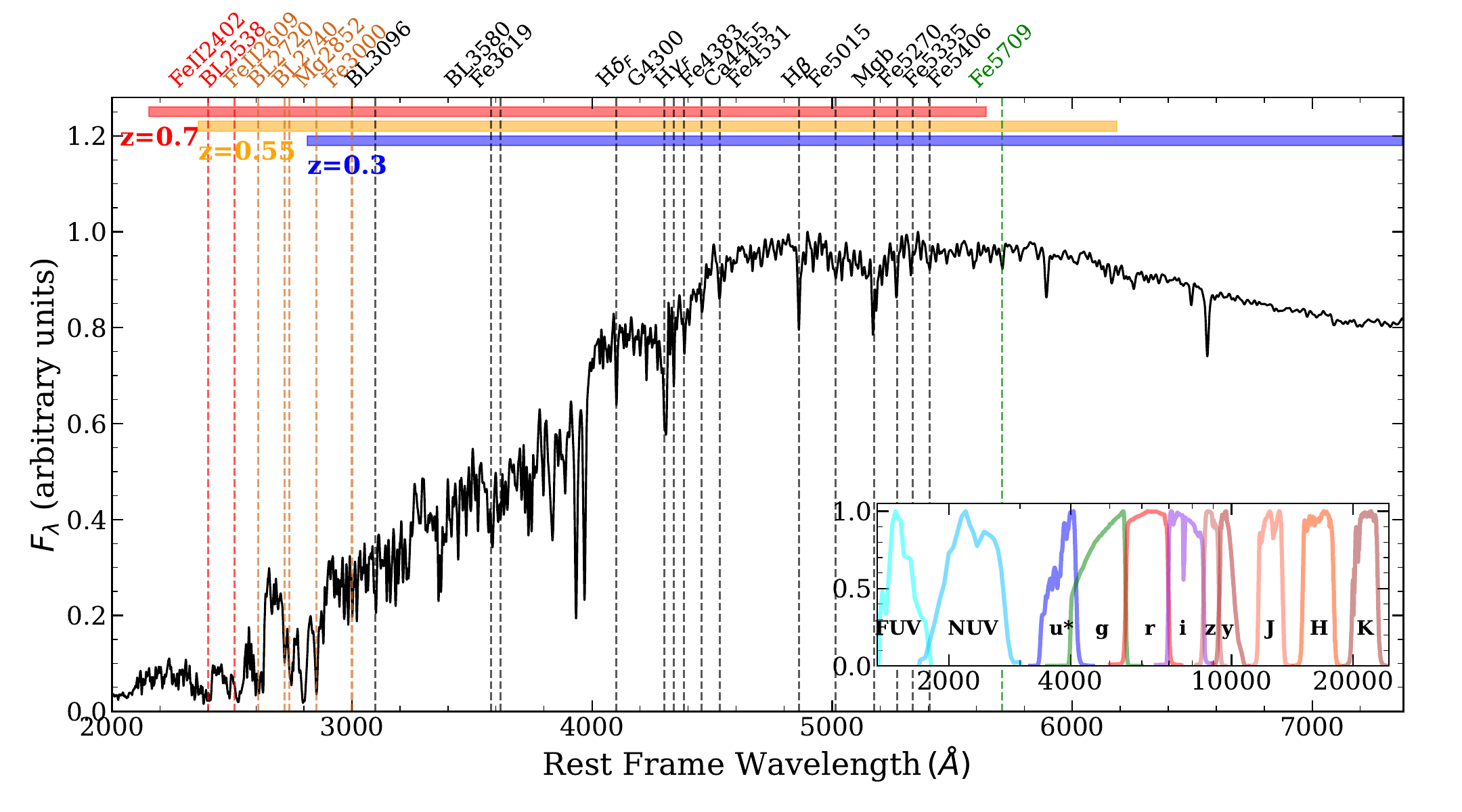}
    \caption{Example of a normalised spectrum for one of our templates. 
    The vertical lines show the central wavelength in the definition of the spectral indices used here. The horizontal bar at the top shows the wavelength coverage available for the WEAVE spectrograph at $z=0.3$ (blue), 0.55 (yellow), and 0.7 (red). The black lines show spectral indices available for all three redshifts, while the green line indicates indices measurable only at $z=0.3$ and 0.55. The orange and red lines show the spectral indices available at $z=0.55$ and 0.7 and only at $z=0.7$, respectively. While not shown here, we also measured the 4000\AA\ break.\ We note that while some spectral indices have their central bandpass within the WEAVE wavelength coverage at a given redshift, they still cannot be used as their blue or red bandpass falls at the very border or outside the wavelength range available to WEAVE at a given redshift \citep[see][and references within for the spectral index definitions]{Ditrani:2023}.The inset shows the filters we used to calculate the observed magnitudes.}
    \label{fig:ML_Fig2}
\end{figure*}

We considered two sets of observables from each template: spectral indices and photometric magnitudes, observables that will be available for the upcoming WEAVE-StePS data. These quantities are computed at three redshifts, $z=0.3$, 0.55, and 0.7. Figure \ref{fig:ML_Fig2} shows a typical spectrum of an old stellar population with all of the observables available. We selected a particular set of indices available in the UV and optical, following the reasoning outlined in \citet{Costantin:2019} and Table 1 of \citet{Ditrani:2023}. To summarise this reasoning, the optical spectral indices were selected specifically as they are sensitive to the age and metallicity of the population but not affected by other physical phenomena, such as the specific elements abundance, the chromospheric emission from the stellar atmosphere, flux calibration, or being highly sensitive to the initial mass function. In the UV part of the spectrum, all spectral indices are considered except Mgwide, which is heavily affected by flux calibration. While D$_n$(4000) is also affected by flux calibration, it is less so compared to Mgwide as it is measured over a shorter wavelength range. However, we still accounted for uncertainties due to flux calibration (see Sect.~\ref{sec:ResNoise}). 

The photometric data consists of the observed magnitudes covering the observed wavelength range from the UV to the NIR (see Sect.~\ref{sec:WEAVE_Photo}). The inset of Fig.~\ref{fig:ML_Fig2} shows the filters that we used (GALEX FUV, GALEX NUV, CFHT u$^\star$, HSC g, r, i, z, y, VISTA J, H, and Ks). 
Observations in all these filters are available in the COSMOS field \citep{Scoville:2007}, which will be observed by WEAVE-StePS \citep{Iovino:2023}.

\subsubsection{Simulated WEAVE-StePS spectra}
\label{sec:WEAVE_Cos}
This section briefly describes how we obtained simulated WEAVE-StePS spectra at each redshift value and S/N value used in our analysis. Full details can be found in \citet{Costantin:2019}. 

To create a set of realistic WEAVE-StePS observed spectra, we started by convolving each template spectrum (see Sect.~\ref{sec:TemplLib}) with a common velocity dispersion of $\sigma_\star=200\,$km/s, well suited to the massive tail of the galaxy stellar mass distribution covered by WEAVE-StePS \citep[see below and][]{Zahid:2016}. The convolved templates are then shifted to the observed wavelengths corresponding to the redshift value considered and trimmed to cover the WEAVE LR-MOS mode wavelength range, including the WEAVE CCD gaps.
These `observed' model spectra are converted from flux units to counts on the CCD, using the total WEAVE throughput, which includes the signal lost due to atmospheric transmission and the optics of the WHT and the WEAVE spectrograph.
The expected noise due to the WEAVE CCD detectors and the Poisson noise of the sky background are added in quadrature to the Poisson noise present in the signal itself in such a way as to obtain the desired S/N in the I band. 
Finally, the spectra are converted back into flux units.\ We note that this methodology results in different S/Ns depending on the considered wavelength. 

In our analysis, we considered three different signal-to-noise ratios and redshifts, chosen according to values expected within WEAVE-StePS \citep{Iovino:2023}. 
Most galaxies are expected to be observed at $z=0.3$, $\mathrm{N_{gal}}\sim 550 \pm 60$ deg$^{-2}$ per $\Delta z=0.1$ and have $\log\,$M$_\star/$M$_\odot \gtrsim 10.2$. The distribution of galaxies observed by the WEAVE-StePS will extend to $z\sim0.8$, but $z=0.7$ with $\mathrm{N_{gal}}\sim 80 \pm 20$ deg$^{-2}$ is the highest redshift that still contains a statistically good sample of galaxies. As a third redshift value we chose $z=0.55$, a representative mid-point value of the redshift distribution, with $\mathrm{N_{gal}}\sim310\pm40$ deg$^{-2}$ per $\Delta z=0.1$.  
At $z=0.55$ we will observe galaxies with stellar mass, $\log\,$M$_\star/$M$_\odot \gtrsim 11.0$, while at 
$z=0.7$, the sample will have $\log\,$M$_\star/$M$_\odot \gtrsim 11.3$.
At $z=0.3$, we have $\sim105\,700$ simulated spectra at each S/N level. 
This number decreases to $\sim 97\, 000$ and $\sim 89\, 000$ at $z=0.55$ and $z=0.7$, respectively, 
as we excluded all galaxies that are older than the age of the Universe.

Using the simulated WEAVE-StePS spectra, we measured the set of UV and optical spectral indices available in the spectral ranges shown in Fig.~\ref{fig:ML_Fig2}. To assess the impact of measurement errors, we generated 1000 Monte Carlo realisations for each spectrum according to the error model described above for a given S/N. 
For each realisation, we computed the spectral indices. We then computed the statistical error on the indices from the standard deviations of the 1000 realisations. We note that D$_n$(4000) will be sensitive to the spectro-photometric calibration; therefore, to account for such systematics, we added a 5\% lower limit on its uncertainty.

\subsubsection{Simulated photometry}
\label{sec:WEAVE_Photo}
From the model templates, we calculated the observed magnitudes in each filter bandpass by transforming the spectra to the corresponding redshift and convolving the model spectrum with the correct filter bandpass. For the uncertainty in these observed magnitudes, we assigned the photometric ancillary data that will be available in WEAVE-StePS to each simulated galaxy \citep{Iovino:2023}. In particular, we considered the photometric catalogues covering the COSMOS field from the UV to NIR. To assign uncertainties to the simulated photometric measurements from our templates, we used the COSMOS catalogue \citep{Laigle:2016} to fit a relation between the observed magnitudes and their formal uncertainties in each band.  We normalised the templates to have I$_{AB}$ = 20.5, which is the WEAVE-StePS magnitude limit.\ We note that we added an error of 0.05\,mag in quadrature to set a lower limit on the uncertainty to account for possible systematics.

\section{Results}
\label{sec:results}
A well-known issue in retrieving the stellar population properties of galaxies is the age-metallicity degeneracy, especially for the older population \citep{Worthey:94}, which is further complicated by the presence of dust. This is largely due to the fact that different combinations of age, metallicity, and dust produce similar spectral shapes, making it impossible to disentangle the components unless we have a sufficient S/N and a large enough wavelength range. Therefore, in this section, we first examine how well ML algorithms retrieve the physical parameters on perfect observables without measurement errors. We then analyse how well the ML algorithms can predict the same physical parameters for our simulated galaxies containing realistic noise. Regarding the spectral indices, we performed this test for S/N levels of 10, 20, and 30. In each case, the noise in the photometric data was fixed to that of the WEAVE-StePS COSMOS ancillary catalogue. For each S/N level, we also tested the retrieval capability of the ML algorithms at three different redshifts, $z = 0.3$, 0.55, and 0.7, respectively.

\subsection{Simulated data with no errors}
\label{sec:SimNoError}
For the initial test case, we used the observables without noise.  We quantified the success of the ML algorithms using either the bias and the Spearman correlation coefficient (scc) when considering noiseless data or the bias, dispersion ($\sigma$), and the fraction of outliers when dealing with noisy data (see the next section).\footnote{We have two sets of statistics, one for perfect simulated data and one for simulated data with noise. We do not quantify $\sigma$ or fraction of outliers with perfect simulated data as this leads to non-physical interpretations of results.} 
The bias is estimated as $\bar{x} = {\rm median}(\Delta x_i)$, where $\Delta x_i = x^{pred}_i - x^{true}_i$. For the dispersion, we used the median absolute deviation (MAD) estimate and it is calculated using the formula
\begin{equation}
    \sigma = 1.4826 \times \text{MAD}(\Delta x_i),
    \label{eq:Dispersion_In}
\end{equation}
where ${\rm MAD}(\Delta x_i) = {\rm median}(|x_i - {\rm median}(x_i)|)$. In the following, we present the dispersion computed in moving average bins, where the bin size is larger than the bin step. The bin and step sizes change depending on the physical parameter, as the range of minimum and maximum values changes.
Finally, we considered a data point an outlier when $\Delta x_i > 2\sigma$. 

\begin{figure*}
    \centering
    \includegraphics[width=\linewidth]{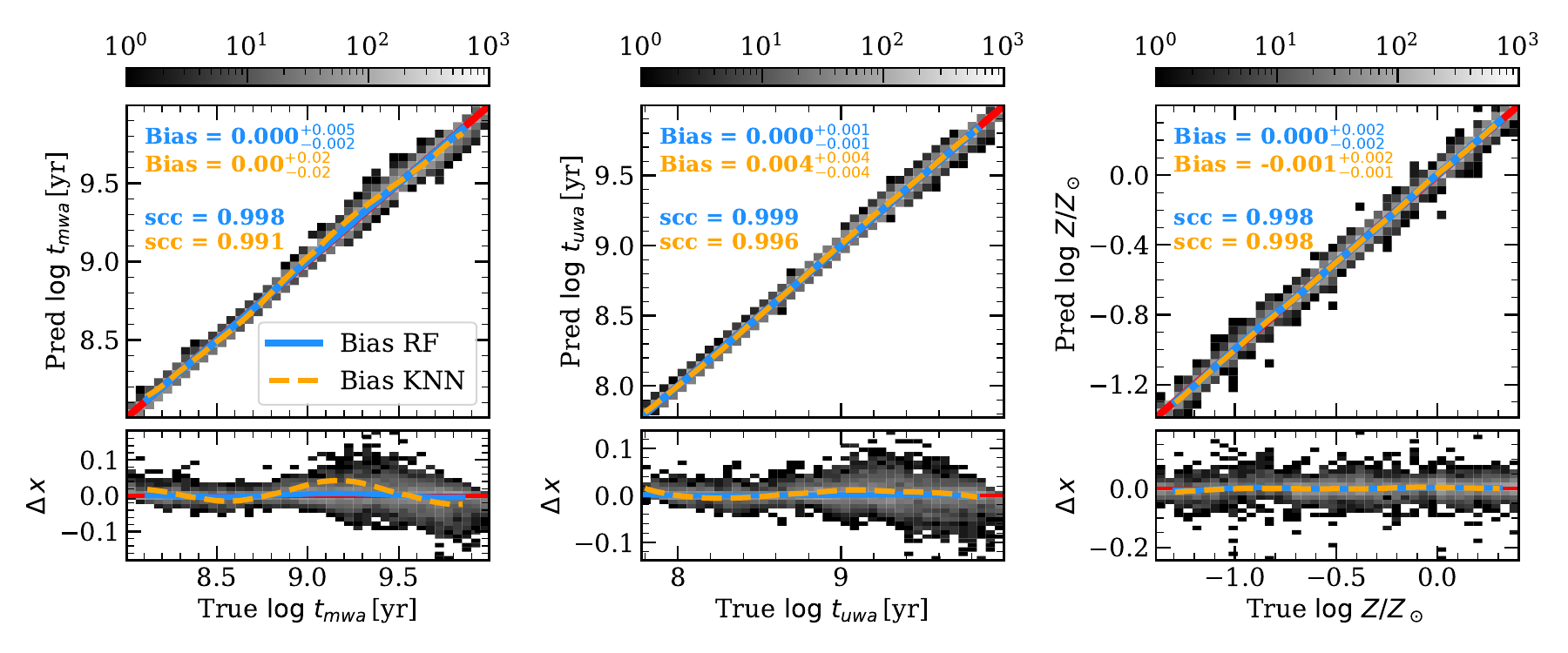}
    \includegraphics[width=\linewidth]{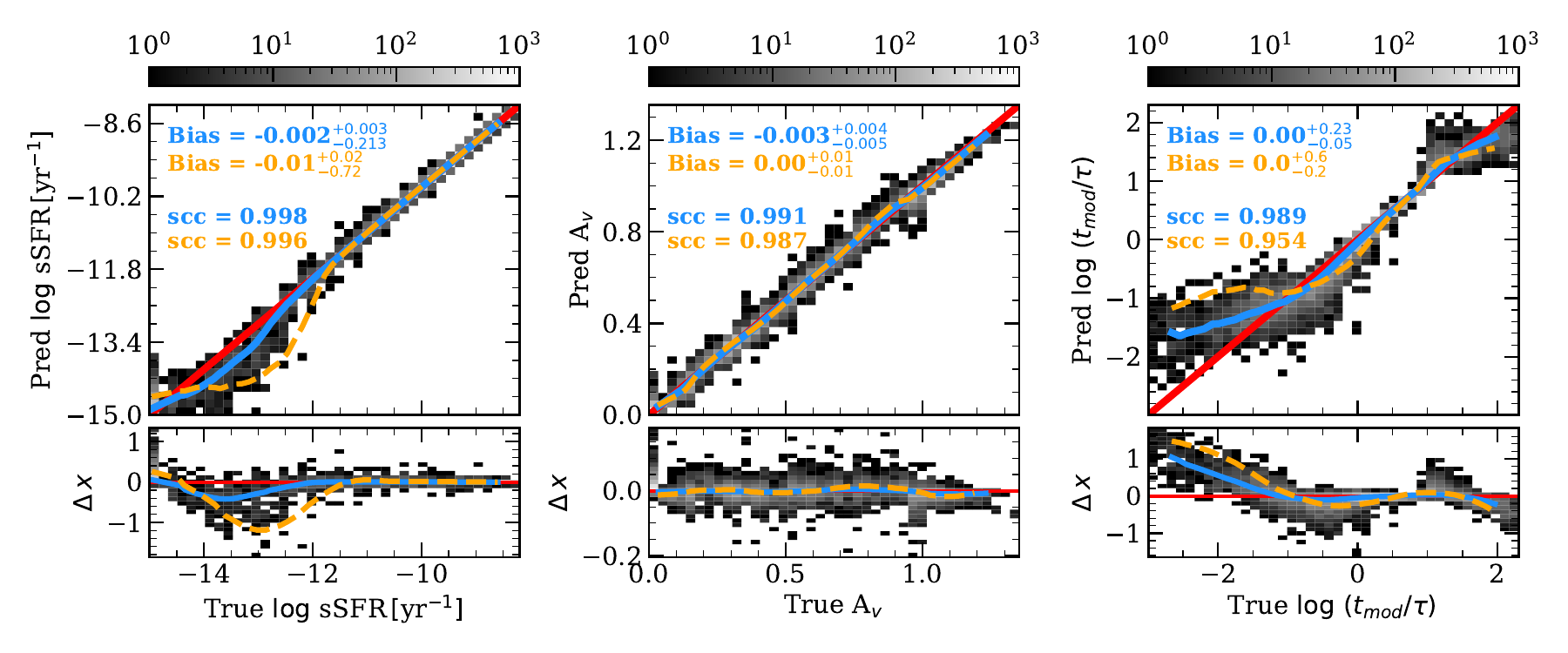}
    \caption{ Retrieval of the physical parameters assuming no noise on data. \textit{Top panels}: Density plots showing predicted vs true values for the 
    mwa (left), uwa (middle), and metallicity (right) derived using RF. The blue lines show the bias computed as a running average with fixed bin step and size. The bin size is six times larger than the bin step, and therefore the same galaxies are present in multiple bins; this ensures we have a sufficient number of galaxies in each bin. The blue labels refer to the median bias and Spearman's correlation coefficient (label scc in the plot) 
    for RF. For comparison purposes, we have included a line showing the bias for KNN and the values of the median bias and the scc in orange.
    The errors quoted with bias are the 25$^{\rm th}$ and 75$^{\rm th}$ percentile. The smaller, lower panels show the difference between predicted and true values. The red line always shows the one-to-one line or the 0 value for the variable considered. The colour bar at the top indicates the number of galaxies within a pixel in the density plot. \textit{Bottom panels}: Same results but for the sSFR (left), dust attenuation, A$_{\rm V}$ (middle), and t$_{\rm mod}/\tau$ (right).
   }
    \label{fig:ML_Fig3}
\end{figure*}

Figure~\ref{fig:ML_Fig3} shows the comparison between predicted values and true values for the mwa
(top left panel) and uwa (top middle panel), metallicity (top right panel), 
sSFR (bottom left panel), A$_{\rm V}$ (bottom middle panel), and $\log (t_{\rm mod}/\tau)$ (bottom right panel) 
assuming no noise on the observables. 
The values quoted are the bias and Spearman's correlation coefficient (scc).\ We note that the quoted bias value is the median bias across all bins. The uncertainty values represent the 25$^{\rm th}$ and 75$^{\rm th}$ percentiles of the distribution, showing the scatter in bias in different bins. We do not present the dispersion values since they are negligible in this case of perfect data. Hence, we also excluded the outlier fractions.
This low scatter is demonstrated by the high scc values calculated for the physical parameters, as discussed below. 
There is a good prediction for both the mwa and uwa, 
metallicity and A$_{\rm V}$, 
as the bias is low, with both the median bias and the percentiles close to 0. 
In addition, the scc is high $\sim 0.99$, indicating a strong correlation between true and predicted values with little scatter. Looking at the difference between predicted and true (bottom sub-panels), we find similar results, but it is evident that RF consistently outperforms KNN. Our retrieval capability of sSFRs is robust at high specific star formation, as the bias values are $\sim -0.002$ and $-0.01$,  but at very low specific star formation, $\log\,$sSFR$\lesssim -12$ and $-11$, there is a clear bias for RF and KNN, respectively, with KNN having a stronger bias. 
This bias may be because there is limited variance in the observed parameters of templates at low sSFRs, thus creating a high degeneracy even for perfect data. Therefore, when averaging over 100 trees/neighbours, the high level of degeneracy in observed parameters for low sSFRs will result in the selection of numerous galaxies with lower sSFRs, thus underestimating the sSFR.

Finally, we have $\log t_{\rm mod}/\tau$, which can be considered a second-order physical 
parameter that indicates the number of e-folds since the peak of star formation. We find that the retrieval of this parameter is biased due to degeneracies that exist at low and high values. At very low values,  
$\log t_{\rm mod}/\tau < -1$, the galaxies are actively forming stars and are
either very young, with $t_{\rm mod}\sim0$\,Gyr, or have a long decay time, $\tau$. 
For these effectively young stellar populations, the observables are relatively insensitive to the precise value of the scale time $\log t_{\rm mod}/\tau$. At the low end, the bias is +0.25 for RF and +0.6 for KNN.
At high values, $\log t_{\rm mod}/\tau > 1$, star formation has ended, and the stellar populations are evolving passively. In this regime, the observables are not strongly sensitive to the scale time, and the bias is $-0.06$ for RF and $-0.2$ for KNN.

\subsection{Simulated data with noise}
\label{sec:ResNoise}
In the following subsections, we explore how the parameter estimation performs on realistic data. Here, we introduce various levels of noise on the spectral indices (S/N$_{\rm I,obs}$=10, 20, and 30) following the steps described in Sect.~\ref{sec:WEAVE_Cos}.\ We note that we do not change the errors in the photometry since it will come from archival data of the COSMOS field with known photometric errors. In addition, we also tested the performance of the algorithms to predict physical parameters at three different redshifts, $z=0.3$, 0.55, and 0.7. At each redshift we set the maximum age of each template to be the age of the Universe ($t_{\rm mod}=10.27\,$Gyr at $z=0.3$, $t_{\rm mod}=8.27\,$Gyr at $z=0.55$ and $t_{\rm mod}=7.34\,$Gyr at $z=0.7$). Finally, while 
the metallicity of the templates covers the range $-1.69\lesssim \log\,Z/Z_\odot \lesssim +0.6$, we restricted it for the testing sample to $-1.39\lesssim \log\,Z/Z_\odot \lesssim +0.4$ to avoid significant bias at the lowest and highest metallicity values.

\subsubsection{Results for $z=0.3$ and S/N=10} 
\label{sec:fidres}
This sample will define the baseline performance. The ML algorithms are trained on the simulated sample with noise. As neither KNN nor RF directly take the uncertainty as inputs, we perturbed the simulated spectrum (with noise) five times, in a Gaussian manner, and re-measured the spectral indices. Similarly, we also perturbed the photometric data five times in accordance with the noise. This procedure grows the training sample five-fold; the testing sample size was not modified, but we also perturbed the testing sample once according to the noise.

\begin{figure*}
    \centering
    \includegraphics[width=0.32\linewidth]{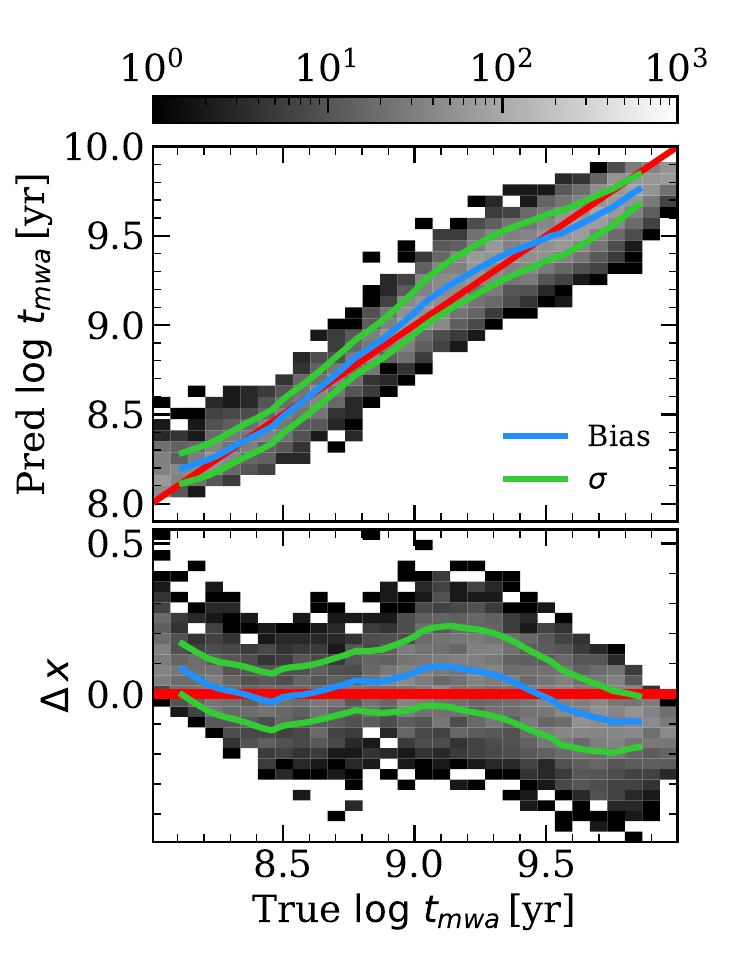}
    \includegraphics[width=0.32\linewidth]{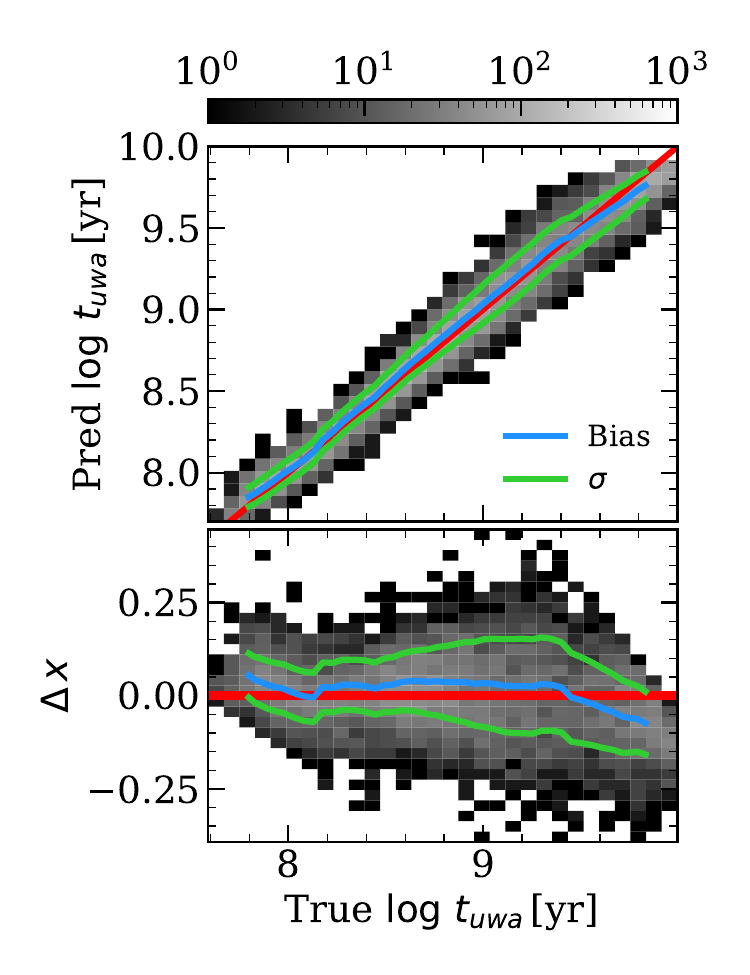}
    \includegraphics[width=0.32\linewidth]{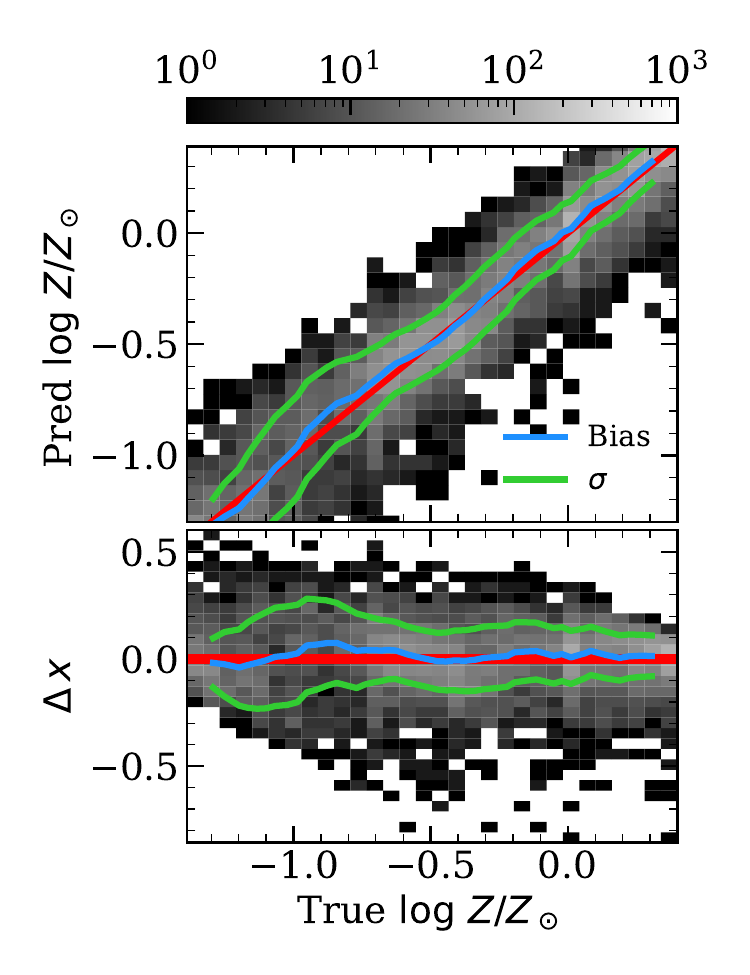}
    
    \includegraphics[width=0.32\linewidth]{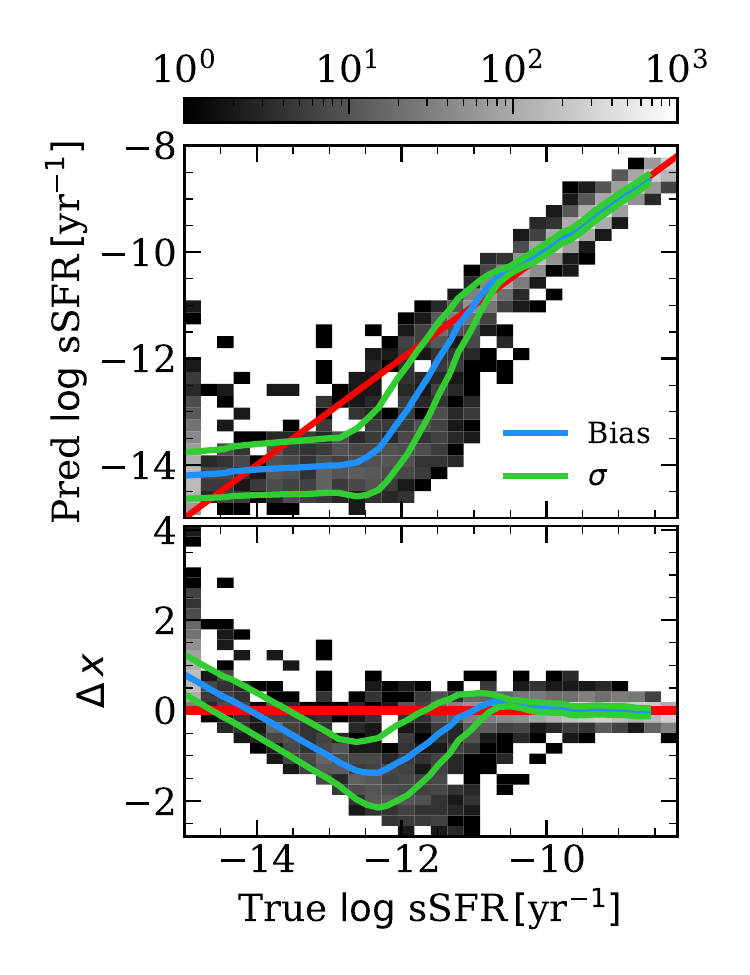}
    \includegraphics[width=0.32\linewidth]{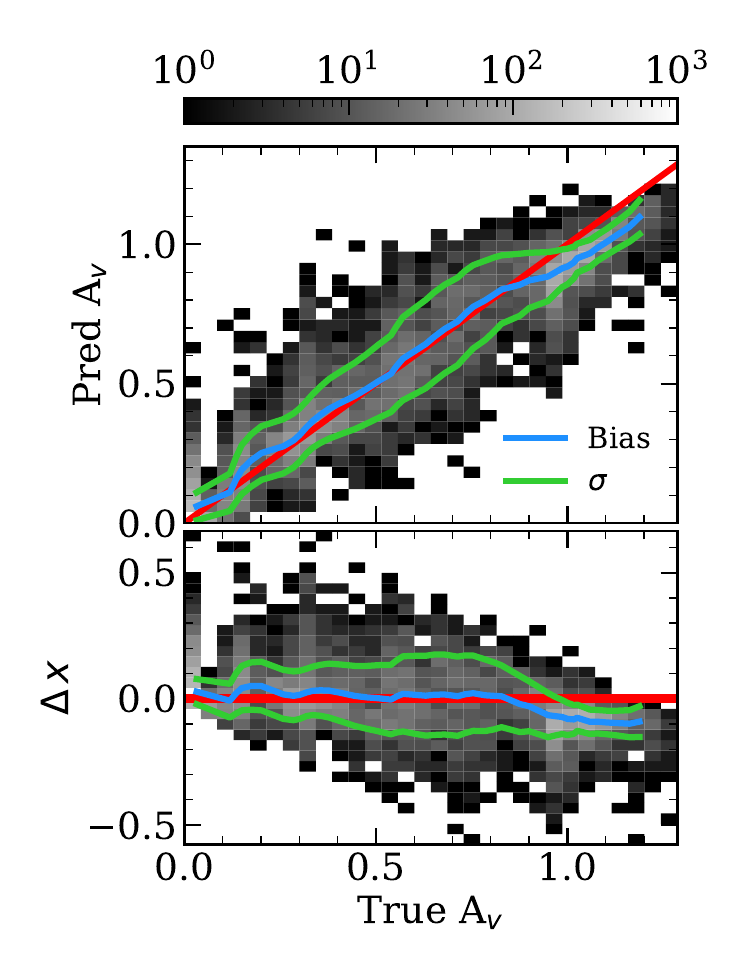}
    \includegraphics[width=0.32\linewidth]{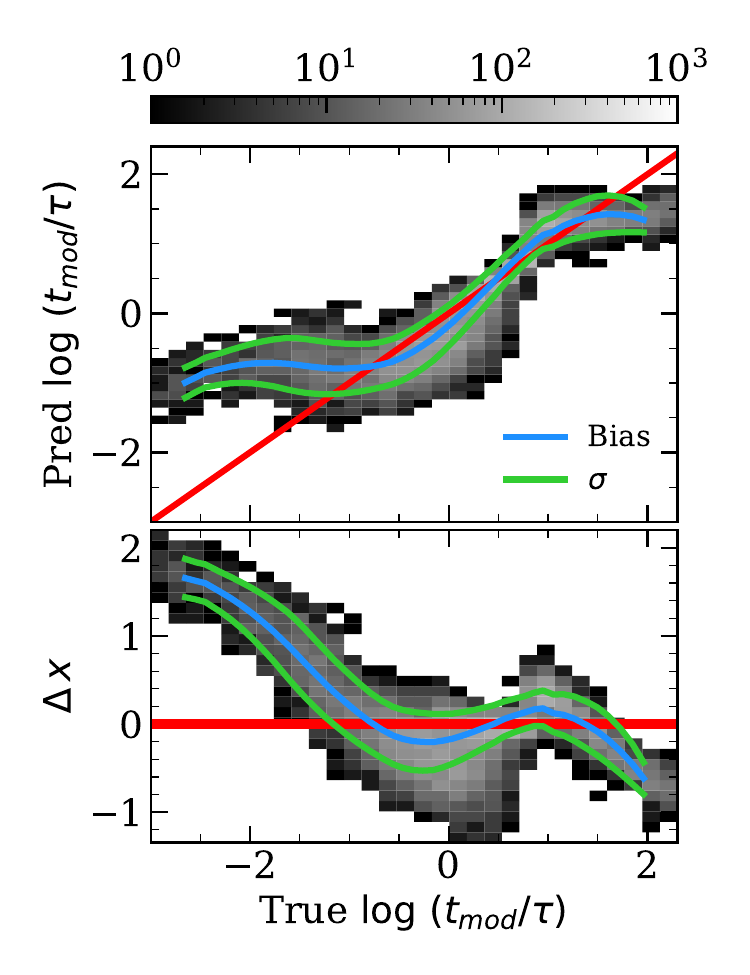}
    \caption{Physical parameters derived using RF, assuming S/N=10 at $z=0.3$. 
    \textit{Top}: Results for the mwa (left), uwa (centre), and metallicity (right). \textit{Bottom}: sSFR, A$_{\rm V}$, and 
    $t_{\rm mod}/\tau$, from left to right. Each sub-panel on the bottom shows the difference between predicted and true values ($\Delta x$). The blue lines show the bias estimated using 
    moving averages. Similarly, the lime-green lines show the $\sigma$. The red 
    line shows the one-to-one line (top sub-panels) or 0 value (bottom sub-panels). Finally, the colour bar shows the number of galaxies in the density plot.}
    \label{fig:fidfig1}
\end{figure*}

Figure~\ref{fig:fidfig1} shows the results obtained for the mwa, uwa, metallicity, sSFR, A$_{\rm V}$, and $\log\,t_{\rm mod}/\tau$ using the RF algorithm.
The retrieval capability worsens compared to when there were no errors on the spectro-photometric observables; however, most of the physical parameters have 
neither a large bias nor scatter. 
For the mwa, the bias reaches 0.1 dex at most.
For the uwa, we find a similar trend; however, in general, both the bias and $\sigma$ are smaller. 
The metallicity shows a negligible bias $<0.1$ dex. This is particularly true for high metallicity, $\log Z/Z_\odot \gtrsim +0.0$, where the bias and the dispersion tend to be the lowest. The A$_{\rm V}$ also has a small bias, straddling the 0 value, except for A$_{\rm V} \gtrsim 1.0$, where there is a general underestimation for the effective dust attenuation, by $\sim 0.1\,$mag.
The sSFR also has low bias and $\sigma$ for $\log\,$sSFR$\gtrsim -11.0$. However, for simulated galaxies considered quiescent, at $\log\,$sSFR$\lesssim -11.0$, we find a strong bias and large $\sigma$. For 
$\log\,$t$_{\rm mod}/\tau$ we generally find high bias and dispersion. Although we observe an increase in the bias for most physical parameters, it is not significantly different from the test case where the observables have no error, as in most instances, the bias is still close to zero. Finally, we observe no clear difference in the relationship between the retrieval capability of the physical parameters and the age group of galaxies, where  we classified the galaxies into three groups based on their stellar populations: young ($t_{\rm mwa} \leq 1 $ Gyr), intermediate ($1 \leq $t$_{\rm mwa} \leq 4,$Gyr), and old (t$_{\rm mwa} \geq 4,$Gyr).\ We note that the main exception to this trend is the metallicity, where the older stellar population exhibits smaller $\sigma$ values for low metallicities in comparison to the younger and intermediate stellar populations. This trend was also observed in a previous study by \citet{Ditrani:2023} using simulated galaxies and also in real data (see Fig.~7 of \citealt{Gallazzi:2005}).

We note that the results discussed in this section use the mean of the 100 trees or of the 100 nearest neighbours -- as given by the output of the \texttt{scikit-learn python} package. While it is possible to extract the median using additional functions for RF, for KNN, this requires modifying the source code itself, which is beyond the scope of this paper. Therefore, we continued to use the mean estimates of 100 trees/neighbours for a fairer comparison between the two algorithms. Nevertheless, we tested how much the results vary when we use the mean and median of 100 trees for RF. Using the median gives results comparable to the mean except at extremes, the minimum and maximum bins along their true value, where the bias is slightly smaller. For example, there is a difference of $\lesssim 0.03\,$dex for the mwa and uwa, and $\lesssim 0.09\,$mag for A$_{\rm V}$ when using the median rather than mean at the boundaries. For metallicity, the difference between bias and $\sigma$ is negligible. Finally, for sSFRs, while the degradation of the sSFR estimates begins at $\log\,$sSFR$\lesssim -11.2$, when using the median, the bias values are larger, as expected, since we imposed a lower limit of $-15.0$. 

Table~\ref{tab:RF_KNN_SN} presents the median bias, $\sigma$ and the fraction of 
outliers for RF and KNN. The uncertainties given are the 
25$^{\rm th}$ and 75$^{\rm th}$ percentiles, showing the variation in both bias and $\sigma$ for 
different bins. While the median bias is comparable to that obtained in the no-noise case, 
$\sim 0.00$, the bias variations are generally larger as seen by the quartile ranges, 
reflecting what is seen in Fig.~\ref{fig:fidfig1}. 
For RF (KNN), both the mwa and uwa have a median 
$\sigma \leq 0.10\,(0.16)\,$dex, and metallicity has a median $\sigma$ of $\sim 0.14\,(0.22)\,$dex. The median bias for the sSFR is relatively low $-$0.03 ($-$0.1)$\,$dex; however, we see that the scatter on the dispersion for different bins is much larger, with 25$^{\rm th}$ and 75$^{\rm th}$ percentile values of $-$0.80 ($-$0.70)$\,$dex and 0.05 (0.20)$\,$dex, respectively. The A$_{\rm V}$ has a lower median bias for RF, whereas for KNN, we see a typical bias of 0.1\,mag. Similarly, KNN has a larger scatter than RF by $\sim0.04\,$mag.  
Finally, $\log\, t_{\rm mod}/\tau$ is found to be the worst constrained parameter as it is the one with the highest bias, $\sim -0.1 (-0.2)\,$dex, and $\sigma$, $\sim 0.28 (0.27)\,$dex. Overall, RF consistently performs better than KNN, as seen in previous literature in other contexts \citep{Thanh:2018}.\ We note that the main exception is in the fraction of outliers, where RF always has a slightly larger outlier fraction. The fraction of outliers is influenced by two factors: how compact the distribution of $\Delta x$ is in a given bin and how Gaussian it is. For RF, we typically find a strongly peaked distribution with asymmetric tails. This leads to a low $\sigma$ but a high fraction of outliers. On the other hand, KNN has a broader peak, resulting in higher $\sigma$, and reduced tails. In this case, the $\Delta x$ distribution is closer to Gaussian, giving a lower fraction of outliers despite having a larger $\sigma$.

{\renewcommand{\arraystretch}{1.30}
\begin{table*}
  \centering \caption{Average statistics in all bins at different combinations of redshifts and S/Ns.}
  \label{tab:RF_KNN_SN}
    \begin{tabular}{lcccccc} 
    \hline
    \multicolumn {1}{c}{} & \multicolumn{3}{c}{Random Forest} &
    \multicolumn{3}{c}{K-Nearest Neighbour}    \\  
    \hline 
    Physical Parameter & Bias & $\sigma$ &  f$_{outlier}$ ($\%$) & Bias & $\sigma$ &  f$_{outlier}$ ($\%$) \\
    \hline
    \multicolumn{7}{c}{$z=0.3$, S/N=10} \\
    \hline
    $\log\, t_{\rm mwa}\,$[yr] &   0.00$_{-0.03}^{+0.05}$ & 0.10$_{-0.01}^{+0.02}$ & 4.94 & $-$0.01$_{-0.07}^{+0.08}$ & 0.16$_{-0.04}^{+0.03}$ & 3.15 \\
    $\log\, t_{\rm uwa}\,$[yr] &  0.01$_{-0.01}^{+0.01}$ & 0.08$_{-0.02}^{+0.02}$ & 5.38 &  $-$0.03$_{-0.02}^{+0.02}$ & 0.11$_{-0.01}^{+0.02}$ & 4.74 \\
    $\log\, Z/Z_\odot$ &  0.01$_{-0.03}^{+0.01}$ & 0.14$_{-0.02}^{+0.02}$ & 7.37 &  0.00$_{-0.01}^{+0.05}$ & 0.20$_{-0.02}^{+0.01}$ & 5.87 \\
    $\log\,$sSFR$\,[\mathrm{yr^{-1}}]$ & $-$0.03$_{-0.80}^{+0.05}$ & 0.4$_{-0.3}^{+0.1}$ & 9.28 &  $-$0.1$_{-0.7}^{+0.2}$ & 0.45$_{-0.26}^{+0.09}$ & 7.90 \\
    A$_{\rm V}$ & 0.00$_{-0.08}^{+0.01}$ & 0.09$_{-0.02}^{+0.04}$ & 11.80 &  $-$0.1$_{-0.2}^{+0.1}$ & 0.13$_{-0.02}^{+0.02}$ & 5.06 \\
    $\log\, t_{\rm mod}/\tau$ & 0.1$_{-0.2}^{+0.7}$ & 0.27$_{-0.06}^{+0.05}$ & 7.40 & 
     0.1$_{-0.3}^{+0.9}$ & 0.23$_{-0.03}^{+0.07}$ & 5.08 \\
    \hline
    \multicolumn{7}{c}{$z=0.3$, S/N=20} \\
    \hline
    $\log\, t_{\rm mwa}\,$[yr] &  0.01$_{-0.04}^{+0.03}$ & 0.09$_{-0.01}^{+0.03}$ & 5.63 &  $-$0.01$_{-0.07}^{+0.07}$ & 0.14$_{-0.04}^{+0.01}$ & 3.16 \\
    $\log\, t_{\rm uwa}\,$[yr] & 0.00$_{-0.01}^{+0.01}$ & 0.07$_{-0.01}^{+0.03}$ & 6.63 &  $-$0.02$_{-0.01}^{+0.02}$ & 0.09$_{-0.01}^{+0.02}$ & 4.88 \\
    $\log\, Z/Z_\odot$ & 0.01$_{-0.02}^{+0.01}$ & 0.13$_{-0.02}^{+0.02}$ & 7.51 &  0.02$_{-0.02}^{+0.02}$ & 0.16$_{-0.02}^{+0.02}$ & 6.14 \\
    $\log\,$sSFR$\,[\mathrm{yr^{-1}}]$ &  $-$0.03$_{-0.86}^{+0.04}$ & 0.4$_{-0.4}^{+0.1}$ & 9.95 &  $-$0.1$_{-0.9}^{+0.1}$ & 0.42$_{-0.28}^{+0.05}$ & 9.15 \\
    A$_{\rm V}$ &  0.00$_{-0.07}^{+0.01}$ & 0.08$_{-0.02}^{+0.03}$ & 11.63  &  $-$0.01$_{-0.12}^{+0.03}$ & 0.11$_{-0.02}^{+0.02}$ & 6.49 \\
    $\log\, t_{\rm mod}/\tau$ &  0.0$_{-0.2}^{+0.7}$ & 0.27$_{-0.06}^{+0.06}$ & 7.79 &  0.1$_{-0.3}^{+0.9}$ & 0.23$_{-0.04}^{+0.05}$ & 5.50 \\
    \hline
    \multicolumn{7}{c}{$z=0.3$, S/N=30} \\
    \hline
    $\log\, t_{\rm mwa}\,$[yr] &  0.00$_{-0.03}^{+0.03}$ & 0.08$_{-0.01}^{+0.02}$ & 6.64 & $-$0.01$_{-0.06}^{+0.06}$ & 0.11$_{-0.02}^{+0.02}$ & 3.19 \\ 
    $\log\, t_{\rm uwa}\,$[yr] & 0.00$_{-0.01}^{+0.01}$ & 0.06$_{-0.01}^{+0.02}$ & 6.98 &  $-$0.01$_{-0.01}^{+0.01}$ & 0.08$_{-0.01}^{+0.02}$ & 4.37 \\
    $\log\, Z/Z_\odot$ & 0.01$_{-0.01}^{+0.01}$ & 0.11$_{-0.01}^{+0.02}$ & 6.11 &  0.01$_{-0.01}^{+0.02}$ & 0.12$_{-0.01}^{+0.02}$ & 5.82 \\
    $\log\,$sSFR$\,[\mathrm{yr^{-1}}]$ & $-$0.03$_{-0.83}^{+0.03}$ & 0.4$_{-0.3}^{+0.1}$ & 10.51 &  $-$0.1$_{-0.9}^{+0.1}$ & 0.4$_{-0.3}^{+0.1}$ & 9.30 \\
    A$_{\rm V}$ &  $-$0.01$_{-0.05}^{+0.01}$ & 0.08$_{-0.02}^{+0.02}$ & 11.32 &  $-$0.02$_{-0.09}^{+0.02}$ &
 0.10$_{-0.02}^{+0.02}$ & 6.77 \\
    $\log\, t_{\rm mod}/\tau$ &  0.0$_{-0.2}^{+0.6}$ & 0.26$_{-0.06}^{+0.06}$ & 8.18 &  0.1$_{-0.3}^{+0.9}$ & 0.22$_{-0.04}^{+0.04}$ & 5.46 \\
    \hline
    \multicolumn{7}{c}{$z=0.55$, S/N=10} \\
    \hline
    $\log\, t_{\rm mwa}\,$[yr] &  0.00$_{-0.03}^{+0.03}$ & 0.10$_{-0.01}^{+0.01}$ & 5.67 &  0.01$_{-0.08}^{+0.06}$ & 0.15$_{-0.04}^{+0.03}$ & 3.63 \\
    $\log\, t_{\rm uwa}\,$[yr] & 0.00$_{-0.01}^{+0.00}$ & 0.08$_{-0.01}^{+0.03}$ & 5.72  &  $-$0.03$_{-0.02}^{+0.04}$ & 0.11$_{-0.01}^{+0.03}$ & 4.97 \\
    $\log\, Z/Z_\odot$ & 0.01$_{-0.02}^{+0.02}$ & 0.17$_{-0.03}^{+0.03}$ & 8.20 &  0.02$_{-0.02}^{+0.08}$ & 0.24$_{-0.02}^{+0.01}$ & 4.73 \\
    $\log\,$sSFR$\,[\mathrm{yr^{-1}}]$ & 0.00$_{-0.79}^{+0.02}$ & 0.3$_{-0.2}^{+0.2}$ & 9.86 &  0.0$_{-0.9}^{+0.2}$ & 0.5$_{-0.3}^{+0.1}$ & 8.17 \\
    A$_{\rm V}$ & 0.01$_{-0.07}^{+0.01}$ & 0.08$_{-0.02}^{+0.03}$ & 12.18 & $-$0.01$_{-0.17}^{+0.09}$ &
 0.13$_{-0.01}^{+0.01}$ & 4.74 \\
    $\log\, t_{\rm mod}/\tau$ &  0.0$_{-0.2}^{+0.7}$ & 0.25$_{-0.05}^{+0.07}$ & 8.73 & 0.0$_{-0.3}^{+0.9}$ & 0.22$_{-0.02}^{+0.07}$ & 5.40 \\
    \hline
    \multicolumn{7}{c}{z=0.7, S/N=10} \\
    \hline
    $\log\, t_{\rm mwa}\,$[yr] & 0.00$_{-0.02}^{+0.03}$ & 0.09$_{-0.01}^{+0.01}$ & 5.93 &  0.00$_{-0.10}^{+0.06}$ & 0.13$_{-0.04}^{+0.02}$ & 4.01 \\
    $\log\, t_{\rm uwa}\,$[yr] & 0.00$_{-0.01}^{+0.01}$ & 0.07$_{-0.00}^{+0.02}$ & 6.02 &  $-$0.03$_{-0.02}^{+0.05}$ & 0.10$_{-0.02}^{+0.03}$ & 4.39 \\
    $\log\, Z/Z_\odot$ & 0.02$_{-0.02}^{+0.01}$ & 0.16$_{-0.02}^{+0.03}$ & 8.32 &  0.04$_{-0.07}^{+0.09}$ & 0.24$_{-0.01}^{+0.00}$ & 5.00 \\
    $\log\,$sSFR$\,[\mathrm{yr^{-1}}]$ &  0.00$_{-0.61}^{+0.02}$ & 0.3$_{-0.2}^{+0.3}$ & 9.39 &  0.0$_{-0.9}^{+0.1}$ & 0.5$_{-0.4}^{+0.1}$ & 9.25 \\
    A$_{\rm V}$ & 0.01$_{-0.07}^{+0.01}$ & 0.07$_{-0.02}^{+0.04}$ & 13.24 &  $-$0.01$_{-0.16}^{+0.08}$ & 0.13$_{-0.02}^{+0.01}$ & 6.11 \\
    $\log\, t_{\rm mod}/\tau$ &  0.0$_{-0.2}^{+0.7}$ & 0.29$_{-0.08}^{+0.04}$ & 9.01 &  0.0$_{-0.3}^{+0.9}$ & 0.22$_{-0.04}^{+0.06}$ & 5.49 \\
    \hline
    \end{tabular}
    \tablefoot{The left and right set of columns show results for RF and KNN, respectively. The first three sets of rows show results at $z=0.3$ for S/N=10, S/N=20, and S/N=30. The bottom two sets of rows show results for S/N=10 at $z=0.55$ and $z=0.7$. The leftmost columns indicate the physical parameters of interest for each set of rows. We estimate the bias, the $\sigma$, and the percentage of outliers for each algorithm and physical parameter. The uncertainty quoted is the 25$^{\rm th}$ and 75$^{\rm th}$ percentiles.}
\end{table*}
}

\subsubsection{Higher S/N at $z=0.3$}
\label{sec:res_snr}
\begin{figure*}
    \centering
    \includegraphics[width=0.32\linewidth]{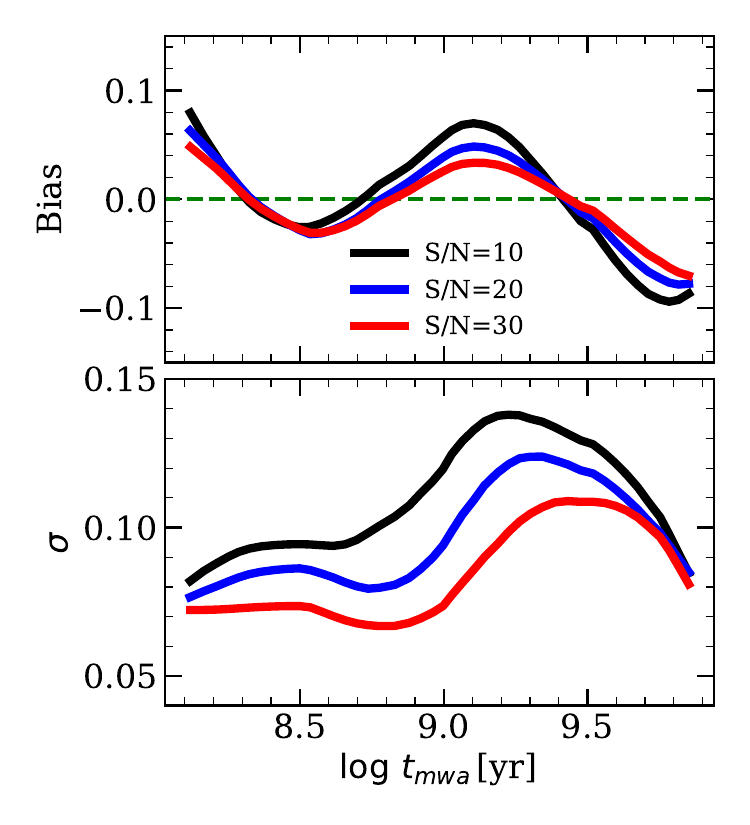}
    \includegraphics[width=0.32\linewidth]{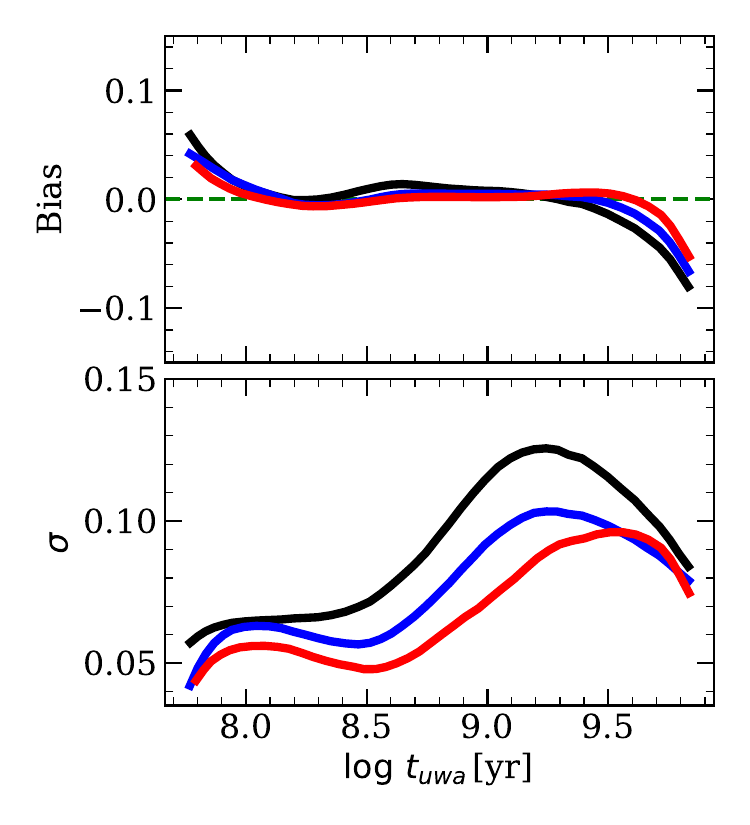}
    \includegraphics[width=0.32\linewidth]{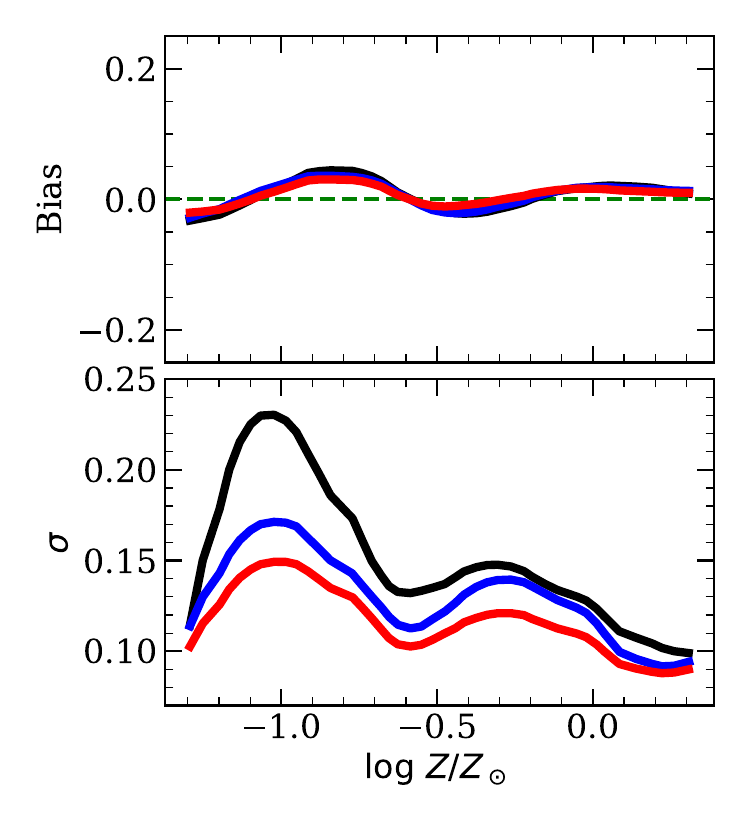}

    \includegraphics[width=0.32\linewidth]{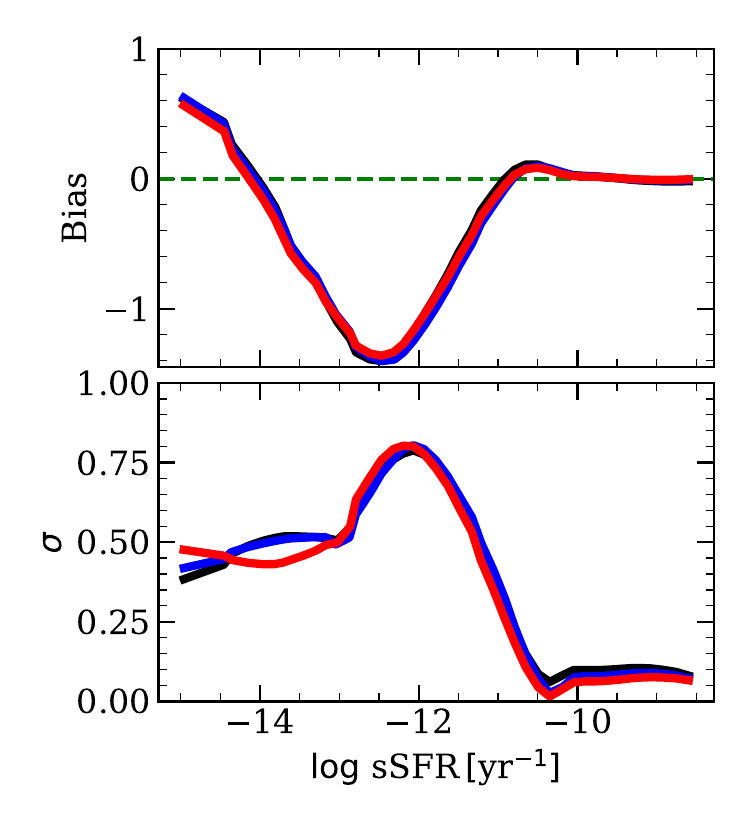}
    \includegraphics[width=0.32\linewidth]{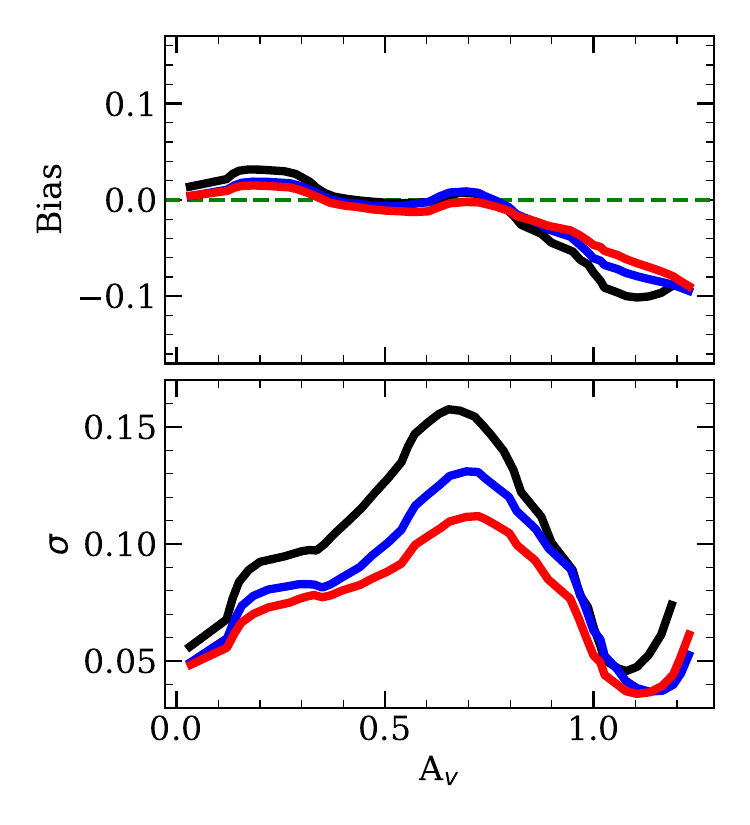}
    \includegraphics[width=0.32\linewidth]{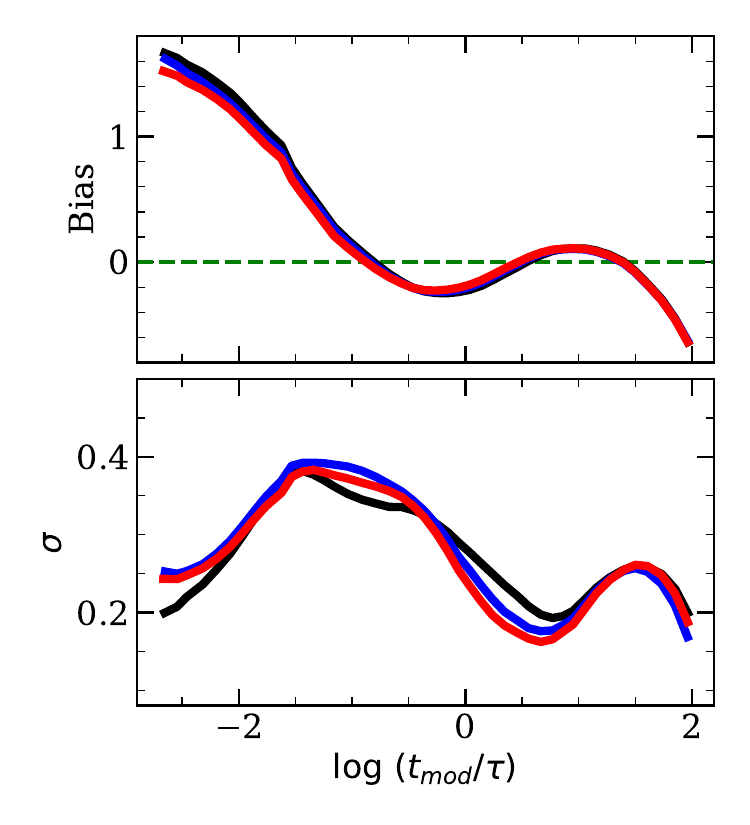}
    \caption{Change in bias (sub-top panels) and $\sigma$ (sub-bottom panels) as a function of 
    the true physical parameter value for RF at $z=0.3$. \textit{Top panels:} Results of, from left to right, 
    the mwa, uwa, and metallicity for S/N$=10$ (black), 20 (blue), and 30 (red). Note that while 
    the S/Ns for the spectral indices change, the uncertainties assumed for the photometric data remain unchanged for the different cases.
    For the bias plot, the dashed green line indicates the zero point.
    \textit{Bottom panels:} Results for the sSFR, A$_{\rm V}$, and $t_{\rm mod}/\tau$.}
    \label{fig:BiasSig_SN}
\end{figure*}

In this section, we analyse how the previous results improve when spectral data are affected by lower noise values. 
For this test, we considered S/N$=20$ and 30 and show S/N$=10$ for comparison. Figure~\ref{fig:BiasSig_SN} shows the bias and $\sigma$, obtained using RF, as a function of the true physical parameter values.  
The black, blue and red lines show 
the values for S/N$=10, 20,$ and $30$, respectively. Overall, we find that improving the 
S/N of the spectra does not significantly improve the bias for most of the physical parameters. 
Regardless of the physical parameter or the true value of the parameter we are considering, 
the difference in bias between S/Ns is $\lesssim 0.05\,$dex for the age, metallicity, and sSFR and $\lesssim 0.05\,$ mag for A$_{\rm V}$. The only exception to this is for $\log\, t_{\rm mod}/\tau$, where for the lowest value, $\log\,t_{\rm mod}/\tau \lesssim -1.2$, we see that the bias (slightly) decreases at higher S/Ns. Compared to the bias, increasing the S/N decreases $\sigma$ for most of the physical parameters. For the mwa and uwa, the largest difference in $\sigma$ is seen for the simulated galaxies in the range $8.5\leq \log\,t \leq9.5\,$yr. We see a decrease in $\sigma$ for metallicity as we increase the S/N; this is significant at low metallicity $\log$Z/Z$_\odot \lesssim -0.5$. Additionally, there is a gradual decrease in the difference between $\sigma$ of different S/Ns as we increase in metallicity. A$_{\rm V}$ shows a gradual improvement in constraint capability with respect to the S/N, as we see a decrease in $\sigma$ with an increase in the S/N in the range $0.15\lesssim$A$_{\rm V} \lesssim 0.8$. For sSFR, improvement in the S/N results in comparable $\sigma$. Finally, we also see no strong trend between the S/N and $\sigma$ for $\log\,t_{\rm mod}/\tau$. Overall, while there is no clear improvement in bias, we find a clear decrease in $\sigma$ for all physical parameters except the $t_{\rm mod}/\tau$ and sSFR at higher S/Ns. 

Table~\ref{tab:RF_KNN_SN} also lists the median bias, $\sigma$, and the fraction of 
outliers for both RF and KNN for S/N=20 and 30 at $z=0.3$. The uncertainties give the 
25$^{\rm th}$ and 75$^{\rm th}$ percentiles of the bias and $\sigma$. 
Similar to the figure, we find minimal change in the median bias for all of the physical parameters, $\sim 0.0\,$dex or mag, irrespective of the S/N.\ We note that while the sSFR and $\log\,t_{\rm mod}/\tau$ have median bias $\sim 0.0\,$dex, they have very high scatter on the bias at different bins, as shown by the percentiles. For example, for the sSFR we find a bias of $\sim -0.86\,$dex at different S/Ns as shown by the 25$^{\rm th}$ percentile value. Similarly, for $t_{\rm mod}/\tau$ we find the 25$^{\rm th}$ and $75^{\rm th}$ percentile values for bias to be $\sim-0.2$ and 0.7\,dex, respectively. The $\sigma$ shows a gradual decrease with increasing S/N for most parameters, by $\sim0.02$\, dex for age and $\sim 0.01\,$mag for A$_{\rm V}$, $\sim0.03\,$dex for metallicity. For the sSFR and $\log\,t_{\rm mod}/\tau$, while we see differences in the curves, we do not see a strong trend when using average statistics. Generally, RF has (marginally) lower bias than KNN, by $\sim 0.02\,$dex. Similarly, RF has a lower dispersion than KNN for most physical parameters, varying between $\sim 0.02$ to 0.6$\,$dex. Again, the main exception to such a trend is seen for $\log\, t_{\rm mod}/\tau$, where KNN has a lower dispersion by $\sim 0.04\,$dex. Finally, KNN generally has fewer outliers than RF owing to the Gaussian-like distribution of the differences between predicted and true values.

\subsubsection{Samples at different redshifts}
\label{sec:res_redshift}
\begin{figure*}
    \centering
    \includegraphics[width=0.32\linewidth]{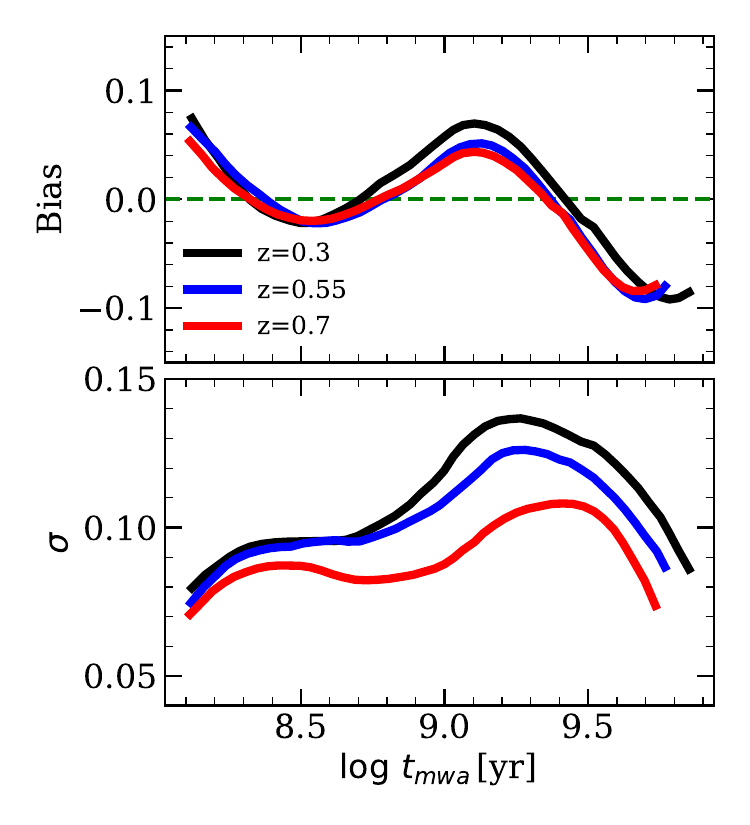}
    \includegraphics[width=0.32\linewidth]{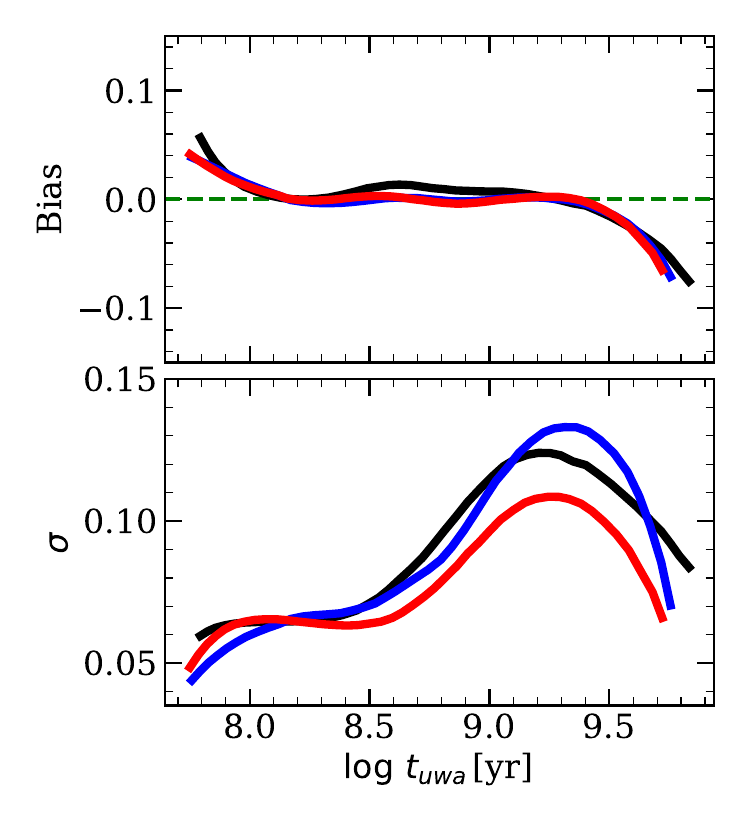}
    \includegraphics[width=0.32\linewidth]{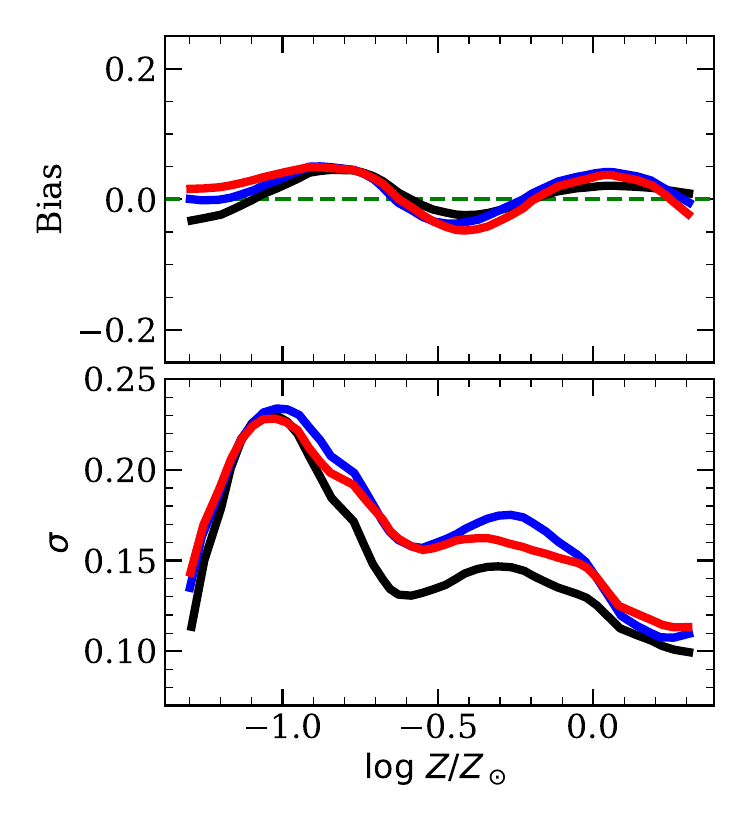}

    \includegraphics[width=0.32\linewidth]{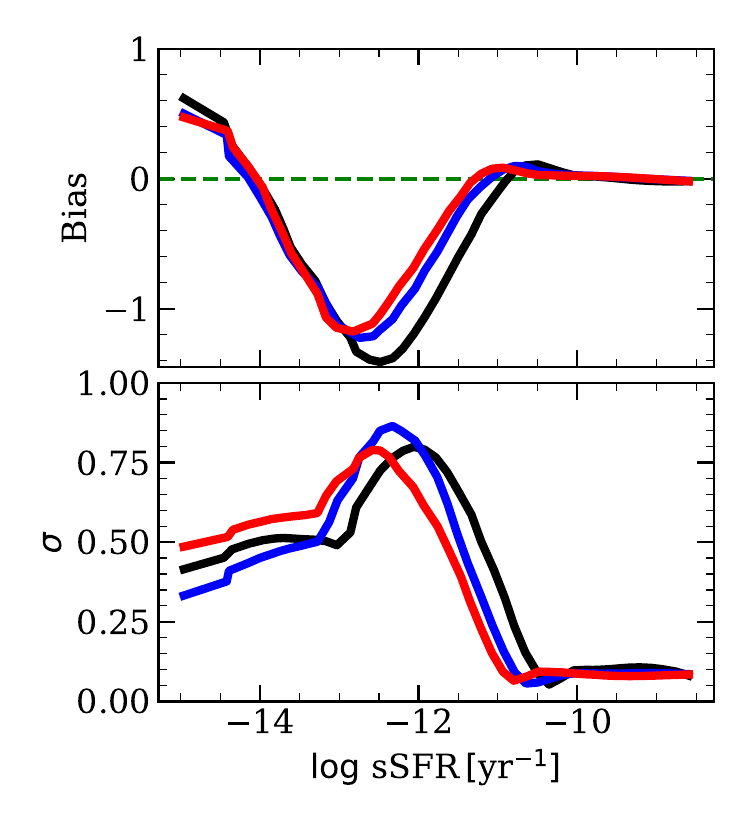}
    \includegraphics[width=0.32\linewidth]{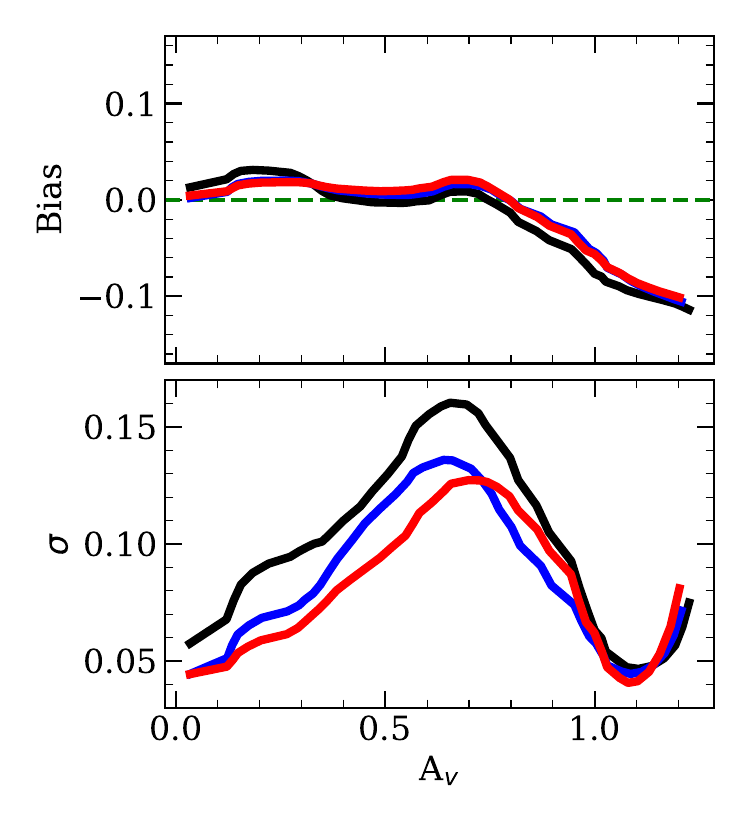}
    \includegraphics[width=0.32\linewidth]{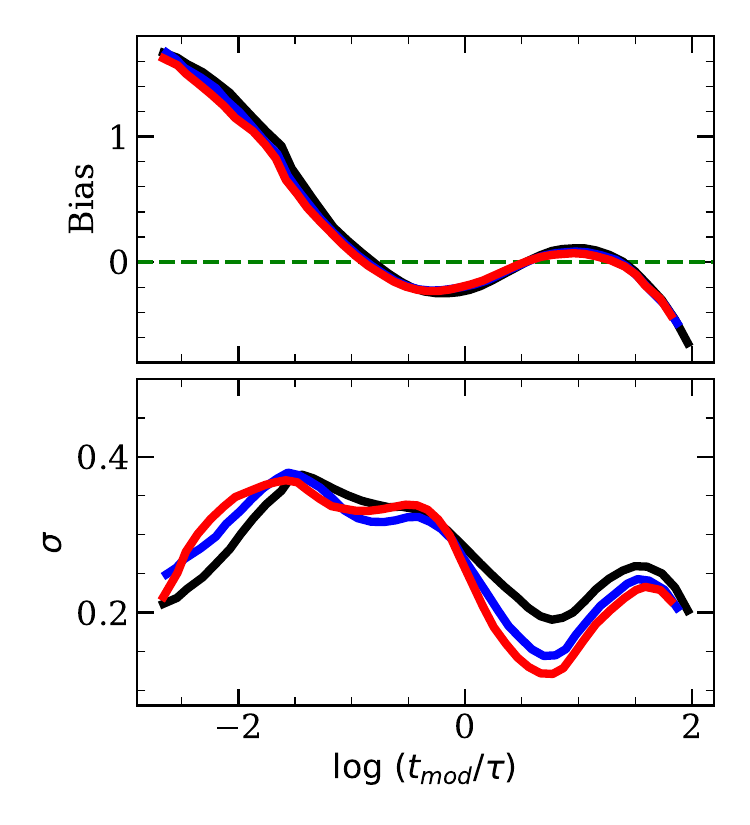}
    \caption{Same as Fig.~\ref{fig:BiasSig_SN} but 
    for S/N=10 at $z=0.3$ (black), 0.55 (blue), and 0.7 (red). Again, these results are obtained with RF.}
    \label{fig:BiasSig_RedSh}
\end{figure*}
In the following section, we quantify differences in retrieving galaxy physical parameters at redshift higher than 0.3 for S/N=10. 
Increasing the galaxy redshift affects both the spectroscopic and the photometric information. Indeed, the higher the redshift, the bluer the observed rest-frame spectral window, thus including a different set of indices. 
In addition to having access to more UV spectral indices, the same spectral indices at different redshifts will possess different S/Ns because they will fall in different observed wavelength regions of the spectrograph. Furthermore, the same photometric filters will sample bluer rest-frame spectral windows. This results in the same simulated galaxy having different magnitudes and uncertainties at different redshifts.

Figure~\ref{fig:BiasSig_RedSh} shows the bias and $\sigma$ as a function of the true physical parameter for $z=0.3, 0.55$ and $0.7$ in black, blue and red lines, respectively, at S/N=10. 
These results are for the RF algorithm. We find no significant differences in the bias for the mwa and uwa, metallicity, or A$_{\rm V}$.\ We note that the maximum age of the simulated galaxies is lower at higher redshifts due to the younger age of the Universe. There is an improvement in the prediction of the sSFR with increasing redshift, as there is a notable reduction in the bias, where the underestimation of the sSFR starts at $-11.5$, at $z=0.7$, rather than at $-11.0$. This is due to two factors: at higher redshifts we have a lower number of quiescent galaxies. 
Therefore, when averaging over 100 trees, we sample over a lower number of galaxies with sSFR$=-15.0$, thus reducing the bias in the estimate of the sSFR. Furthermore, and more importantly, at higher redshifts we have access to a larger number of UV spectral indices, and the observed magnitudes sample a bluer region of the spectra, which are more sensitive to the recent SFH and provide a better constraint on the sSFR \citep{Vaz:2016,RLF:2016,Salvador:2020}. 
For $t_{\rm mod}/\tau$, we see only minor improvements in bias as the redshift increases. In contrast to bias, there is a clear improvement in $\sigma$ for both mwa and uwa estimations with an increase in redshift. This effect is most evident for older ages, $\log\, t_{\rm mwa}\,$(yr)$\gtrsim 9.0$ and $\log\, t_{\rm uwa}\,$(yr)$\gtrsim 8.5$. 
The better estimation of age is also visible in \citet{Costantin:2019}, which uses a more traditional Bayesian approach. This trend is likely due to the larger number of UV spectral indices and the bluer rest frame magnitudes available at higher redshifts. In contrast, the $\sigma$ for metallicity are comparable at different redshifts. This could be due to higher redshift observations probing less of the IR part of the spectrum, which, combined with optical observed magnitudes, is known to help constrain metallicity. 
For $0.15\lesssim\,$A$_{\rm V} \lesssim 0.8\,$mag, we find a larger $\sigma$ at $z=0.3$ than at higher redshifts.\ We note that at $z=0.55$ and $z=0.7$, the $\sigma$ values are comparable. At higher redshift, the observed magnitudes sample the bluer part of the spectrum, which is known to be more affected by dust, hence the slight improvement in our A$_{\rm V}$ prediction. Finally, the $\sigma$ of $\log\, t_{\rm mod}/\tau$ shows no strong correlation with any redshift.

Table~\ref{tab:RF_KNN_SN} also lists the bias, $\sigma$ and percentage of outliers for the various physical parameters at $z=0.55$ and 0.7. This supports the results found in Fig.~\ref{fig:BiasSig_RedSh}, where the bias is comparable between redshifts for most parameters. The main exception is for sSFR, where while the median value is similar, the 25$^{\rm th}$ percentile value is closer to 0 by 0.19$\,$dex at $z=0.7$ than at 0.3. The median of $\sigma$ is marginally lower for both the mwa and uwa. Again for the sSFR, we see a decrease in $\sigma$ of 0.1$\,$dex between $z=0.3$ to 0.7. In contrast, the metallicity $\sigma$ values are comparable, a difference of only 0.03$\,$dex as we increase the redshift. There is no clear trend in regards to $t_{\rm mod}/\tau$. In comparison, KNN shows a decrease in median $\sigma$ for a mwa (uwa) of $0.03\,(0.01)\,$dex between $z=0.3$ and 0.7. The median $\sigma$ for KNN does not have any strong correlation with redshift for the sSFR, A$_{\rm V}$, or $\log t_{\rm mod}/\tau$; however, overall they display the same behaviour as RF. In contrast, we see a negligible increase in the median $\sigma$ of $0.04\,$dex for metallicity between $z=0.3$ and 0.7. While RF and KNN follow the same trend with an increase in $z$, RF again finds a better constraint than KNN.

\section{A test case: Red, green, and blue galaxies}
\label{sec:Class}
As an example of how quantities retrieved from ML algorithms can be applied to real data, we consider here the use of sSFRs to classify galaxies into three categories: BC, mostly star-forming galaxies, GV, galaxies in the process of quenching, and RS, mostly  quiescent galaxies \citep[][and references within]{Salim:14}. Once we group simulated galaxies into BC, GV, and RS, we check for completeness of our classification for different S/Ns, at each redshift.\ We note that here we do not look at the purity of our classification as this is heavily dependent on both the distribution and fraction of galaxies in BC, GV, and RS. While completeness might also be affected by the actual distribution of galaxies, as objects close to the border are more likely to be misclassified than those at the centre, it is likely more robust against the fraction of populations in each region.

While there are several ways to define BC, GV, and RS, we utilise the definition based on sSFR here. We adopt a cut-off value for the sSFR that evolves with cosmic time, as this has been observed in both hydrodynamical simulations and observations  \citep{Fritz:2014, Trayford:2016, Phill:19, Wright:2019, Jian:2020}. To define the cut-off boundary between star-forming and quiescent galaxies, we use the formula:  $\log\,($sSFR(yr$^{-1}$))$ = -11.0 +0.5z$ \citep{Furlong:2015}.
We then consider 0.75\,dex above (below) the cut-off point to define the border between BC (RS) and GV,  resulting in a GV width of 1.5\,dex in the sSFR.
Table~\ref{tab:GV_limits} shows the GV limits obtained by such a method, where $\log\,$sSFR$_{\mathrm{GV, RS}}$ is the border between GV and RS, and $\log\,$sSFR$_{\mathrm{BC, GV}}$ is the boundary between BC and GV.  

\begin{table}
  \centering \caption{sSFR boundary between BC and GV (sSFR$_\text{BC, GV}$)
  and GV and RS (sSFR$_\text{GV, RS}$) for the three redshifts explored in this work, i.e. $z=0.3, 0.55$, and $0.7$.}\label{tab:GV_limits}
    \begin{tabular}{lcr} 
    \hline
     $z$ & $\log\,$ sSFR$_\text{BC, GV}$ & $\log\,$ sSFR$_\text{GV, RS}$  \\
     \hline
     0.3 &  -10.1 & -11.6 \\
     0.55 & -10.0 & -11.5 \\
     0.7 &  -9.9 & -11.4 \\
     \hline
    \end{tabular}
\end{table}
 
Completeness is defined as 
\begin{equation}
    \text{Completeness}  = \frac{\text{TP}}{\text{TP + FN}},
\end{equation}
where TP is the number of true positives, that is, simulated galaxies correctly identified to be in BC, GV, or RS; the FN is the number of false negatives, that is, galaxies misclassified as belonging to the `wrong' region. 
Table~\ref{tab:Comp_ML} shows the completeness obtained with RF and KNN at each redshift for different S/Ns. 

\begin{table}
  \centering \caption{Completeness of BC, GV, and RS.}\label{tab:Comp_ML}
  \resizebox{\columnwidth}{!}{%
    \begin{tabular}{lcccccc} 
    \hline
    \multicolumn {1}{c}{} & \multicolumn{3}{c}{Random Forest} &
    \multicolumn{3}{c}{K-Nearest Neighbour}    \\  
    \hline 
    \multicolumn {7}{c}{SN=10 (Completeness)} \\
    \hline
    Redshift & z=0.3 & z=0.55  & z=0.7 & z=0.3 & z=0.55  & z=0.7 \\
    \hline
    BC & 0.99 & 0.99 & 0.99 & 0.99 & 0.99  & 0.99 \\
    GV & 0.78 & 0.86 & 0.88 & 0.52 & 0.55  & 0.64 \\
    RS & 0.99 & 0.98 & 0.98 & 0.97 & 0.96  & 0.96 \\
    \hline
    \multicolumn {7}{c}{SN=20 (Completeness)} \\
    \hline
    BC & 0.99 & 0.99 & 0.99 & 0.99 & 0.99 & 0.99 \\
    GV & 0.80 & 0.88 & 0.90 & 0.60 & 0.69 & 0.75 \\
    RS & 0.99 & 0.98 & 0.97 & 0.98 & 0.98 & 0.97 \\
    \hline
    \multicolumn {7}{c}{SN=30 (Completeness)} \\
    \hline
    BC & 0.99 & 0.99 & 0.99 & 0.99 & 0.99 & 0.99 \\ 
    GV & 0.81 & 0.88 & 0.92 & 0.67 & 0.76 & 0.83 \\
    RS & 0.99 & 0.98 & 0.98 & 0.98 & 0.98 & 0.98 \\
    \hline
    \end{tabular}
    }
    \tablefoot{The left and right set of columns show result for RF and KNN. Top, middle and bottom sets of rows shows completeness at  S/N=10, 20, and 30. We tabulate results at each of three redshifts, $z=0.3$, 0.55, and 0.7. }
\end{table}

The completeness of galaxies in BC and RS is consistently very high ($\gtrsim 0.96$) irrespective of the S/N or ML algorithm we use. On the other hand, for GV galaxies, completeness is lower, although RF still performs relatively well, with completeness $\gtrsim 0.75$.\ We note that not only does the GV completeness improve with increasing S/N, but we also find higher values with increasing redshift, with an increase of $\sim$10\% from $z=0.3$ to 0.7. For KNN, while there is also an improvement with S/N and redshift, we find that the overall completeness is $\lesssim 0.65$,  at the lowest S/N and redshift. This stems from the significant bias in predicting sSFRs with KNN (see Sect.~3.2), where we consistently underestimate the true value.  The underestimation leads to misclassifying a GV galaxy into an RS one, while the opposite scenario is unlikely.

\begin{figure}
    \centering
    \includegraphics[width=\linewidth]{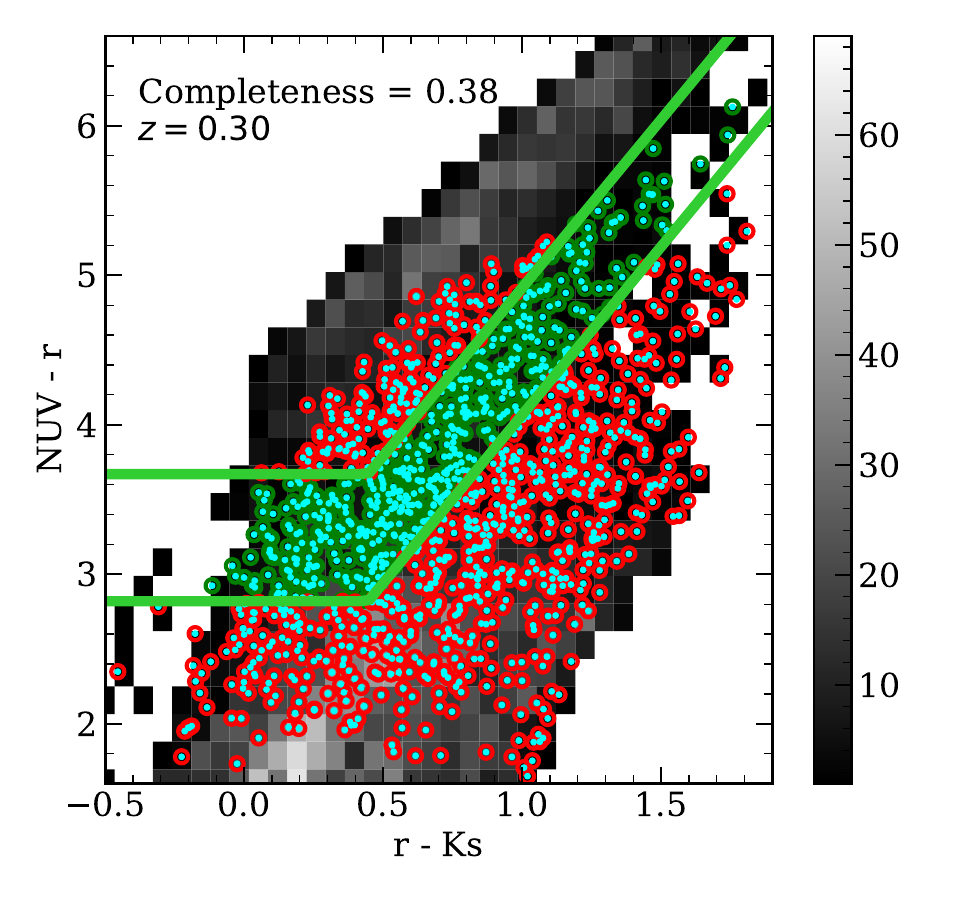}
    \caption{Distribution of galaxy templates in the colour--colour diagram. The lime-green lines define the demarcation used to select GV galaxies \citep{Moutard:2020}. Cyan dots are GV galaxies, identified using the true sSFR at $z=0.3$. Green (red) circles around the cyan dots show galaxies that are correctly identified (misidentified) as GV galaxies according to colour--colour selection. The colour bar on the side shows the number of galaxies.}
    \label{fig:CC_GV}
\end{figure}

To compare our results with those obtained with other methods, we also compute the completeness of GV classification using a NUVrKs diagram \citep{Arnouts:2013}. The density plot in Fig.~\ref{fig:CC_GV} shows the distribution of simulated galaxies in colour-colour space. The completeness for this method is 0.38, $\sim$0.40 ($0.30$) lower than for RF (KNN) at the same redshift.\ We note that at different redshifts, we have different numbers of galaxies, and also, the definition of demarcation lines varies \citep[see equation 1 of ][]{Moutard:2020}. Therefore, we repeated the analysis at $z=0.55$ and $z=0.7$, finding that the completeness of NUVrKs-selected GV galaxies drops to 0.36 and 0.32, respectively. 
Methods that use the sSFR to define GV galaxies are expected to perform better than simple colour selections: the sSFR is the physical parameter that defines GV galaxies, and it spans a wide range of rest-frame NUVrKs colours. 
Therefore, a simple colour-colour selection of GV galaxies is prone to large incompleteness. On the contrary, the use of ML to estimate the sSFR with photometry and spectral indices has small uncertainties, leading to high completeness.

\section{Discussion}
\label{sec:Disc}
We highlight the capabilities of two ML algorithms to retrieve the physical properties of galaxies based on photometric and spectroscopic information. We analysed how results vary for simulated galaxy spectra at different redshifts, with S/N$_{\rm I,obs}$ of 10, 20, and 30, respectively. In the following, we discuss some caveats and possible limitations of our assumptions (Sect.~\ref{sec:MoS}). 
In addition, we compare our results with those obtained using a more traditional Bayesian approach (Sect.~\ref{sec:BayesComp}). Finally, since spectroscopic data can have a wide range of S/Ns, we analyse how much our results vary when the training and testing data have different S/Ns (Sect.~\ref{sec:sn_comp}).  

\subsection{Photometry versus spectroscopy}
\label{sec:MoS}

Both photometry and spectroscopy provide valuable constraints on the physical parameters of galaxies. Therefore, many state-of-the-art spectral fitting algorithms combine the two types of information \citep{Carnall:2018,Johnson:21, Cappellari:2023}. In this section, we discuss the ability of ML algorithms to retrieve physical parameters of galaxies based on spectroscopy alone or with spectro-photometry information, discussing limitations due to observations and stellar population models.

\begin{figure}
    \centering
    \includegraphics[width=\linewidth]{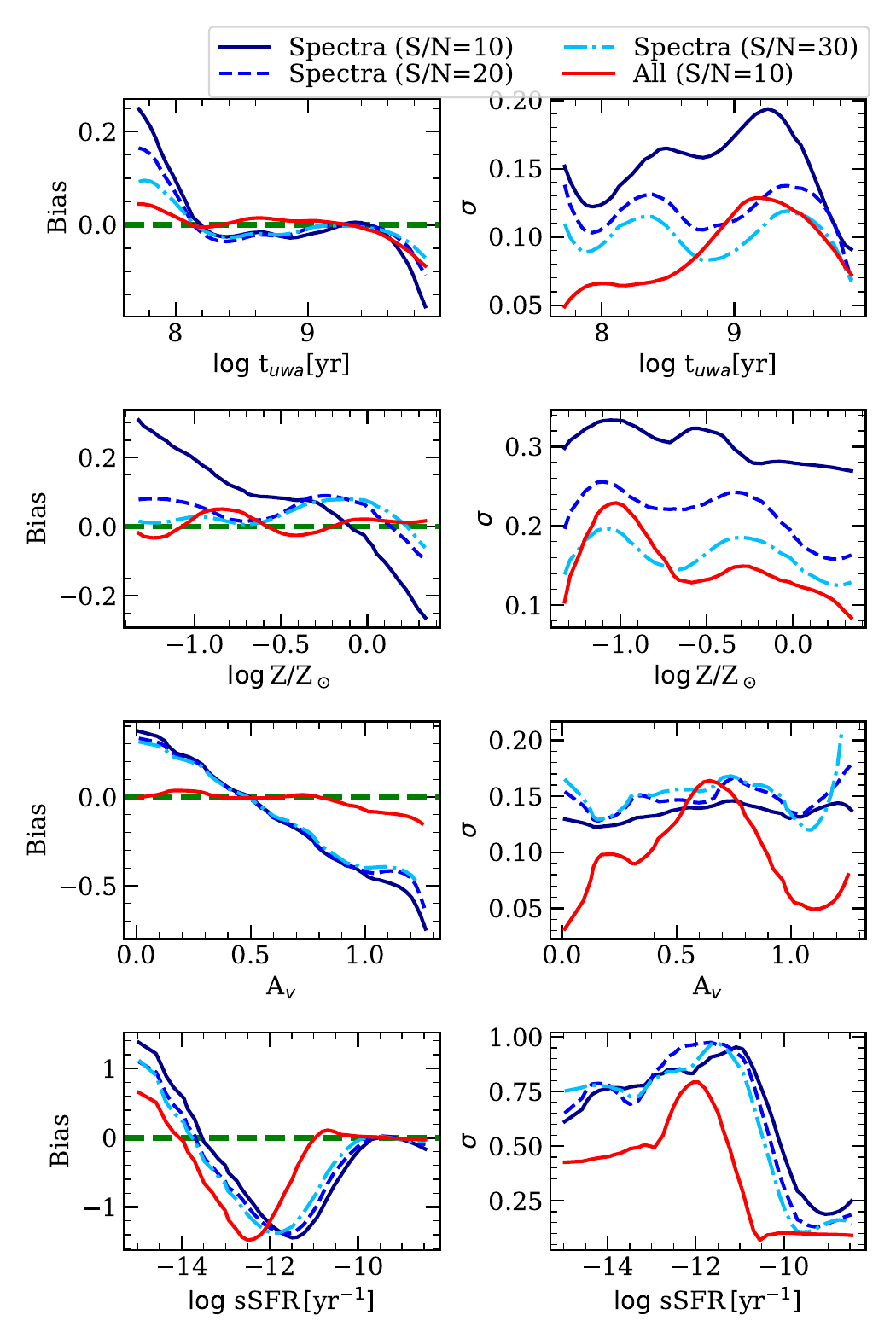}
    \caption{
    Bias (left) and $\sigma$ (right) of the uwa, metallicity, A$_{\rm V}$, and sSFR retrieved using either spectral indices only, at different S/Ns, or both spectroscopic and photometric information. The solid dark blue, dashed blue,
    and dash-dotted  sky-blue curves show results for S/N=10, 20, and 30, respectively. For comparison, 
    the red curves show results obtained using both observed magnitudes and spectral indices at S/N=10. The dashed green line in the left panels indicates zero bias. The results are for the RF algorithm at $z=0.3$.
    }
    \label{fig:LS_Mag_Comp}
\end{figure}

Figure~\ref{fig:LS_Mag_Comp} shows how well we can retrieve the uwa, metallicity, A$_{\rm V}$, and sSFR using only spectroscopic information at $z=0.3$ (i; blue curves), an I-band S/N (per \AA) of 10, 20, and 30; (ii; red curves), and both photometry (all bands) and spectroscopy, the latter with S/N=10. Panels in the left (right) column of Fig.~\ref{fig:LS_Mag_Comp} show the bias (dispersion, $\sigma$) on retrieved quantities. 

For age and metallicity, at S/N=10, the predictions generally give the highest bias values (especially at the extremes of the distribution) and $\sigma$. At S/N=20, we find improvement in bias and $\sigma$ --- most noticeable for younger ages. However, the $\sigma$ for spectra is still larger but with differences of less than 0.1$\,$dex. At S/N=30, the bias and $\sigma$ are comparable both when using spectroscopic information only and photometry plus spectroscopy, although the latter method still provides lower $\sigma$ values for ages younger than $\log t_{\rm uwa}[\rm yr]\sim 8$.\ We note that, while not shown, the mwa shows the the same trend as the uwa.

In contrast to these trends, there is limited constraining capability on A$_{\rm V}$ solely based on spectral information, independent of the S/N, as spectral indices are measured over short wavelength ranges and are therefore insensitive to dust.\ We note the low constraining ability is due to D$_n$(4000) being sensitive to dust \citep{MacAruthur:05}. 
For the sSFR, we find a noticeable difference when using only spectral or all information; for the former, the increase in bias and $\sigma$, towards lower sSFR values, starts at $\log\,$sSFR$\sim -10$ rather than at $-11$. While increasing the S/N tends to decrease the bias and $\sigma$, the degradation of constraining capability for $\log\,$sSFR remains at about $-10$.

While we find that the inclusion of photometry gives us a higher constraining capability for the A$_{\rm V}$ and, more importantly, for the sSFR, we warn the reader that the uncertainties considered here do not account for all sources of systematics --- as we consider only formal photometric errors from the COSMOS catalogue \citep{Laigle:2016}. While we set a lower limit on photometric errors by adding 0.05 mag in quadrature to the formal errors on magnitudes, the true uncertainties, due to systematics, may be larger for several reasons. A significant source of systematics on photometry is dust, as the reddening correction depends on the adopted dust model. Different dust models may lead to large variations on measured magnitudes, with such systematics propagating to colours, thus affecting the retrieval of physical parameters \citep{Salim:2020, Pacifici:2023}.

\begin{figure}
    \centering
    \includegraphics[width=\linewidth]{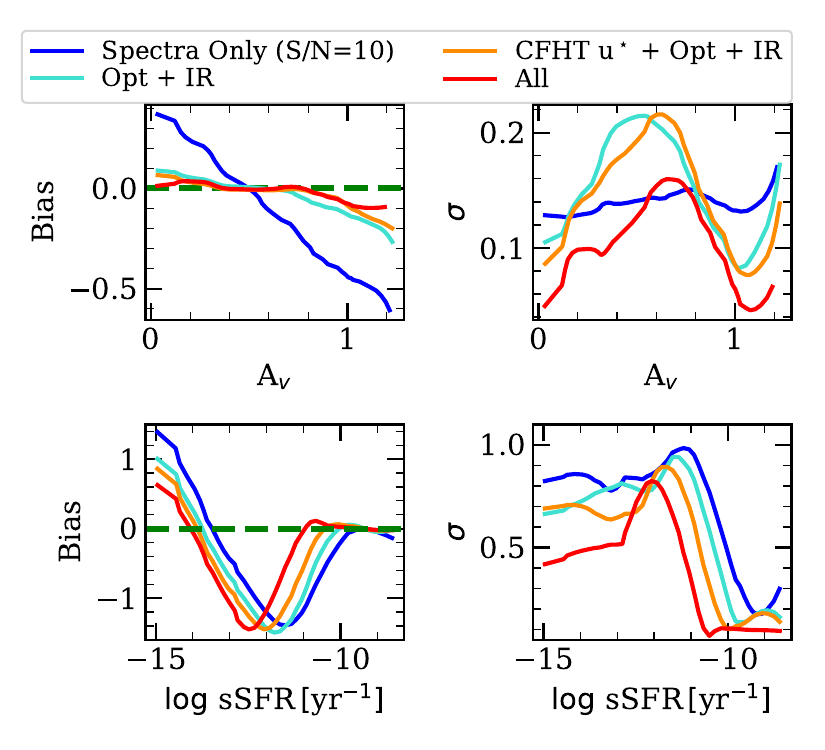}
    \caption{Same as Fig.~\ref{fig:LS_Mag_Comp} but also showing results when considering different observed magnitude combinations at $z=0.3$. Note for all results shown, we also use all available spectral indices, at $z=0.3$, with S/N=10. The turquoise lines show results for observed optical (these consist of HSC g, r, i, z, and y filters) and IR magnitudes (VISTA H, J, and Ks). The dark orange lines show results for observed optical, IR, and CFHT u$^\star$ magnitudes.}
    \label{fig:Comparison_Fig}
\end{figure}

While NIR wavebands are less affected by dust, they are also susceptible to systematics due to differences among different stellar population models. To illustrate this point, we compare magnitudes obtained from BC16 and EMILES \citep{Vaz:2015} simple stellar population (SSP) models. 
We considered only ages $\gtrsim 1$~Gyr, where EMILES stellar population models are `safe' in the NIR spectral range ($\lambda \gtrsim 8950\,$\AA , for all metallicities; see \citealt{Vaz:2015}). While magnitude differences between the two sets of models are small in the optical (less than a few percent), large differences are found in the NIR. In particular, for an age of $\sim 1$\,Gyr at solar metallicity, we find differences of $\gtrsim 0.2\,$ mag in the observed H and Ks bands at $z=0.3$. Such systematic differences are due to EMILES models relying on the empirical library Indo-US library \citep{Valdes:2004} whereas BC16 uses the semi-empirical library BaSeL 3.1 \citep{Westera:2002}.
Systematics in the models also affect the UV part of the spectra. In the UV, EMILES models rely on fully empirical stellar library, whereas BC16 is purely theoretical. The comparison of  CFHT u$^\star$ magnitudes shows a maximum difference of 0.26$\,$mag for a SSP age of $\sim$1.4\,Gyr, at solar metallicity. These systematics may lead to higher uncertainties than what we have assumed on our magnitudes. Another possible source of systematics is that spectroscopic and photometric observations usually cover different galaxy regions. Both, slit and aperture spectroscopy usually cover the central parts of the galaxies, while magnitudes obtained from photometry are integrated in larger apertures. This may lead to discrepancies in the estimates of various physical parameters, depending on whether spectroscopic or photometric data are used. Last but not least, one should bear in mind that while optical photometry is affected by the age-metallicity degeneracy, for an unresolved stellar population, spectral indices are able to partially break it \citep[e.g.][]{Worthey:94, Arimoto:1996}. The combinations of UV and optical \citep{Kaviraj:2007} or optical and IR \citep{Carter:2009} colours have been shown to be successful at partially breaking this degeneracy.\ However, they suffer from other issues; for example, the UV and optical combination suffers from age--dust degeneracy. 

In addition to systematics, one usually has limited access to far-UV (FUV) and near-UV (NUV) photometric data, because of the faint flux of galaxies at these wavelengths. For example, within the WEAVE-StePS, in the CFHTLS-W4 field, only 4$\%$ (20$\%$) of the sources have FUV (NUV) photometry, while for the ELAIS field, only 10$\%$ of galaxies have GALEX data. In contrast, for I$_{\mathrm{AB}} < 20.5$ at $z>0.3$, the COSMOS field has 60$\%$ and 70$\%$ of galaxies with FUV and NUV photometry, respectively. Finally, even when FUV photometry is available, its interpretation for old stellar populations is not trivial, as this spectral range includes contributions from stars that are difficult to model \citep[e.g. very young stars and/or hot evolved stars; see][]{LeCras:2016, Salvador:2020}.

Despite such issues, photometry is a reasonable source of information for a sensible estimation of the A$_{\rm V}$ and sSFR. Without photometry, our estimates of the sSFR worsen significantly, affecting the completeness. For instance, when using only spectral indices at $z=0.3$, the GV completeness drops to 0.43, 0.51, and 0.56 for S/N=10, 20, and 30, respectively. Although these values are relatively low, they are still higher than that obtained from the NUVrKs diagram (i.e. 0.38; see Sect.~\ref{sec:Class}). A relevant question is what happens to the constraining capability on A$_{\rm V}$ and sSFR, when only limited photometric information is available. We address this point in Fig.~\ref{fig:Comparison_Fig}, showing the bias and $\sigma$ on the A$_{\rm V}$ and sSFR, for simulated galaxies at $z=0.3$. The blue and red lines in Fig.~\ref{fig:Comparison_Fig} are the same as in Fig.~\ref{fig:LS_Mag_Comp}: they show constraints from spectroscopy only, with S/N$=10$, and photometry plus spectroscopy, respectively. The figure also shows results when spectroscopy is combined with (i) only optical and IR photometry (turquoise lines); and (ii) optical, IR, and u-band photometry (dark orange lines).
 Access to photometric information improves the accuracy of the A$_{\rm V}$ and, more importantly, the sSFR. When only optical and IR photometry are used, there is a marginal improvement in the estimate of the sSFR, leading to a relatively low completeness of 0.52 for GV galaxies,  comparable to that for spectroscopy only, with S/N=20 (see above). On the other hand, adding CFHT u$^\star$ to optical and IR magnitudes (see the dark orange lines in Fig.~\ref{fig:Comparison_Fig}) leads to a more significant improvement of sSFR estimates. The deterioration now starts at sSFR $\sim-10.5$, increasing GV completeness up to 0.62. This is particularly relevant for the WEAVE-StePS, where in the CFHTLS-W4 and COSMOS fields 99$\%$ and 70$\%$ of galaxies have u-band magnitudes at $z>0.3$ and I$_{AB}$<20.5.

\subsection{\protect{Comparison with Bayesian methodology}}
\label{sec:BayesComp}
In this section, we compare results obtained from ML techniques with those from a more traditional Bayesian approach following the method outlined in \citet{Gallazzi:2005} and \citet{Zibetti:2017}.
To make a fair comparison with previous works presenting results for WEAVE-StePS simulated data~\citep{Costantin:2019, Ditrani:2023}, we consider only UV and optical spectral 
indices, assuming S/N$_{\rm I,obs}=10$, excluding the photometric information. 
Under the assumption that errors  are normally distributed, the goodness of the fit for a given statistical model is given by 
\begin{equation}
    \chi^2 = \sum_i \left[\frac{O_i - M_i}{\sigma_i}\right]^2,
    \label{eq:Chi2_eq}
\end{equation}
where $O_i$ are observed values (i.e. index line strengths), $\sigma_i$ their errors, and $M_i$ are
model values. Equation~\ref{eq:Chi2_eq} can be converted into a posterior probability distribution function, defined as $\mathcal{L} \propto e^{-\chi^2/2}$. 
The predicted value of a given quantity is estimated by marginalising the probability distribution function with respect to it, taking median values (see \citealt{Ditrani:2023}).\ We note that for a fair comparison, we also use median values for RF, namely we consider the median rather than the mean of 100 trees from the output of RF algorithm.

For a given physical parameter, we grouped simulated galaxies into groups of 1200 objects according to the true values. For each 
bin, we calculated the bias and $\sigma$. Table~\ref{tab:BayesTab} shows quantities for the age and metallicity from both the RF algorithm and the Bayesian approach. 
\begin{table}
  \centering \caption{Comparison of bias and $\sigma$ of metallicity and age obtained using RF and Bayesian approach.}\label{tab:BayesTab}
    \begin{tabular}{lcccc} 
    \hline
    \multicolumn {1}{c}{} & \multicolumn{2}{c}{RF} & \multicolumn{2}{c}{Bayesian}  \\  
    \hline
    $\log Z/Z_\odot$ & Bias & $\sigma$ & Bias & $\sigma$ \\
    \hline
        $-$1.17 & 0.13 & 0.37 & 0.07 & 0.35 \\
        $-$0.67 & 0.10 & 0.29 & 0.07 & 0.27 \\
        $-$0.40 & 0.07 & 0.35 & 0.06 & 0.34 \\
        $-$0.05 & 0.00 & 0.32 & 0.05 & 0.32 \\
        0.22 & $-$0.08 & 0.24 &  0.04 & 0.23 \\
        0.38 & $-$0.19 & 0.23 & $-$0.14 & 0.20 \\
    \hline
    Age$_u$(Gyr) &  Bias & $\sigma$ & Bias & $\sigma$\\
    \hline
        0.09 & 0.02 & 0.03 & 0.02 & 0.03 \\
        0.30 & 0.00 & 0.11 & $-$0.01 & 0.10 \\
        0.61 & $-$0.03 & 0.20 & $-$0.04 & 0.19 \\
        1.37 & $-$0.07 & 0.52 & $-$0.04 & 0.47 \\
        4.19 & $-$0.23 & 1.34 & $-$0.25 & 1.24 \\
        7.94 & $-$2.55 & 1.35 &  $-$2.37 & 1.27 \\
    \hline
    \end{tabular}
    \tablefoot{Top and bottom rows tabulates the bias and $\sigma$ for metallicity and age across different bins.}
\end{table}
We find the largest differences in bias between the two methods for bins that contain either the oldest metal-poor populations ($\log\,Z/Z_\odot = -1.17$) or the oldest metal-rich populations ($\log\,Z/Z_\odot = 0.38$), particularly those at the boundaries of the parameter space. For the lowest and highest metallicity bins, the difference in bias (between the RF and the Bayesian methods)  are $0.06$ and $0.05\,$ dex, respectively (see Table~\ref{tab:BayesTab}). Similarly for the uwa, we have the largest difference, of 0.18$\,$Gyr, in the oldest age bin, Age$_u \sim7.94\,$Gyr. In contrast to the bias, we find comparable results between both methods in their $\sigma$. While both the Bayesian and RF approaches perform similarly, the main advantage of RF is that it is orders of magnitude faster than the Bayesian approach, especially when the training has been completed. Such a large difference in time is due to the fact that ML methods have a lower computational complexity than Bayesian inference. Once the tree is constructed, the RF simply goes down the different branches to  retrieve the physical parameter, while Bayesian inference requires a comparison with each template. Previous works in the literature have also noted differences in computational time between various types of ML algorithms and classical methods \citep{SS:2017, Dominguez:2018, Davidzon:2019}. 

Finally, there have been studies that use more complex ML algorithms in astronomy for different purposes. For example, \citet{MS:2023} used an artificial neural network (ANN) and RF to classify objects into quasars, galaxies or stars, finding ANN to outperform RF, given proper calibration. In addition, \citet{Hunt:2024} carried out a similar analysis to us, where they predicted the average ages of galaxies from the GAMA survey using ANN. They trained the ANN to predict average ages based on the optical spectral indices, finding results that are comparable to ours.

\subsection{Varying the signal-to-noise of the training and testing samples}
\label{sec:sn_comp}

Throughout Sect.~\ref{sec:results}, we assumed the spectra in both the training and testing samples have the same S/N. For a given S/N, we included five random realisations of each galaxy in the training sample. 
The benefits of such a methodology are outlined by \citet{Shy2022}, namely that it becomes possible for the ML algorithms to incorporate the uncertainties on the observables. However, when dealing with real data, one may expect to observe a large number of spectra (e.g. $\sim\, 25000$ target galaxies from WEAVE-StePS) with a wide range of signal-to-noise ratios. Since training the ML algorithm for the S/N of each galaxy would be extremely time-consuming,  we tested the effect of setting the S/N of the training sample to a different value than that of the testing sample. In the following, we discuss the impact of this test on our predictions of age and metallicity.

\begin{figure}
    \centering
    \includegraphics[width=\linewidth]{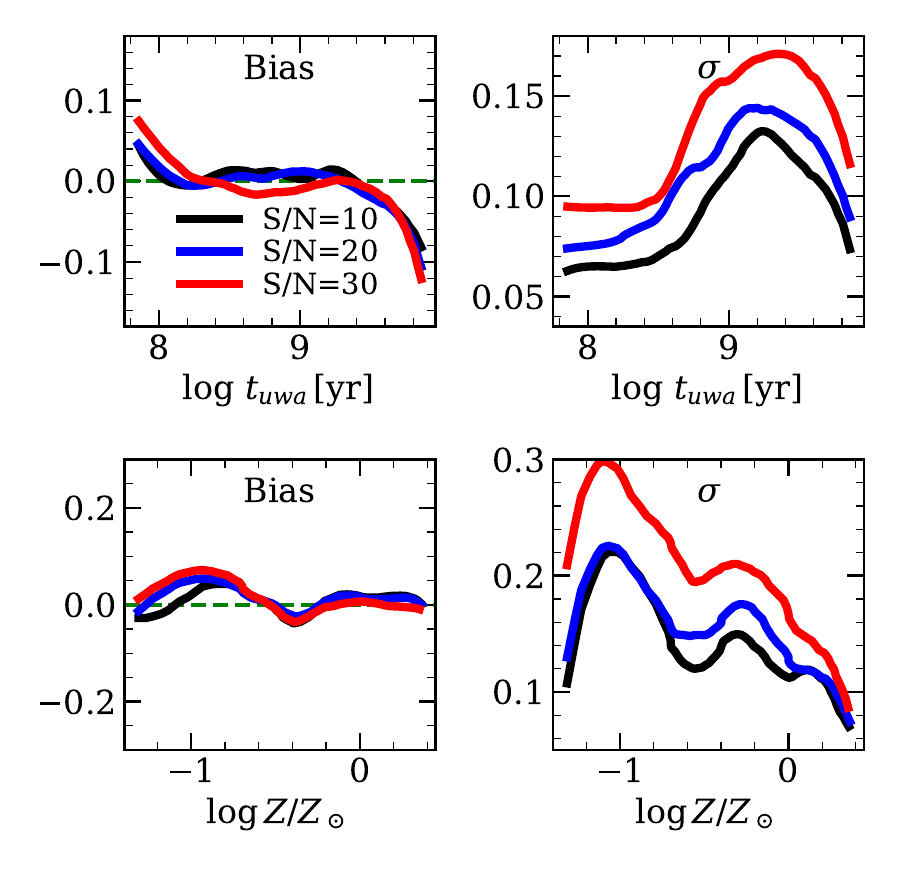}
    \caption{Bias (left panels) and $\sigma$ (right panels) of the uwa (top panels)
    and metallicity (bottom panels) as a function of true values. Black, blue, and red 
    lines correspond to training samples 
    with S/N=10, 20, and 30, respectively. Note that for the testing sample,   we set
    S/N=10 in all three cases.}
    \label{fig:Comp_Meth_Fig}
\end{figure}

Figure~\ref{fig:Comp_Meth_Fig} shows the change in bias and $\sigma$ for both age and metallicity when the training sample has a S/N of 10, 20, and 30 (see the black, blue, and red lines, respectively), and the testing sample has a S/N of 10. There are no significant variations in the age bias with an increase in the difference in S/N between the training and testing samples. The same applies to metallicity, although for S/N=30, we have a slightly higher bias (within $0.03$~dex) at the lowest metallicity (see the red and black curves in the lower-left panel of Fig.~\ref{fig:Comp_Meth_Fig}). On the contrary, we find a clear trend of  $\sigma$ with the S/N of the training sample. For age and metallicity, the $\sigma$ increases as the difference in S/N between training and testing samples increases, implying a deterioration in the estimate 
of physical parameters. The difference in $\sigma$ between S/N=10 and 30 amounts to $\sim0.05\,$dex ($\sim0.10\,$dex) for the uwa (metallicity), whereas the difference between S/N=10 and 20 is considerably smaller, $\leq 0.02\,$dex, for both quantities. This suggests that for galaxies with S/Ns close to 10, assuming a fixed S/N=10 for the training sample, should have negligible impact, for most of the applications, on the estimate of physical parameters. This is actually the case for the WEAVE-StePS, where one expects most of the source to have spectra with S/Ns in the range from 8.8 to 16.2 (see Fig.~9 of \citealt{Iovino:2023}).  Figure~\ref{fig:Comp_Meth_Fig} suggests that, in this case, assuming S/N=10 to create the training sample will give results fully consistent with those shown in Fig.~\ref{fig:fidfig1}, where both the training and testing samples were assumed to have the same S/N. This result is encouraging and opens up new opportunities in the analysis of massive datasets using ML techniques.

\section{Summary and conclusion}
\label{sec:Sum}
With the increased number of upcoming spectroscopic surveys, the astrophysics community will have access to a large number of high-resolution spectra with a wider spectral coverage than ever before, coupled to rich photometric information. These observations span from the local Universe to high redshifts and will cover both the UV and optical rest frames \citep{Euclid:2023, Iovino4MOST, Iovino:2023}. With access to such a large quantity of data, there is an increasing need to find faster ways to extract accurate estimates of physical parameters.

In this work, for the first time, we tested the predictive capabilities of two easy-to-implement ML algorithms (KNN and RF) in estimating the average age (both the mwa and uwa), metallicity, sSFR, A$_v$, and t$_{\rm mod}/\tau$ using photometric and spectral information. We measured UV and optical spectral indices, which are sensitive to age and metallicity, in template spectra and obtained synthetic photometry in the representative UV, optical, and NIR bands. 

First, we analysed how well we can retrieve the physical parameters, assuming no uncertainty in our observed measurements. We then produced realistic simulations of WEAVE-StePS-like spectra for S/N=10, 20, and 30 at redshifts $z$=0.3, 0.55, and 0.7 following the method outlined in \citet{Costantin:2019}. We also assigned uncertainties to the magnitudes calculated for each galaxy template using the photometric ancillary data. We find that, in general, both RF and KNN are able to accurately predict most of the parameters, with a low bias, for all S/Ns and at different redshifts. The main exception was for the sSFR and $t_{\rm mod}/\tau$. For the sSFR, while the median bias was $\sim -0.03$, the lower percentiles ranged from $\sim 0.61$, at $z=0.7$, to $\sim 0.86$, at $z=0.3$ (see Table~\ref{tab:RF_KNN_SN}).\ We note that this large scatter in the bias is due to an underestimation of the sSFR, specifically below $\lesssim - 11.0$ (see Figs.~\ref{fig:BiasSig_SN} and \ref{fig:BiasSig_RedSh}). The $\sigma$ were also relatively low for RF (KNN); for the age we found a maximum $\sigma \sim 0.10\, (0.16)\,$dex for S/N=10 at $z=0.3$, which decreased to $\sim 0.08\, (0.11)\,$dex for S/N=30. At higher redshifts, we find a similar constraint. For metallicity, the $\sigma$ on the estimates vary from $\lesssim 0.11 - 0.16\, (0.12 - 0.25)\,$dex for different S/Ns at $z=0.3$.  At higher redshifts for S/N=10, we find a slightly worse constraint, $\sigma \sim (0.17)\, 0.24\,$dex. The A$_{\rm V}$ has a typical $\sigma$ of $0.09\, (0.10-0.13)\,$ mag. Again, the sSFR has a different constraining power depending on the value. For star-forming galaxies, $\log\,$sSFR$ \gtrsim -10$, we find a typical $\sigma$ of $\lesssim 0.1\, (0.2)\,$dex, which increases to  $0.3 - 0.4\,$dex at lower values, depending on the redshift. For RF and KNN, we find no strong trend between redshifts, S/Ns, and constraint ability for $t_{\rm mod}/\tau$.

In addition, the retrieved sSFR was utilised to classify galaxies into three categories, BC, GV and RS, and it was checked for completeness. The completeness was satisfactory for RF within each region, with GV values of $\gtrsim 0.75$ at $z=0.3$. At a redshift of $z=0.7$ we find a much higher completeness, $\sim 0.90$. In comparison, the completeness in BC and RS was much higher, $\gtrsim 0.96$. While KNN performed similarly well for BC and RS, the GV completeness was lower and ranged from $\sim 0.52$--$0.67$ and $\sim 0.64$--$0.84$ at $z=0.3$ and 0.7, respectively. Although KNN had a lower completeness, it can still be used to accurately estimate the sSFR, thus giving a higher completeness than what we found for colour--colour selection, $\sim 0.35$ (see Sect.~\ref{sec:Class}). Finally, with a more traditional approach, Bayesian statistics, we observe outcomes similar to those of the ML algorithms. Nevertheless, considering the computational time, the ML algorithm is considerably faster than the traditional method. This is especially true since most of the computational time needed for ML is for training, which is only required once. Once trained, the prediction of physical parameters for any new set of galaxies ($\sim 9600$) can be achieved within seconds. 
Therefore, ML techniques are an excellent and efficient tool that can be used to exploit the high-quality data that will be available from WEAVE-StePS, and the huge quantity of data from upcoming large spectroscopic surveys from the MUltiplexed Survey Telescope \citep[MUST;][]{Zhang:2023}, the MaunaKea Spectroscopic Explorer \citep[MSE;][]{MSE:2019}, and the Wide-Field Spectroscopic Telescope \cite[WST;][]{WST:2024}.

\begin{acknowledgements}  
Funding for the WEAVE facility has been provided by UKRI STFC, the University of Oxford, NOVA, NWO, Instituto de Astrofísica de Canarias (IAC), the Isaac Newton Group partners (STFC, NWO, and Spain, led by 
the IAC), INAF, CNRS-INSU, the Observatoire de Paris, Région Île-de-France, CONCYT through INAOE, Konkoly Observatory (CSFK), Max-Planck-Institut für Astronomie (MPIA Heidelberg), Lund University, the Leibniz
Institute for Astrophysics Potsdam (AIP), the Swedish Research Council, the European Commission, and the University of Pennsylvania.

The WEAVE Survey Consortium consists of the ING, its three partners, represented by UKRI STFC, NWO, and the IAC, NOVA, INAF, GEPI, INAOE, and individual WEAVE Participants. Please see the relevant footnotes
for the WEAVE website\footnote{\url{https://ingconfluence.ing.iac.es/confluence//display/WEAV/The+WEAVE+Project}} and for the full list of granting agencies and grants supporting WEAVE
\footnote{\url{https://ingconfluence.ing.iac.es/confluence/display/WEAV/WEAVE+Acknowledgements}}.

This work makes use of data from the European Space Agency (ESA) mission {\it Gaia} (\url{https://www.cosmos.esa.int/gaia}), processed by the {\it Gaia}

Data Processing and Analysis Consortium (DPAC, \url{https://www.cosmos.esa.int/web/gaia/dpac/consortium}). Funding for the DPAC has been provided by national institutions, in particular, the institutions part
icipating in the {\it Gaia} Multilateral Agreement.

This work makes use of data from the VIMOS VLT Deep Survey, the VIPERS-MLS database and the HST-COSMOS database, operated by CeSAM/Laboratoire d'Astrophysique de Marseille, France.

This work makes use of observations obtained with MegaPrime/MegaCam, a joint project of CFHT and CEA/IRFU, at the Canada-France-Hawaii Telescope (CFHT), which is operated by the National Research Council (NRC)
 of Canada, the Institut National des Science de l'Univers of the Centre National de la Recherche Scientifique (CNRS) of France, and the University of Hawaii.

This work makes use of data products produced at Terapix available at the Canadian Astronomy Data Centre as part of the Canada-France-Hawaii Telescope Legacy Survey, a collaborative project of NRC and CNRS.

J.A. acknowledges financial support from INAF-WEAVE funds, program 1.05.03.04.05 and INAF-OABrera funds, program 1.05.01.01. 

A.I., F.D., M.L.  acknowledge financial support from INAF Mainstream grant 2019 WEAVE StePS 1.05.01.86.16 and INAF Large Grant 2022 WEAVE StePS 1.05.12.01.11.

R.G.B. and R.G.D. acknowledge financial support from the Severo Ochoa grant CEX2021-001131-S funded by MCIN/AEI/ 10.13039/501100011033 and to PID2022-141755NB-I00.
 
A.F.M. acknowledges support from RYC2021-031099-I and PID2021-123313NA-I00 of MICIN/AEI/10.13039/501100011033/FEDER,UE,NextGenerationEU/PR

M.Bi. acknowledges support from STFC grant numbers ST/N021702/1

G.B., M.Bo., F.R.D., A.I., F.L.B., M.L., P.M., B.P., C.T., D.V., S.Z., and A.R.G. acknowledge financial support from INAF funds, program 1.05.01.86.16 - Mainstream 2019.   

L.C. acknowledges support by grant PIB2021-127718NB-100 and PID2022-139567NB-I00 from the Spanish Ministry of Science and Innovation/State Agency of Research MCIN/AEI/10.13039/501100011033 and by {\it ERDF A way of making Europe}.

Co-funded by the European Union. Views and opinions expressed are however those of the author(s) only and do not necessarily reflect those of the European Union. Neither the European Union nor the granting authority can be held responsible for them. JHK acknowledges grant PID2022-136505NB-I00 funded by MCIN/AEI/10.13039/501100011033 and EU, ERDF.

E.M.C. and A.P. are supported by the MUR grant PRIN INAF 2022 C53D2300085-0006 and Padua University grants DOR 2021-2023.

We also like to thank Daniela Bettoni for providing suggestions to improve this paper.

\end{acknowledgements}
\bibliographystyle{aa}
\bibliography{references_ML_WStePS.bib}

\begin{thebibliography}{91}
\expandafter\ifx\csname natexlab\endcsname\relax\def\natexlab#1{#1}\fi

\bibitem[{{Aihara} {et~al.}(2018){Aihara}, {Arimoto}, {Armstrong}, {Arnouts},
  {Bahcall}, {Bickerton}, {Bosch}, {Bundy}, {Capak}, {Chan}, {Chiba}, {Coupon},
  {Egami}, {Enoki}, {Finet}, {Fujimori}, {Fujimoto}, {Furusawa}, {Furusawa},
  {Goto}, {Goulding}, {Greco}, {Greene}, {Gunn}, {Hamana}, {Harikane},
  {Hashimoto}, {Hattori}, {Hayashi}, {Hayashi}, {He{\l}miniak}, {Higuchi},
  {Hikage}, {Ho}, {Hsieh}, {Huang}, {Huang}, {Ikeda}, {Imanishi}, {Inoue},
  {Iwasawa}, {Iwata}, {Jaelani}, {Jian}, {Kamata}, {Karoji}, {Kashikawa},
  {Katayama}, {Kawanomoto}, {Kayo}, {Koda}, {Koike}, {Kojima}, {Komiyama},
  {Konno}, {Koshida}, {Koyama}, {Kusakabe}, {Leauthaud}, {Lee}, {Lin}, {Lin},
  {Lupton}, {Mandelbaum}, {Matsuoka}, {Medezinski}, {Mineo}, {Miyama},
  {Miyatake}, {Miyazaki}, {Momose}, {More}, {More}, {Moritani}, {Moriya},
  {Morokuma}, {Mukae}, {Murata}, {Murayama}, {Nagao}, {Nakata}, {Niida},
  {Niikura}, {Nishizawa}, {Obuchi}, {Oguri}, {Oishi}, {Okabe}, {Okamoto},
  {Okura}, {Ono}, {Onodera}, {Onoue}, {Osato}, {Ouchi}, {Price}, {Pyo}, {Sako},
  {Sawicki}, {Shibuya}, {Shimasaku}, {Shimono}, {Shirasaki}, {Silverman},
  {Simet}, {Speagle}, {Spergel}, {Strauss}, {Sugahara}, {Sugiyama}, {Suto},
  {Suyu}, {Suzuki}, {Tait}, {Takada}, {Takata}, {Tamura}, {Tanaka}, {Tanaka},
  {Tanaka}, {Tanaka}, {Terai}, {Terashima}, {Toba}, {Tominaga}, {Toshikawa},
  {Turner}, {Uchida}, {Uchiyama}, {Umetsu}, {Uraguchi}, {Urata}, {Usuda},
  {Utsumi}, {Wang}, {Wang}, {Wong}, {Yabe}, {Yamada}, {Yamanoi}, {Yasuda},
  {Yeh}, {Yonehara}, \& {Yuma}}]{Aihara:2018}
{Aihara}, H., {Arimoto}, N., {Armstrong}, R., {et~al.} 2018, \pasj, 70, S4

\bibitem[{Altman(1992)}]{NSA:1992}
Altman, N.~S. 1992, The American Statistician, 46, 175

\bibitem[{{Angthopo} {et~al.}(2019){Angthopo}, {Ferreras}, \&
  {Silk}}]{Angthopo:2019}
{Angthopo}, J., {Ferreras}, I., \& {Silk}, J. 2019, \mnras, 488, L99

\bibitem[{{Angthopo} {et~al.}(2020){Angthopo}, {Ferreras}, \&
  {Silk}}]{Angthopo:2020}
{Angthopo}, J., {Ferreras}, I., \& {Silk}, J. 2020, \mnras, 495, 2720

\bibitem[{{Arimoto}(1996)}]{Arimoto:1996}
{Arimoto}, N. 1996, in Astronomical Society of the Pacific Conference Series,
  Vol.~98, From Stars to Galaxies: the Impact of Stellar Physics on Galaxy
  Evolution, ed. C.~{Leitherer}, U.~{Fritze-von-Alvensleben}, \& J.~{Huchra},
  287

\bibitem[{{Arnouts} {et~al.}(2013){Arnouts}, {Le Floc'h}, {Chevallard},
  {Johnson}, {Ilbert}, {Treyer}, {Aussel}, {Capak}, {Sanders}, {Scoville},
  {McCracken}, {Milliard}, {Pozzetti}, \& {Salvato}}]{Arnouts:2013}
{Arnouts}, S., {Le Floc'h}, E., {Chevallard}, J., {et~al.} 2013, \aap, 558, A67

\bibitem[{{Baldry} {et~al.}(2004){Baldry}, {Glazebrook}, {Brinkmann},
  {Ivezi{\'c}}, {Lupton}, {Nichol}, \& {Szalay}}]{2004Bald}
{Baldry}, I.~K., {Glazebrook}, K., {Brinkmann}, J., {et~al.} 2004, ApJ, 600,
  681

\bibitem[{{Ball} {et~al.}(2008){Ball}, {Brunner}, {Myers}, {Strand}, {Alberts},
  \& {Tcheng}}]{Ball:2008}
{Ball}, N.~M., {Brunner}, R.~J., {Myers}, A.~D., {et~al.} 2008, \apj, 683, 12

\bibitem[{{Bennett} {et~al.}(2014){Bennett}, {Larson}, {Weiland}, \&
  {Hinshaw}}]{Bennett:2014}
{Bennett}, C.~L., {Larson}, D., {Weiland}, J.~L., \& {Hinshaw}, G. 2014, \apj,
  794, 135

\bibitem[{{Bonjean} {et~al.}(2019){Bonjean}, {Aghanim}, {Salom{\'e}}, {Beelen},
  {Douspis}, \& {Soubri{\'e}}}]{Bonjean:2019}
{Bonjean}, V., {Aghanim}, N., {Salom{\'e}}, P., {et~al.} 2019, \aap, 622, A137

\bibitem[{{Breiman}(2001)}]{Breiman:2001}
{Breiman}, L. 2001, Machine Learning, 45, 5

\bibitem[{{Brinchmann} {et~al.}(2004){Brinchmann}, {Charlot}, {White},
  {Tremonti}, {Kauffmann}, {Heckman}, \& {Brinkmann}}]{Brinchmann:2004}
{Brinchmann}, J., {Charlot}, S., {White}, S.~D.~M., {et~al.} 2004, \mnras, 351,
  1151

\bibitem[{{Bruzual} \& {Charlot}(2003)}]{BC03}
{Bruzual}, G. \& {Charlot}, S. 2003, MNRAS, 344, 1000

\bibitem[{{Cappellari}(2017)}]{2017pPXF}
{Cappellari}, M. 2017, MNRAS, 466, 798

\bibitem[{{Cappellari}(2023)}]{Cappellari:2023}
{Cappellari}, M. 2023, \mnras, 526, 3273

\bibitem[{{Carnall} {et~al.}(2018){Carnall}, {McLure}, {Dunlop}, \&
  {Dav{\'e}}}]{Carnall:2018}
{Carnall}, A.~C., {McLure}, R.~J., {Dunlop}, J.~S., \& {Dav{\'e}}, R. 2018,
  \mnras, 480, 4379

\bibitem[{{Carter} {et~al.}(2009){Carter}, {Smith}, {Percival}, {Baldry},
  {Collins}, {James}, {Salaris}, {Simpson}, {Stott}, \&
  {Mobasher}}]{Carter:2009}
{Carter}, D., {Smith}, D. J.~B., {Percival}, S.~M., {et~al.} 2009, \mnras, 397,
  695

\bibitem[{{Chabrier}(2003)}]{Chab:03}
{Chabrier}, G. 2003, \pasp, 115, 763

\bibitem[{{Charlot} \& {Fall}(2000)}]{Charlot2000}
{Charlot}, S. \& {Fall}, S.~M. 2000, \apj, 539, 718

\bibitem[{{Cid Fernandes} {et~al.}(2005){Cid Fernandes}, {Mateus}, {Sodr{\'e}},
  {Stasi{\'n}ska}, \& {Gomes}}]{SLight}
{Cid Fernandes}, R., {Mateus}, A., {Sodr{\'e}}, L., {Stasi{\'n}ska}, G., \&
  {Gomes}, J.~M. 2005, MNRAS, 358, 363

\bibitem[{{Costantin} {et~al.}(2019){Costantin}, {Iovino}, {Zibetti},
  {Longhetti}, {Gallazzi}, {Mercurio}, {Lonoce}, {Balcells}, {Bolzonella},
  {Busarello}, {Dalton}, {Ferr{\'e}-Mateu}, {Garc{\'\i}a-Benito}, {Gargiulo},
  {Haines}, {Jin}, {La Barbera}, {McGee}, {Merluzzi}, {Morelli}, {Murphy},
  {Peralta de Arriba}, {Pizzella}, {Poggianti}, {Pozzetti},
  {S{\'a}nchez-Bl{\'a}zquez}, {Talia}, {Tortora}, {Trager}, {Vazdekis},
  {Vergani}, \& {Vulcani}}]{Costantin:2019}
{Costantin}, L., {Iovino}, A., {Zibetti}, S., {et~al.} 2019, \aap, 632, A9

\bibitem[{{Crain} {et~al.}(2015){Crain}, {Schaye}, {Bower}, {Furlong},
  {Schaller}, {Theuns}, {Dalla Vecchia}, {Frenk}, {McCarthy}, {Helly},
  {Jenkins}, {Rosas-Guevara}, {White}, \& {Trayford}}]{Crain:15}
{Crain}, R.~A., {Schaye}, J., {Bower}, R.~G., {et~al.} 2015, MNRAS, 450, 1937

\bibitem[{{Dalton} {et~al.}(2016){Dalton}, {Trager}, {Abrams}, {Bonifacio},
  {Aguerri}, {Middleton}, {Benn}, {Dee}, {Say{\`e}de}, {Lewis}, {Pragt},
  {Pico}, {Walton}, {Rey}, {Allende Prieto}, {Pe{\~n}ate}, {Lhome},
  {Ag{\'o}cs}, {Alonso}, {Terrett}, {Brock}, {Gilbert}, {Schallig}, {Ridings},
  {Guinouard}, {Verheijen}, {Tosh}, {Rogers}, {Lee}, {Steele}, {Stuik},
  {Tromp}, {Jask{\'o}}, {Carrasco}, {Farcas}, {Kragt}, {Lesman}, {Kroes},
  {Mottram}, {Bates}, {Rodriguez}, {Gribbin}, {Delgado}, {Herreros}, {Martin},
  {Cano}, {Navarro}, {Irwin}, {Lewis}, {Gonzalez Solares}, {Murphy}, {Worley},
  {Bassom}, {O'Mahoney}, {Bianco}, {Zurita}, {ter Horst}, {Molinari}, {Lodi},
  {Guerra}, {Martin}, {Vallenari}, {Salasnich}, {Baruffolo}, {Jin}, {Hill},
  {Smith}, {Drew}, {Poggianti}, {Pieri}, {Dominquez Palmero}, \&
  {Farina}}]{Dalton:2016}
{Dalton}, G., {Trager}, S., {Abrams}, D.~C., {et~al.} 2016, in Society of
  Photo-Optical Instrumentation Engineers (SPIE) Conference Series, Vol. 9908,
  Ground-based and Airborne Instrumentation for Astronomy VI, ed. C.~J.
  {Evans}, L.~{Simard}, \& H.~{Takami}, 99081G

\bibitem[{{Dalton} {et~al.}(2012){Dalton}, {Trager}, {Abrams}, {Carter},
  {Bonifacio}, {Aguerri}, {MacIntosh}, {Evans}, {Lewis}, {Navarro}, {Agocs},
  {Dee}, {Rousset}, {Tosh}, {Middleton}, {Pragt}, {Terrett}, {Brock}, {Benn},
  {Verheijen}, {Cano Infantes}, {Bevil}, {Steele}, {Mottram}, {Bates},
  {Gribbin}, {Rey}, {Rodriguez}, {Delgado}, {Guinouard}, {Walton}, {Irwin},
  {Jagourel}, {Stuik}, {Gerlofsma}, {Roelfsma}, {Skillen}, {Ridings},
  {Balcells}, {Daban}, {Gouvret}, {Venema}, \& {Girard}}]{Dalton:2012}
{Dalton}, G., {Trager}, S.~C., {Abrams}, D.~C., {et~al.} 2012, in Society of
  Photo-Optical Instrumentation Engineers (SPIE) Conference Series, Vol. 8446,
  Ground-based and Airborne Instrumentation for Astronomy IV, ed. I.~S.
  {McLean}, S.~K. {Ramsay}, \& H.~{Takami}, 84460P

\bibitem[{{Davidzon} {et~al.}(2019){Davidzon}, {Laigle}, {Capak}, {Ilbert},
  {Masters}, {Hemmati}, {Apostolakos}, {Coupon}, {de la Torre}, {Devriendt},
  {Dubois}, {Kashino}, {Paltani}, \& {Pichon}}]{Davidzon:2019}
{Davidzon}, I., {Laigle}, C., {Capak}, P.~L., {et~al.} 2019, \mnras, 489, 4817

\bibitem[{{de Graaff} {et~al.}(2021){de Graaff}, {Bezanson}, {Franx}, {van der
  Wel}, {Holden}, {van de Sande}, {Bell}, {D'Eugenio}, {Maseda}, {Muzzin},
  {Sobral}, {Straatman}, \& {Wu}}]{Graaff:2021}
{de Graaff}, A., {Bezanson}, R., {Franx}, M., {et~al.} 2021, \apj, 913, 103

\bibitem[{{Ditrani} {et~al.}(2023){Ditrani}, {Longhetti}, {La Barbera},
  {Iovino}, {Costantin}, {Zibetti}, {Gallazzi}, {Fossati}, {Angthopo},
  {Ascasibar}, {Poggianti}, {S{\'a}nchez-Bl{\'a}zquez}, {Balcells}, {Bianconi},
  {Bolzonella}, {Cassar{\`a}}, {Cucciati}, {Dalton}, {Ferr{\'e}-Mateu},
  {Garc{\'\i}a-Benito}, {Granett}, {Gullieuszik}, {Ikhsanova}, {Jin}, {Knapen},
  {McGee}, {Mercurio}, {Morelli}, {Moretti}, {Murphy}, {Pizzella}, {Pozzetti},
  {Spiniello}, {Tortora}, {Trager}, {Vazdekis}, {Vergani}, \&
  {Vulcani}}]{Ditrani:2023}
{Ditrani}, F.~R., {Longhetti}, M., {La Barbera}, F., {et~al.} 2023, \aap, 677,
  A93

\bibitem[{{Dom{\'\i}nguez S{\'a}nchez} {et~al.}(2018){Dom{\'\i}nguez
  S{\'a}nchez}, {Huertas-Company}, {Bernardi}, {Tuccillo}, \&
  {Fischer}}]{Dominguez:2018}
{Dom{\'\i}nguez S{\'a}nchez}, H., {Huertas-Company}, M., {Bernardi}, M.,
  {Tuccillo}, D., \& {Fischer}, J.~L. 2018, \mnras, 476, 3661

\bibitem[{{Donnari} {et~al.}(2019){Donnari}, {Pillepich}, {Nelson},
  {Vogelsberger}, {Genel}, {Weinberger}, {Marinacci}, {Springel}, \&
  {Hernquist}}]{Donnari:2019}
{Donnari}, M., {Pillepich}, A., {Nelson}, D., {et~al.} 2019, \mnras, 485, 4817

\bibitem[{{Euclid Collaboration: Bisigello} {et~al.}(2023){Euclid
  Collaboration: Bisigello}, {Conselice}, {Baes}, {Bolzonella}, {Brescia},
  {Cavuoti}, {Cucciati}, {Humphrey}, {Hunt}, {Maraston}, {Pozzetti}, {Tortora},
  {van Mierlo}, {Aghanim}, {Auricchio}, {Baldi}, {Bender}, {Bodendorf},
  {Bonino}, {Branchini}, {Brinchmann}, {Camera}, {Capobianco}, {Carbone},
  {Carretero}, {Castander}, {Castellano}, {Cimatti}, {Congedo}, {Conversi},
  {Copin}, {Corcione}, {Courbin}, {Cropper}, {Da Silva}, {Degaudenzi},
  {Douspis}, {Dubath}, {Duncan}, {Dupac}, {Dusini}, {Farrens}, {Ferriol},
  {Frailis}, {Franceschi}, {Franzetti}, {Fumana}, {Garilli}, {Gillard},
  {Gillis}, {Giocoli}, {Grazian}, {Grupp}, {Guzzo}, {Haugan}, {Holmes},
  {Hormuth}, {Hornstrup}, {Jahnke}, {K{\"u}mmel}, {Kermiche}, {Kiessling},
  {Kilbinger}, {Kohley}, {Kunz}, {Kurki-Suonio}, {Ligori}, {Lilje}, {Lloro},
  {Maiorano}, {Mansutti}, {Marggraf}, {Markovic}, {Marulli}, {Massey},
  {Maurogordato}, {Medinaceli}, {Meneghetti}, {Merlin}, {Meylan}, {Moresco},
  {Moscardini}, {Munari}, {Niemi}, {Padilla}, {Paltani}, {Pasian}, {Pedersen},
  {Pettorino}, {Polenta}, {Poncet}, {Popa}, {Raison}, {Renzi}, {Rhodes},
  {Riccio}, {Rix}, {Romelli}, {Roncarelli}, {Rosset}, {Rossetti}, {Saglia},
  {Sapone}, {Sartoris}, {Schneider}, {Scodeggio}, {Secroun}, {Seidel},
  {Sirignano}, {Sirri}, {Stanco}, {Tallada-Cresp{\'\i}}, {Tavagnacco},
  {Taylor}, {Tereno}, {Toledo-Moreo}, {Torradeflot}, {Tutusaus}, {Valentijn},
  {Valenziano}, {Vassallo}, {Wang}, {Zacchei}, {Zamorani}, {Zoubian},
  {Andreon}, {Bardelli}, {Boucaud}, {Colodro-Conde}, {Di Ferdinando},
  {Graci{\'a}-Carpio}, {Lindholm}, {Maino}, {Mei}, {Scottez}, {Sureau},
  {Tenti}, {Zucca}, {Borlaff}, {Ballardini}, {Biviano}, {Bozzo}, {Burigana},
  {Cabanac}, {Cappi}, {Carvalho}, {Casas}, {Castignani}, {Cooray}, {Coupon},
  {Courtois}, {Cuby}, {Davini}, {De Lucia}, {Desprez}, {Dole}, {Escartin},
  {Escoffier}, {Farina}, {Fotopoulou}, {Ganga}, {Garcia-Bellido}, {George},
  {Giacomini}, {Gozaliasl}, {Hildebrandt}, {Hook}, {Huertas-Company}, {Kansal},
  {Keihanen}, {Kirkpatrick}, {Loureiro}, {Mac{\'\i}as-P{\'e}rez},
  {Magliocchetti}, {Mainetti}, {Marcin}, {Martinelli}, {Martinet}, {Metcalf},
  {Monaco}, {Morgante}, {Nadathur}, {Nucita}, {Patrizii}, {Peel}, {Potter},
  {Pourtsidou}, {P{\"o}ntinen}, {Reimberg}, {S{\'a}nchez}, {Sakr}, {Schirmer},
  {Sefusatti}, {Sereno}, {Stadel}, {Teyssier}, {Valieri}, {Valiviita}, \&
  {Viel}}]{Euclid:2023}
{Euclid Collaboration: Bisigello}, L., {Conselice}, C.~J., {Baes}, M., {et~al.}
  2023, \mnras, 520, 3529

\bibitem[{{Faber} {et~al.}(2007){Faber}, {Willmer}, {Wolf}, {Koo}, {Weiner},
  {Newman}, {Im}, {Coil}, {Conroy}, {Cooper}, {Davis}, {Finkbeiner}, {Gerke},
  {Gebhardt}, {Groth}, {Guhathakurta}, {Harker}, {Kaiser}, {Kassin},
  {Kleinheinrich}, {Konidaris}, {Kron}, {Lin}, {Luppino}, {Madgwick},
  {Meisenheimer}, {Noeske}, {Phillips}, {Sarajedini}, {Schiavon}, {Simard},
  {Szalay}, {Vogt}, \& {Yan}}]{Faber:07}
{Faber}, S.~M., {Willmer}, C.~N.~A., {Wolf}, C., {et~al.} 2007, ApJ, 665, 265

\bibitem[{{Fritz} {et~al.}(2014){Fritz}, {Scodeggio}, {Ilbert}, {Bolzonella},
  {Davidzon}, {Coupon}, {Garilli}, {Guzzo}, {Zamorani}, {Abbas}, {Adami},
  {Arnouts}, {Bel}, {Bottini}, {Branchini}, {Cappi}, {Cucciati}, {De Lucia},
  {de la Torre}, {Franzetti}, {Fumana}, {Granett}, {Iovino}, {Krywult}, {Le
  Brun}, {Le F{\`e}vre}, {Maccagni}, {Ma{\l}ek}, {Marulli}, {McCracken},
  {Paioro}, {Polletta}, {Pollo}, {Schlagenhaufer}, {Tasca}, {Tojeiro},
  {Vergani}, {Zanichelli}, {Burden}, {Di Porto}, {Marchetti}, {Marinoni},
  {Mellier}, {Moscardini}, {Nichol}, {Peacock}, {Percival}, {Phleps}, \&
  {Wolk}}]{Fritz:2014}
{Fritz}, A., {Scodeggio}, M., {Ilbert}, O., {et~al.} 2014, \aap, 563, A92

\bibitem[{{Furlong} {et~al.}(2015){Furlong}, {Bower}, {Theuns}, {Schaye},
  {Crain}, {Schaller}, {Dalla Vecchia}, {Frenk}, {McCarthy}, {Helly},
  {Jenkins}, \& {Rosas-Guevara}}]{Furlong:2015}
{Furlong}, M., {Bower}, R.~G., {Theuns}, T., {et~al.} 2015, \mnras, 450, 4486

\bibitem[{{Gallazzi} {et~al.}(2005){Gallazzi}, {Charlot}, {Brinchmann},
  {White}, \& {Tremonti}}]{Gallazzi:2005}
{Gallazzi}, A., {Charlot}, S., {Brinchmann}, J., {White}, S. D.~M., \&
  {Tremonti}, C.~A. 2005, \mnras, 362, 41

\bibitem[{{Gavazzi} {et~al.}(2002){Gavazzi}, {Bonfanti}, {Sanvito}, {Boselli},
  \& {Scodeggio}}]{Gavazzi:2002}
{Gavazzi}, G., {Bonfanti}, C., {Sanvito}, G., {Boselli}, A., \& {Scodeggio}, M.
  2002, \apj, 576, 135

\bibitem[{{Guzzo} {et~al.}(2014){Guzzo}, {Scodeggio}, {Garilli}, {Granett},
  {Fritz}, {Abbas}, {Adami}, {Arnouts}, {Bel}, {Bolzonella}, {Bottini},
  {Branchini}, {Cappi}, {Coupon}, {Cucciati}, {Davidzon}, {De Lucia}, {de la
  Torre}, {Franzetti}, {Fumana}, {Hudelot}, {Ilbert}, {Iovino}, {Krywult}, {Le
  Brun}, {Le F{\`e}vre}, {Maccagni}, {Ma{\l}ek}, {Marulli}, {McCracken},
  {Paioro}, {Peacock}, {Polletta}, {Pollo}, {Schlagenhaufer}, {Tasca},
  {Tojeiro}, {Vergani}, {Zamorani}, {Zanichelli}, {Burden}, {Di Porto},
  {Marchetti}, {Marinoni}, {Mellier}, {Moscardini}, {Nichol}, {Percival},
  {Phleps}, \& {Wolk}}]{Guzzo:2014}
{Guzzo}, L., {Scodeggio}, M., {Garilli}, B., {et~al.} 2014, \aap, 566, A108

\bibitem[{{Hopkins} {et~al.}(2013)}]{2013GAMA}
{Hopkins}, A.~M. {et~al.} 2013, MNRAS, 430, 2047

\bibitem[{{Hunt} {et~al.}(2024){Hunt}, {Pimbblet}, \& {Benoit}}]{Hunt:2024}
{Hunt}, L.~J., {Pimbblet}, K.~A., \& {Benoit}, D.~M. 2024, \mnras, 529, 479

\bibitem[{{Ilbert} {et~al.}(2006){Ilbert}, {Arnouts}, {McCracken},
  {Bolzonella}, {Bertin}, {Le F{\`e}vre}, {Mellier}, {Zamorani}, {Pell{\`o}},
  {Iovino}, {Tresse}, {Le Brun}, {Bottini}, {Garilli}, {Maccagni}, {Picat},
  {Scaramella}, {Scodeggio}, {Vettolani}, {Zanichelli}, {Adami}, {Bardelli},
  {Cappi}, {Charlot}, {Ciliegi}, {Contini}, {Cucciati}, {Foucaud}, {Franzetti},
  {Gavignaud}, {Guzzo}, {Marano}, {Marinoni}, {Mazure}, {Meneux}, {Merighi},
  {Paltani}, {Pollo}, {Pozzetti}, {Radovich}, {Zucca}, {Bondi}, {Bongiorno},
  {Busarello}, {de La Torre}, {Gregorini}, {Lamareille}, {Mathez}, {Merluzzi},
  {Ripepi}, {Rizzo}, \& {Vergani}}]{Ilbert:2006}
{Ilbert}, O., {Arnouts}, S., {McCracken}, H.~J., {et~al.} 2006, \aap, 457, 841

\bibitem[{{Iovino} {et~al.}(2023{\natexlab{a}}){Iovino}, {Mercurio},
  {Gallazzi}, {La Barbera}, {Longhetti}, {Tortora}, {Zibetti}, {Belfiore},
  {Bianconi}, {Busarello}, {Corsini}, {Costantin}, {De Lucia}, {De Propris},
  {D'Eugenio}, {Fontanot}, {Garc{\'\i}a-Benito}, {Hirschmann}, {Haines},
  {Mannucci}, {McGee}, {Merluzzi}, {Morelli}, {Moretti}, {Pasquali},
  {Poggianti}, {Pozzetti}, {Rodighiero}, {S{\'a}nchez-Bl{\'a}zquez}, {van der
  Wel}, {Vazdekis}, {Vulcani}, {Zanella}, {Annunziatella}, {Concas},
  {Cassar{\`a}}, {Cresci}, {Curti}, {de Lorenzo-C{\'a}ceres}, {Mateu},
  {Delgado}, {Mancini}, {Pacifici}, {Perez-Montero}, {Pizzella},
  {Perez-Gonzalez}, {Trager}, \& {Vergani}}]{Iovino4MOST}
{Iovino}, A., {Mercurio}, A., {Gallazzi}, A.~R., {et~al.} 2023{\natexlab{a}},
  The Messenger, 190, 22

\bibitem[{{Iovino} {et~al.}(2023{\natexlab{b}}){Iovino}, {Poggianti},
  {Mercurio}, {Longhetti}, {Bolzonella}, {Busarello}, {Gullieuszik}, {La
  Barbera}, {Merluzzi}, {Morelli}, {Tortora}, {Vergani}, {Zibetti}, {Haines},
  {Costantin}, {Ditrani}, {Pozzetti}, {Angthopo}, {Balcells}, {Bardelli},
  {Benn}, {Bianconi}, {Cassar{\`a}}, {Corsini}, {Cucciati}, {Dalton},
  {Ferr{\'e}-Mateu}, {Fossati}, {Gallazzi}, {Garc{\'\i}a-Benito}, {Granett},
  {Gonz{\'a}lez Delgado}, {Ikhsanova}, {Iodice}, {Jin}, {Knapen}, {McGee},
  {Moretti}, {Murphy}, {Peralta de Arriba}, {Pizzella},
  {S{\'a}nchez-Bl{\'a}zquez}, {Spiniello}, {Talia}, {Trager}, {Vazdekis},
  {Vulcani}, \& {Zucca}}]{Iovino:2023}
{Iovino}, A., {Poggianti}, B.~M., {Mercurio}, A., {et~al.} 2023{\natexlab{b}},
  \aap, 672, A87

\bibitem[{{Jian} {et~al.}(2020){Jian}, {Lin}, {Koyama}, {Tanaka}, {Umetsu},
  {Hsieh}, {Higuchi}, {Oguri}, {More}, {Komiyama}, {Kodama}, {Nishizawa}, \&
  {Chang}}]{Jian:2020}
{Jian}, H.-Y., {Lin}, L., {Koyama}, Y., {et~al.} 2020, \apj, 894, 125

\bibitem[{{Jin} {et~al.}(2024){Jin}, {Trager}, {Dalton}, {Aguerri}, {Drew},
  {Falc{\'o}n-Barroso}, {G{\"a}nsicke}, {Hill}, {Iovino}, {Pieri}, {Poggianti},
  {Smith}, {Vallenari}, {Abrams}, {Aguado}, {Antoja}, {Arag{\'o}n-Salamanca},
  {Ascasibar}, {Babusiaux}, {Balcells}, {Barrena}, {Battaglia}, {Belokurov},
  {Bensby}, {Bonifacio}, {Bragaglia}, {Carrasco}, {Carrera}, {Cornwell},
  {Dom{\'\i}nguez-Palmero}, {Duncan}, {Famaey}, {Fari{\~n}a}, {Gonzalez},
  {Guest}, {Hatch}, {Hess}, {Hoskin}, {Irwin}, {Knapen}, {Koposov}, {Kuchner},
  {Laigle}, {Lewis}, {Longhetti}, {Lucatello}, {M{\'e}ndez-Abreu}, {Mercurio},
  {Molaeinezhad}, {Mongui{\'o}}, {Morrison}, {Murphy}, {Peralta de Arriba},
  {P{\'e}rez}, {P{\'e}rez-R{\`a}fols}, {Pic{\'o}}, {Raddi}, {Romero-G{\'o}mez},
  {Royer}, {Siebert}, {Seabroke}, {Som}, {Terrett}, {Thomas}, {Wesson},
  {Worley}, {Alfaro}, {Allende Prieto}, {Alonso-Santiago}, {Amos}, {Ashley},
  {Balaguer-N{\'u}{\~n}ez}, {Balbinot}, {Bellazzini}, {Benn}, {Berlanas},
  {Bernard}, {Best}, {Bettoni}, {Bianco}, {Bishop}, {Blomqvist}, {Boeche},
  {Bolzonella}, {Bonoli}, {Bosma}, {Britavskiy}, {Busarello}, {Caffau},
  {Cantat-Gaudin}, {Castro-Ginard}, {Couto}, {Carbajo-Hijarrubia}, {Carter},
  {Casamiquela}, {Conrado}, {Corcho-Caballero}, {Costantin}, {Deason}, {de
  Burgos}, {De Grandi}, {Di Matteo}, {Dom{\'\i}nguez-G{\'o}mez}, {Dorda},
  {Drake}, {Dutta}, {Erkal}, {Feltzing}, {Ferr{\'e}-Mateu}, {Feuillet},
  {Figueras}, {Fossati}, {Franciosini}, {Frasca}, {Fumagalli}, {Gallazzi},
  {Garc{\'\i}a-Benito}, {Gentile Fusillo}, {Gebran}, {Gilbert}, {Gledhill},
  {Gonz{\'a}lez Delgado}, {Greimel}, {Guarcello}, {Guerra}, {Gullieuszik},
  {Haines}, {Hardcastle}, {Harris}, {Haywood}, {Helmi}, {Hernandez}, {Herrero},
  {Hughes}, {Ir{\v{s}}i{\v{c}}}, {Jablonka}, {Jarvis}, {Jordi}, {Kondapally},
  {Kordopatis}, {Krogager}, {La Barbera}, {Lam}, {Larsen}, {Lemasle}, {Lewis},
  {Lhom{\'e}}, {Lind}, {Lodi}, {Longobardi}, {Lonoce}, {Magrini}, {Ma{\'\i}z
  Apell{\'a}niz}, {Marchal}, {Marco}, {Martin}, {Matsuno}, {Maurogordato},
  {Merluzzi}, {Miralda-Escud{\'e}}, {Molinari}, {Monari}, {Morelli}, {Mottram},
  {Naylor}, {Negueruela}, {O{\~n}orbe}, {Pancino}, {Peirani}, {Peletier},
  {Pozzetti}, {Rainer}, {Ramos}, {Read}, {Rossi}, {R{\"o}ttgering},
  {Rubi{\~n}o-Mart{\'\i}n}, {Sabater}, {San Juan}, {Sanna}, {Schallig},
  {Schiavon}, {Schultheis}, {Serra}, {Shimwell}, {Sim{\'o}n-D{\'\i}az},
  {Smith}, {Sordo}, {Sorini}, {Soubiran}, {Starkenburg}, {Steele}, {Stott},
  {Stuik}, {Tolstoy}, {Tortora}, {Tsantaki}, {Van der Swaelmen}, {van Weeren},
  {Vergani}, {Verheijen}, {Verro}, {Vink}, {Vioque}, {Walcher}, {Walton},
  {Wegg}, {Weijmans}, {Williams}, {Wilson}, {Wright}, {Xylakis-Dornbusch},
  {Youakim}, {Zibetti}, \& {Zurita}}]{Jin:2024}
{Jin}, S., {Trager}, S.~C., {Dalton}, G.~B., {et~al.} 2024, \mnras, 530, 2688

\bibitem[{{Johnson} {et~al.}(2021){Johnson}, {Leja}, {Conroy}, \&
  {Speagle}}]{Johnson:21}
{Johnson}, B.~D., {Leja}, J., {Conroy}, C., \& {Speagle}, J.~S. 2021, \apjs,
  254, 22

\bibitem[{{Kaviraj} {et~al.}(2007){Kaviraj}, {Rey}, {Rich}, {Yoon}, \&
  {Yi}}]{Kaviraj:2007}
{Kaviraj}, S., {Rey}, S.~C., {Rich}, R.~M., {Yoon}, S.~J., \& {Yi}, S.~K. 2007,
  \mnras, 381, L74

\bibitem[{{Koekemoer} {et~al.}(2011){Koekemoer}, {Faber}, {Ferguson}, {Grogin},
  {Kocevski}, {Koo}, {Lai}, {Lotz}, {Lucas}, {McGrath}, {Ogaz}, {Rajan},
  {Riess}, {Rodney}, {Strolger}, {Casertano}, {Castellano}, {Dahlen},
  {Dickinson}, {Dolch}, {Fontana}, {Giavalisco}, {Grazian}, {Guo}, {Hathi},
  {Huang}, {van der Wel}, {Yan}, {Acquaviva}, {Alexander}, {Almaini}, {Ashby},
  {Barden}, {Bell}, {Bournaud}, {Brown}, {Caputi}, {Cassata}, {Challis},
  {Chary}, {Cheung}, {Cirasuolo}, {Conselice}, {Roshan Cooray}, {Croton},
  {Daddi}, {Dav{\'e}}, {de Mello}, {de Ravel}, {Dekel}, {Donley}, {Dunlop},
  {Dutton}, {Elbaz}, {Fazio}, {Filippenko}, {Finkelstein}, {Frazer}, {Gardner},
  {Garnavich}, {Gawiser}, {Gruetzbauch}, {Hartley}, {H{\"a}ussler},
  {Herrington}, {Hopkins}, {Huang}, {Jha}, {Johnson}, {Kartaltepe},
  {Khostovan}, {Kirshner}, {Lani}, {Lee}, {Li}, {Madau}, {McCarthy},
  {McIntosh}, {McLure}, {McPartland}, {Mobasher}, {Moreira}, {Mortlock},
  {Moustakas}, {Mozena}, {Nandra}, {Newman}, {Nielsen}, {Niemi}, {Noeske},
  {Papovich}, {Pentericci}, {Pope}, {Primack}, {Ravindranath}, {Reddy},
  {Renzini}, {Rix}, {Robaina}, {Rosario}, {Rosati}, {Salimbeni}, {Scarlata},
  {Siana}, {Simard}, {Smidt}, {Snyder}, {Somerville}, {Spinrad}, {Straughn},
  {Telford}, {Teplitz}, {Trump}, {Vargas}, {Villforth}, {Wagner}, {Wandro},
  {Wechsler}, {Weiner}, {Wiklind}, {Wild}, {Wilson}, {Wuyts}, \&
  {Yun}}]{Koekemoer:2011}
{Koekemoer}, A.~M., {Faber}, S.~M., {Ferguson}, H.~C., {et~al.} 2011, \apjs,
  197, 36

\bibitem[{{Laigle} {et~al.}(2016){Laigle}, {McCracken}, {Ilbert}, {Hsieh},
  {Davidzon}, {Capak}, {Hasinger}, {Silverman}, {Pichon}, {Coupon}, {Aussel},
  {Le Borgne}, {Caputi}, {Cassata}, {Chang}, {Civano}, {Dunlop}, {Fynbo},
  {Kartaltepe}, {Koekemoer}, {Le F{\`e}vre}, {Le Floc'h}, {Leauthaud}, {Lilly},
  {Lin}, {Marchesi}, {Milvang-Jensen}, {Salvato}, {Sanders}, {Scoville},
  {Smolcic}, {Stockmann}, {Taniguchi}, {Tasca}, {Toft}, {Vaccari}, \&
  {Zabl}}]{Laigle:2016}
{Laigle}, C., {McCracken}, H.~J., {Ilbert}, O., {et~al.} 2016, \apjs, 224, 24

\bibitem[{{Le Cras} {et~al.}(2016){Le Cras}, {Maraston}, {Thomas}, \&
  {York}}]{LeCras:2016}
{Le Cras}, C., {Maraston}, C., {Thomas}, D., \& {York}, D.~G. 2016, \mnras,
  461, 766

\bibitem[{{Li} {et~al.}(2022){Li}, {Napolitano}, {Feng}, {Li}, {Amaro}, {Xie},
  {Tortora}, {Bilicki}, {Brescia}, {Cavuoti}, \& {Radovich}}]{LiR:2022}
{Li}, R., {Napolitano}, N.~R., {Feng}, H., {et~al.} 2022, \aap, 666, A85

\bibitem[{{Lilly} {et~al.}(2009){Lilly}, {Le Brun}, {Maier}, {Mainieri},
  {Mignoli}, {Scodeggio}, {Zamorani}, {Carollo}, {Contini}, {Kneib}, {Le
  F{\`e}vre}, {Renzini}, {Bardelli}, {Bolzonella}, {Bongiorno}, {Caputi},
  {Coppa}, {Cucciati}, {de la Torre}, {de Ravel}, {Franzetti}, {Garilli},
  {Iovino}, {Kampczyk}, {Kovac}, {Knobel}, {Lamareille}, {Le Borgne}, {Pello},
  {Peng}, {P{\'e}rez-Montero}, {Ricciardelli}, {Silverman}, {Tanaka}, {Tasca},
  {Tresse}, {Vergani}, {Zucca}, {Ilbert}, {Salvato}, {Oesch}, {Abbas},
  {Bottini}, {Capak}, {Cappi}, {Cassata}, {Cimatti}, {Elvis}, {Fumana},
  {Guzzo}, {Hasinger}, {Koekemoer}, {Leauthaud}, {Maccagni}, {Marinoni},
  {McCracken}, {Memeo}, {Meneux}, {Porciani}, {Pozzetti}, {Sanders},
  {Scaramella}, {Scarlata}, {Scoville}, {Shopbell}, \&
  {Taniguchi}}]{Lilly:2009}
{Lilly}, S.~J., {Le Brun}, V., {Maier}, C., {et~al.} 2009, \apjs, 184, 218

\bibitem[{{L{\'o}pez Fern{\'a}ndez} {et~al.}(2016){L{\'o}pez Fern{\'a}ndez},
  {Cid Fernandes}, {Gonz{\'a}lez Delgado}, {Vale Asari}, {P{\'e}rez},
  {Garc{\'\i}a-Benito}, {de Amorim}, {Lacerda}, {Cortijo-Ferrero}, \&
  {S{\'a}nchez}}]{RLF:2016}
{L{\'o}pez Fern{\'a}ndez}, R., {Cid Fernandes}, R., {Gonz{\'a}lez Delgado},
  R.~M., {et~al.} 2016, \mnras, 458, 184

\bibitem[{{MacArthur}(2005)}]{MacAruthur:05}
{MacArthur}, L.~A. 2005, \apj, 623, 795

\bibitem[{{Mainieri} {et~al.}(2024){Mainieri}, {Anderson}, {Brinchmann},
  {Cimatti}, {Ellis}, {Hill}, {Kneib}, {McLeod}, {Opitom}, {Roth},
  {Sanchez-Saez}, {Smilljanic}, {Tolstoy}, {Bacon}, {Randich}, {Adamo},
  {Annibali}, {Arevalo}, {Audard}, {Barsanti}, {Battaglia}, {Bayo Aran},
  {Belfiore}, {Bellazzini}, {Bellini}, {Beltran}, {Berni}, {Bianchi}, {Biazzo},
  {Bisero}, {Bisogni}, {Bland-Hawthorn}, {Blondin}, {Bodensteiner}, {Boffin},
  {Bonito}, {Bono}, {Bouche}, {Bowman}, {Braga}, {Bragaglia}, {Branchesi},
  {Brucalassi}, {Bryant}, {Bryson}, {Busa}, {Camera}, {Carbone}, {Casali},
  {Casali}, {Casasola}, {Castro}, {Catelan}, {Cavallo}, {Chiappini}, {Cioni},
  {Colless}, {Colzi}, {Contarini}, {Couch}, {D'Ammando}, {d'Assignies D.},
  {D'Orazi}, {da Silva}, {Dainotti}, {Damiani}, {Danielski}, {De Cia}, {de
  Jong}, {Dhawan}, {Dierickx}, {Driver}, {Dupletsa}, {Escoffier}, {Escorza},
  {Fabrizio}, {Fiorentino}, {Fontana}, {Fontani}, {Forero Sanchez}, {Franois},
  {Galindo-Guil}, {Gallazzi}, {Galli}, {Garcia}, {Garcia-Rojas}, {Garilli},
  {Grand}, {Guarcello}, {Hazra}, {Helmi}, {Herrero}, {Iglesias}, {Ilic},
  {Irsic}, {Ivanov}, {Izzo}, {Jablonka}, {Joachimi}, {Kakkad}, {Kamann},
  {Koposov}, {Kordopatis}, {Kovacevic}, {Kraljic}, {Kuncarayakti}, {Kwon}, {La
  Forgia}, {Lahav}, {Laigle}, {Lazzarin}, {Leaman}, {Leclercq}, {Lee}, {Lee},
  {Lehnert}, {Lira}, {Loffredo}, {Lucatello}, {Magrini}, {Maguire}, {Mahler},
  {Zahra Majidi}, {Malavasi}, {Mannucci}, {Marconi}, {Martin}, {Marulli},
  {Massari}, {Matsuno}, {Mattheee}, {McGee}, {Merc}, {Merle}, {Miglio},
  {Migliorini}, {Minchev}, {Minniti}, {Miret-Roig}, {Monreal Ibero}, {Montano},
  {Montet}, {Moresco}, {Moretti}, {Moscardini}, {Moya}, {Mueller},
  {Nanayakkara}, {Nicholl}, {Nordlander}, {Onori}, {Padovani}, {Pala}, {Panda},
  {Pandey-Pommier}, {Pasquini}, {Pawlak}, {Pessi}, {Pisani}, {Popovic},
  {Prisinzano}, {Raddi}, {Rainer}, {Rebassa-Mansergas}, {Richard}, {Rigault},
  {Rocher}, {Romano}, {Rosati}, {Sacco}, {Sanchez-Janssen}, {Sander},
  {Sanders}, {Sargent}, {Sarpa}, {Schimd}, {Schipani}, {Sefusatti}, {Smith},
  {Spina}, {Steinmetz}, {Tacchella}, {Tautvaisiene}, {Theissen}, {Thomas},
  {Ting}, {Travouillon}, {Tresse}, {Trivedi}, {Tsantaki}, {Tsedrik}, {Urrutia},
  {Valenti}, {Van der Swaelmen}, {Van Eck}, {Verdiani}, {Verdier}, {Vergani},
  {Verhamme}, {Vernet}, {Verza}, {Viel}, {Vielzeuf}, {Vietri}, {Vink},
  {Viscasillas Vazquez}, {Wang}, {Weilbacher}, {Wendt}, {Wright}, {Ye},
  {Yeche}, {Yu}, {Zafar}, {Zibetti}, {Ziegler}, \& {Zinchenko}}]{WST:2024}
{Mainieri}, V., {Anderson}, R.~I., {Brinchmann}, J., {et~al.} 2024, arXiv
  e-prints, arXiv:2403.05398

\bibitem[{{Mart{\'\i}nez-Solaeche} {et~al.}(2023){Mart{\'\i}nez-Solaeche},
  {Queiroz}, {Gonz{\'a}lez Delgado}, {Rodrigues}, {Garc{\'\i}a-Benito},
  {P{\'e}rez-R{\`a}fols}, {Raul Abramo}, {D{\'\i}az-Garc{\'\i}a}, {Pieri},
  {Chaves-Montero}, {Hern{\'a}n-Caballero}, {Rodr{\'\i}guez-Mart{\'\i}n},
  {Bonoli}, {Morrison}, {M{\'a}rquez}, {V{\'\i}lchez},
  {Fern{\'a}ndez-Ontiveros}, {Marra}, {Alcaniz}, {Benitez}, {Cenarro},
  {Crist{\'o}bal-Hornillos}, {Dupke}, {Ederoclite}, {L{\'o}pez-Sanjuan},
  {Mar{\'\i}n-Franch}, {Mendes de Oliveira}, {Moles}, {Sodr{\'e}}, {Taylor},
  {Varela}, \& {V{\'a}zquez Rami{\'o}}}]{MS:2023}
{Mart{\'\i}nez-Solaeche}, G., {Queiroz}, C., {Gonz{\'a}lez Delgado}, R.~M.,
  {et~al.} 2023, \aap, 673, A103

\bibitem[{{Martins} {et~al.}(2005){Martins}, {Gonz{\'a}lez Delgado},
  {Leitherer}, {Cervi{\~n}o}, \& {Hauschildt}}]{Martins:05}
{Martins}, L.~P., {Gonz{\'a}lez Delgado}, R.~M., {Leitherer}, C.,
  {Cervi{\~n}o}, M., \& {Hauschildt}, P. 2005, \mnras, 358, 49

\bibitem[{{McCracken} {et~al.}(2012){McCracken}, {Milvang-Jensen}, {Dunlop},
  {Franx}, {Fynbo}, {Le F{\`e}vre}, {Holt}, {Caputi}, {Goranova}, {Buitrago},
  {Emerson}, {Freudling}, {Hudelot}, {L{\'o}pez-Sanjuan}, {Magnard}, {Mellier},
  {M{\o}ller}, {Nilsson}, {Sutherland}, {Tasca}, \& {Zabl}}]{McCrack:2012}
{McCracken}, H.~J., {Milvang-Jensen}, B., {Dunlop}, J., {et~al.} 2012, \aap,
  544, A156

\bibitem[{{Momcheva} {et~al.}(2016){Momcheva}, {Brammer}, {van Dokkum},
  {Skelton}, {Whitaker}, {Nelson}, {Fumagalli}, {Maseda}, {Leja}, {Franx},
  {Rix}, {Bezanson}, {Da Cunha}, {Dickey}, {F{\"o}rster Schreiber},
  {Illingworth}, {Kriek}, {Labb{\'e}}, {Ulf Lange}, {Lundgren}, {Magee},
  {Marchesini}, {Oesch}, {Pacifici}, {Patel}, {Price}, {Tal}, {Wake}, {van der
  Wel}, \& {Wuyts}}]{Momcheva:2016}
{Momcheva}, I.~G., {Brammer}, G.~B., {van Dokkum}, P.~G., {et~al.} 2016, \apjs,
  225, 27

\bibitem[{{Moutard} {et~al.}(2020){Moutard}, {Malavasi}, {Sawicki}, {Arnouts},
  \& {Tripathi}}]{Moutard:2020}
{Moutard}, T., {Malavasi}, N., {Sawicki}, M., {Arnouts}, S., \& {Tripathi}, S.
  2020, \mnras, 495, 4237

\bibitem[{{Nelson} {et~al.}(2019){Nelson}, {Springel}, {Pillepich},
  {Rodriguez-Gomez}, {Torrey}, {Genel}, {Vogelsberger}, {Pakmor}, {Marinacci},
  {Weinberger}, {Kelley}, {Lovell}, {Diemer}, \& {Hernquist}}]{Nelson:2019}
{Nelson}, D., {Springel}, V., {Pillepich}, A., {et~al.} 2019, Computational
  Astrophysics and Cosmology, 6, 2

\bibitem[{{Oke}(1974)}]{Oke:1974}
{Oke}, J.~B. 1974, \apjs, 27, 21

\bibitem[{{Pacifici} {et~al.}(2023){Pacifici}, {Iyer}, {Mobasher}, {da Cunha},
  {Acquaviva}, {Burgarella}, {Calistro Rivera}, {Carnall}, {Chang}, {Chartab},
  {Cooke}, {Fairhurst}, {Kartaltepe}, {Leja}, {Ma{\l}ek}, {Salmon}, {Torelli},
  {Vidal-Garc{\'\i}a}, {Boquien}, {Brammer}, {Brown}, {Capak}, {Chevallard},
  {Circosta}, {Croton}, {Davidzon}, {Dickinson}, {Duncan}, {Faber}, {Ferguson},
  {Fontana}, {Guo}, {Haeussler}, {Hemmati}, {Jafariyazani}, {Kassin}, {Larson},
  {Lee}, {Mantha}, {Marchi}, {Nayyeri}, {Newman}, {Pandya}, {Pforr}, {Reddy},
  {Sanders}, {Shah}, {Shahidi}, {Stevans}, {Triani}, {Tyler}, {Vanderhoof}, {de
  la Vega}, {Wang}, \& {Weston}}]{Pacifici:2023}
{Pacifici}, C., {Iyer}, K.~G., {Mobasher}, B., {et~al.} 2023, \apj, 944, 141

\bibitem[{Pedregosa {et~al.}(2011)Pedregosa, Varoquaux, Gramfort, Michel,
  Thirion, Grisel, Blondel, Prettenhofer, Weiss, Dubourg, Vanderplas, Passos,
  Cournapeau, Brucher, Perrot, \& Duchesnay}]{scikit-learn}
Pedregosa, F., Varoquaux, G., Gramfort, A., {et~al.} 2011, Journal of Machine
  Learning Research, 12, 2825

\bibitem[{{Phillipps} {et~al.}(2019){Phillipps}, {Bremer}, {Hopkins}, {De
  Propris}, {Taylor}, {James}, {Davies}, {Cluver}, {Driver}, {Eales},
  {Holwerda}, {Kelvin}, \& {Sansom}}]{Phill:19}
{Phillipps}, S., {Bremer}, M.~N., {Hopkins}, A.~M., {et~al.} 2019, MNRAS, 485,
  5559

\bibitem[{{Salim}(2014)}]{Salim:14}
{Salim}, S. 2014, Serb. Astron. J., 189, 1

\bibitem[{{Salim} \& {Narayanan}(2020)}]{Salim:2020}
{Salim}, S. \& {Narayanan}, D. 2020, \araa, 58, 529

\bibitem[{{Salvador-Rusi{\~n}ol} {et~al.}(2020){Salvador-Rusi{\~n}ol},
  {Vazdekis}, {La Barbera}, {Beasley}, {Ferreras}, {Negri}, \& {Dalla
  Vecchia}}]{Salvador:2020}
{Salvador-Rusi{\~n}ol}, N., {Vazdekis}, A., {La Barbera}, F., {et~al.} 2020,
  Nature Astronomy, 4, 252

\bibitem[{{S{\'a}nchez-Bl{\'a}zquez} {et~al.}(2006){S{\'a}nchez-Bl{\'a}zquez},
  {Peletier}, {Jim{\'e}nez-Vicente}, {Cardiel}, {Cenarro},
  {Falc{\'o}n-Barroso}, {Gorgas}, {Selam}, \& {Vazdekis}}]{MILES:2006}
{S{\'a}nchez-Bl{\'a}zquez}, P., {Peletier}, R.~F., {Jim{\'e}nez-Vicente}, J.,
  {et~al.} 2006, \mnras, 371, 703

\bibitem[{{Schawinski} {et~al.}(2014){Schawinski}, {Urry}, {Simmons},
  {Fortson}, {Kaviraj}, {Keel}, {Lintott}, {Masters}, {Nichol}, {Sarzi},
  {Skibba}, {Treister}, {Willett}, {Wong}, \& {Yi}}]{Schawinski:2014}
{Schawinski}, K., {Urry}, C.~M., {Simmons}, B.~D., {et~al.} 2014, \mnras, 440,
  889

\bibitem[{{Schaye} {et~al.}(2015)}]{Schaye:15}
{Schaye}, J. {et~al.} 2015, MNRAS, 446, 521

\bibitem[{{Scoville} {et~al.}(2007){Scoville}, {Aussel}, {Brusa}, {Capak},
  {Carollo}, {Elvis}, {Giavalisco}, {Guzzo}, {Hasinger}, {Impey}, {Kneib},
  {LeFevre}, {Lilly}, {Mobasher}, {Renzini}, {Rich}, {Sanders}, {Schinnerer},
  {Schminovich}, {Shopbell}, {Taniguchi}, \& {Tyson}}]{Scoville:2007}
{Scoville}, N., {Aussel}, H., {Brusa}, M., {et~al.} 2007, \apjs, 172, 1

\bibitem[{{Shy} {et~al.}(2022){Shy}, {Tak}, {Feigelson}, {Timlin}, \&
  {Babu}}]{Shy2022}
{Shy}, S., {Tak}, H., {Feigelson}, E.~D., {Timlin}, J.~D., \& {Babu}, G.~J.
  2022, \aj, 164, 6

\bibitem[{{Simet} {et~al.}(2021){Simet}, {Chartab}, {Lu}, \&
  {Mobasher}}]{Simet:2021}
{Simet}, M., {Chartab}, N., {Lu}, Y., \& {Mobasher}, B. 2021, \apj, 908, 47

\bibitem[{{Somerville} \& {Dav{\'e}}(2015)}]{Somerville:2015}
{Somerville}, R.~S. \& {Dav{\'e}}, R. 2015, \araa, 53, 51

\bibitem[{Springel {et~al.}(2017)Springel, Pakmor, Pillepich, Weinberger,
  Nelson, Hernquist, Vogelsberger, Genel, Torrey, Marinacci, \&
  Naiman}]{SpringTNG:17}
Springel, V., Pakmor, R., Pillepich, A., {et~al.} 2017, MNRAS, 475, 676

\bibitem[{{Stensbo-Smidt} {et~al.}(2017){Stensbo-Smidt}, {Gieseke}, {Igel},
  {Zirm}, \& {Steenstrup Pedersen}}]{SS:2017}
{Stensbo-Smidt}, K., {Gieseke}, F., {Igel}, C., {Zirm}, A., \& {Steenstrup
  Pedersen}, K. 2017, \mnras, 464, 2577

\bibitem[{{Strateva} {et~al.}(2001)}]{Strateva:01}
{Strateva}, I. {et~al.} 2001, AJ, 122, 1861

\bibitem[{{Thanh Noi} \& {Kappas}(2018)}]{Thanh:2018}
{Thanh Noi}, P. \& {Kappas}, M. 2018

\bibitem[{{The MSE Science Team} {et~al.}(2019){The MSE Science Team},
  {Babusiaux}, {Bergemann}, {Burgasser}, {Ellison}, {Haggard}, {Huber},
  {Kaplinghat}, {Li}, {Marshall}, {Martell}, {McConnachie}, {Percival},
  {Robotham}, {Shen}, {Thirupathi}, {Tran}, {Yeche}, {Yong}, {Adibekyan},
  {Silva Aguirre}, {Angelou}, {Asplund}, {Balogh}, {Banerjee}, {Bannister},
  {Barr{\'\i}a}, {Battaglia}, {Bayo}, {Bechtol}, {Beck}, {Beers}, {Bellinger},
  {Berg}, {Bestenlehner}, {Bilicki}, {Bitsch}, {Bland-Hawthorn}, {Bolton},
  {Boselli}, {Bovy}, {Bragaglia}, {Buzasi}, {Caffau}, {Cami}, {Carleton},
  {Casagrande}, {Cassisi}, {Catelan}, {Chang}, {Cortese}, {Damjanov}, {Davies},
  {de Grijs}, {de Rosa}, {Deason}, {di Matteo}, {Drlica-Wagner}, {Erkal},
  {Escorza}, {Ferrarese}, {Fleming}, {Font-Ribera}, {Freeman}, {G{\"a}nsicke},
  {Gabdeev}, {Gallagher}, {Gandolfi}, {Garc{\'\i}a}, {Gaulme}, {Geha},
  {Gennaro}, {Gieles}, {Gilbert}, {Gordon}, {Goswami}, {Greco}, {Grillmair},
  {Guiglion}, {H{\'e}nault-Brunet}, {Hall}, {Handler}, {Hansen}, {Hathi},
  {Hatzidimitriou}, {Haywood}, {Hern{\'a}ndez Santisteban}, {Hillenbrand},
  {Hopkins}, {Howlett}, {Hudson}, {Ibata}, {Ili{\'c}}, {Jablonka}, {Ji},
  {Jiang}, {Juneau}, {Karakas}, {Karinkuzhi}, {Kim}, {Kong}, {Konstantopoulos},
  {Krogager}, {Lagos}, {Lallement}, {Laporte}, {Lebreton}, {Lee}, {Lewis},
  {Lianou}, {Liu}, {Lodieu}, {Loveday}, {M{\'e}sz{\'a}ros}, {Makler}, {Mao},
  {Marchesini}, {Martin}, {Mateo}, {Melis}, {Merle}, {Miglio}, {Gohar
  Mohammad}, {Molaverdikhani}, {Monier}, {Morel}, {Mosser}, {Nataf}, {Necib},
  {Neilson}, {Newman}, {Nierenberg}, {Nord}, {Noterdaeme}, {O'Dea}, {Oshagh},
  {Pace}, {Palanque-Delabrouille}, {Pandey}, {Parker}, {Pawlowski}, {Peter},
  {Petitjean}, {Petric}, {Placco}, {Popovi{\'c}}, {Price-Whelan}, {Prsa},
  {Ravindranath}, {Rich}, {Ruan}, {Rybizki}, {Sakari}, {Sanderson}, {Schiavon},
  {Schimd}, {Serenelli}, {Siebert}, {Siudek}, {Smiljanic}, {Smith}, {Sobeck},
  {Starkenburg}, {Stello}, {Szab{\'o}}, {Szabo}, {Taylor}, {Thanjavur},
  {Thomas}, {Tollerud}, {Toonen}, {Tremblay}, {Tresse}, {Tsantaki},
  {Valentini}, {Van Eck}, {Variu}, {Venn}, {Villaver}, {Walker}, {Wang},
  {Wang}, {Wilson}, {Wright}, {Xu}, {Yildiz}, {Zhang}, {Zwintz}, {Anguiano},
  {Bedell}, {Chaplin}, {Collet}, {Cuillandre}, {Duc}, {Flagey}, {Hermes},
  {Hill}, {Kamath}, {Laychak}, {Ma{\l}ek}, {Marley}, {Sheinis}, {Simons},
  {Sousa}, {Szeto}, {Ting}, {Vegetti}, {Wells}, {Babas}, {Bauman}, {Bosselli},
  {C{\^o}t{\'e}}, {Colless}, {Comparat}, {Courtois}, {Crampton}, {Croom},
  {Davies}, {de Grijs}, {Denny}, {Devost}, {di Matteo}, {Driver},
  {Fernandez-Lorenzo}, {Guhathakurta}, {Han}, {Higgs}, {Hill}, {Ho}, {Hopkins},
  {Hudson}, {Ibata}, {Isani}, {Jarvis}, {Johnson}, {Jullo}, {Kaiser}, {Kneib},
  {Koda}, {Koshy}, {Mignot}, {Murowinski}, {Newman}, {Nusser}, {Pancoast},
  {Peng}, {Peroux}, {Pichon}, {Poggianti}, {Richard}, {Salmon}, {Seibert},
  {Shastri}, {Smith}, {Sutaria}, {Tao}, {Taylor}, {Tully}, {van Waerbeke},
  {Vermeulen}, {Walker}, {Willis}, {Willot}, \& {Withington}}]{MSE:2019}
{The MSE Science Team}, {Babusiaux}, C., {Bergemann}, M., {et~al.} 2019, arXiv
  e-prints, arXiv:1904.04907

\bibitem[{{Trayford} {et~al.}(2016){Trayford}, {Theuns}, {Bower}, {Crain},
  {Lagos}, {Schaller}, \& {Schaye}}]{Trayford:2016}
{Trayford}, J.~W., {Theuns}, T., {Bower}, R.~G., {et~al.} 2016, \mnras, 460,
  3925

\bibitem[{{Valdes} {et~al.}(2004){Valdes}, {Gupta}, {Rose}, {Singh}, \&
  {Bell}}]{Valdes:2004}
{Valdes}, F., {Gupta}, R., {Rose}, J.~A., {Singh}, H.~P., \& {Bell}, D.~J.
  2004, \apjs, 152, 251

\bibitem[{{van der Wel} {et~al.}(2016){van der Wel}, {Noeske}, {Bezanson},
  {Pacifici}, {Gallazzi}, {Franx}, {Mu{\~n}oz-Mateos}, {Bell}, {Brammer},
  {Charlot}, {Chauk{\'e}}, {Labb{\'e}}, {Maseda}, {Muzzin}, {Rix}, {Sobral},
  {van de Sande}, {van Dokkum}, {Wild}, \& {Wolf}}]{LEGAC:2016}
{van der Wel}, A., {Noeske}, K., {Bezanson}, R., {et~al.} 2016, ApJ, 223, 29

\bibitem[{{Vazdekis} {et~al.}(2015){Vazdekis}, {Coelho}, {Cassisi},
  {Ricciardelli}, {Falc{\'o}n-Barroso}, {S{\'a}nchez-Bl{\'a}zquez}, {La
  Barbera}, {Beasley}, \& {Pietrinferni}}]{Vaz:2015}
{Vazdekis}, A., {Coelho}, P., {Cassisi}, S., {et~al.} 2015, \mnras, 449, 1177

\bibitem[{{Vazdekis} {et~al.}(2016){Vazdekis}, {Koleva}, {Ricciardelli},
  {R{\"o}ck}, \& {Falc{\'o}n-Barroso}}]{Vaz:2016}
{Vazdekis}, A., {Koleva}, M., {Ricciardelli}, E., {R{\"o}ck}, B., \&
  {Falc{\'o}n-Barroso}, J. 2016, \mnras, 463, 3409

\bibitem[{{Westera} {et~al.}(2002){Westera}, {Lejeune}, {Buser}, {Cuisinier},
  \& {Bruzual}}]{Westera:2002}
{Westera}, P., {Lejeune}, T., {Buser}, R., {Cuisinier}, F., \& {Bruzual}, G.
  2002, \aap, 381, 524

\bibitem[{{Williams} {et~al.}(2009){Williams}, {Quadri}, {Franx}, {van Dokkum},
  \& {Labb{\'e}}}]{2009Will}
{Williams}, R.~J., {Quadri}, R.~F., {Franx}, M., {van Dokkum}, P., \&
  {Labb{\'e}}, I. 2009, ApJ, 691, 1879

\bibitem[{{Worthey}(1994)}]{Worthey:94}
{Worthey}, G. 1994, \apjs, 95, 107

\bibitem[{{Wright} {et~al.}(2019){Wright}, {Lagos}, {Davies}, {Power},
  {Trayford}, \& {Wong}}]{Wright:2019}
{Wright}, R.~J., {Lagos}, C. d.~P., {Davies}, L. J.~M., {et~al.} 2019, \mnras,
  487, 3740

\bibitem[{{York} {et~al.}(2000)}]{SDSS}
{York}, D.~G. {et~al.} 2000, AJ, 120, 1579

\bibitem[{{Zahid} {et~al.}(2016){Zahid}, {Geller}, {Fabricant}, \&
  {Hwang}}]{Zahid:2016}
{Zahid}, H.~J., {Geller}, M.~J., {Fabricant}, D.~G., \& {Hwang}, H.~S. 2016,
  \apj, 832, 203

\bibitem[{{Zhang} {et~al.}(2023){Zhang}, {Jiang}, {Shectman}, {Yang}, {Cai},
  {Shi}, {Huang}, {Lu}, {Zheng}, {Kang}, {Mao}, \& {Huang}}]{Zhang:2023}
{Zhang}, J., {Jiang}, H., {Shectman}, S., {et~al.} 2023, PhotoniX, 4

\bibitem[{{Zibetti} {et~al.}(2017){Zibetti}, {Gallazzi}, {Ascasibar},
  {Charlot}, {Galbany}, {Garc{\'\i}a Benito}, {Kehrig}, {de
  Lorenzo-C{\'a}ceres}, {Lyubenova}, {Marino}, {M{\'a}rquez}, {S{\'a}nchez},
  {van de Ven}, {Walcher}, \& {Wisotzki}}]{Zibetti:2017}
{Zibetti}, S., {Gallazzi}, A.~R., {Ascasibar}, Y., {et~al.} 2017, \mnras, 468,
  1902

\end{thebibliography}

\appendix
\section{Optimising ML algorithms}
\label{sec:OP_ML}

For the optimisation of the ML algorithms, it is necessary to fine-tune the hyperparameters as these will affect the prediction of the physical parameters that we considered. We use both sets of observables (the spectral indices and observed magnitude) as input. For the output, we estimate the physical parameters --- mwa, uwa, t$_{\rm mod}/\tau$, metallicity, A$_{\rm v}$, and sSFR. The hyperparameters are adjusted for multiple runs, and two different statistics, R-squared ($R^2$) and root-mean-square-error (RMSE), are used to assess the effectiveness of the algorithms. The $R^2$ is defined as
\begin{equation}
    R^2 = 1 - \frac{\sum_{i=1}^n \left(x_i^{pred} - x_i^{true}\right)^2}{\sum_{i=1}^n\left(x_i^{true} - \bar{x}_i^{true}\right)^2},
    \label{eq:r2}
\end{equation}
where $\bar{x}_i^{true}$ is mean value of the physical parameter and $n$ is the number of samples. Similarly RMSE is formulated as\begin{equation}
    \rm{RMSE} = \sqrt{\sum_i^n \frac{\left (x_i^{pred} - x_i^{true} \right)^2 }{n}}
    \label{eq:rmse}.
\end{equation}
These statistics are commonly used to evaluate the goodness of fit for regression. The $R^2$ value ranges from 0 to 1, where a value $\sim 1$ indicates a good fit. Unlike $R^2$, RMSE provides an absolute measure of fit, where a lower value indicates a better fit. We note that these statistics are calculated for each physical parameter, such as the mwa and uwa, and that the final statistics quoted are averaged across all physical parameters considered. 

\begin{figure}
    \centering
    \includegraphics[width=\linewidth]{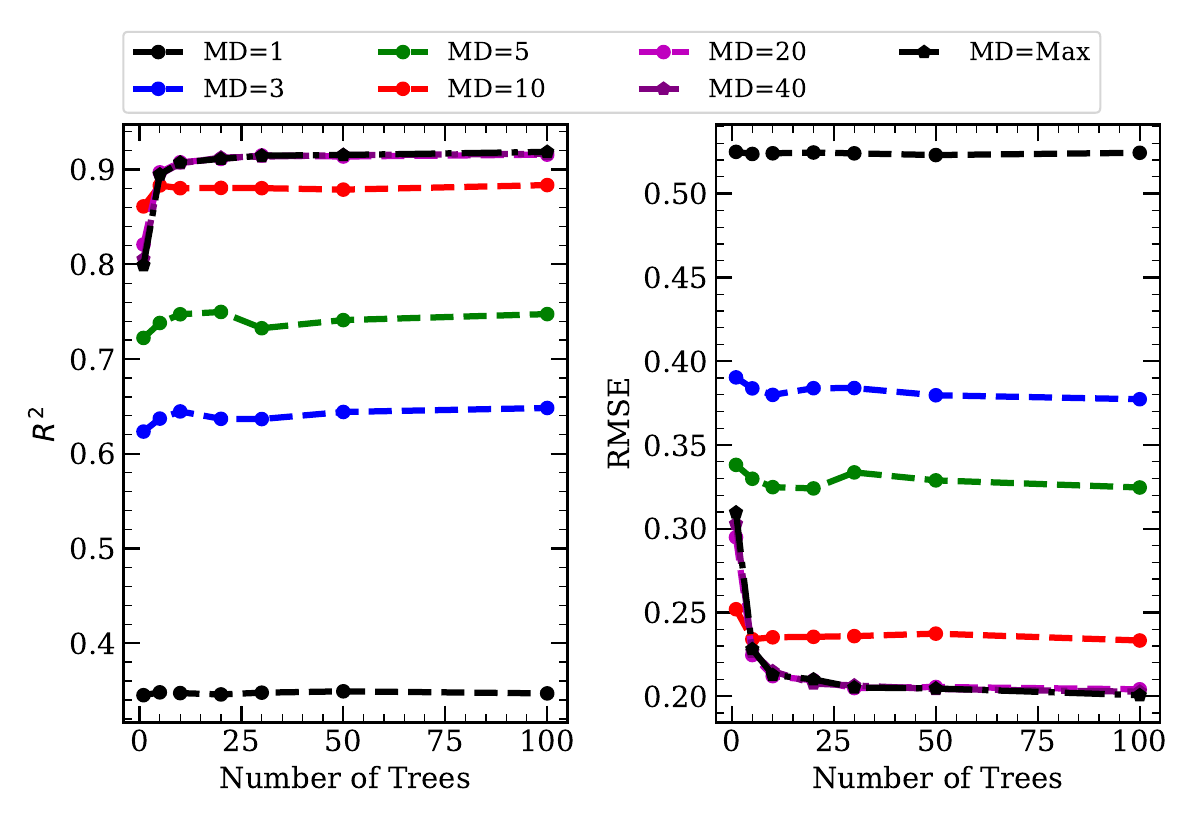}
    \caption{Variation in $R^2$ (left) and RMSE (right) for different numbers of trees and maximum depths (MDs). The number of trees varies (1, 5, 10, 20, 30, 50, and 100), while the MDs we consider are 1 (dashed black line with circles), 3 (dashed blue line with circles), 5 (dashed green line with circles), 10 (dashed red line with circles), 20 (dashed magenta line with circles), 40 (solid purple lines with pentagon markers), and `Max' (solid black lines with pentagon markers). The statistics are averaged over all six parameters that we consider. Note that these runs assume S/N$=$10, and the simulated templates are at $z=0.3$.}
    \label{fig:tree_check}
\end{figure}

\begin{figure}
    \centering
    \includegraphics[width=\linewidth]{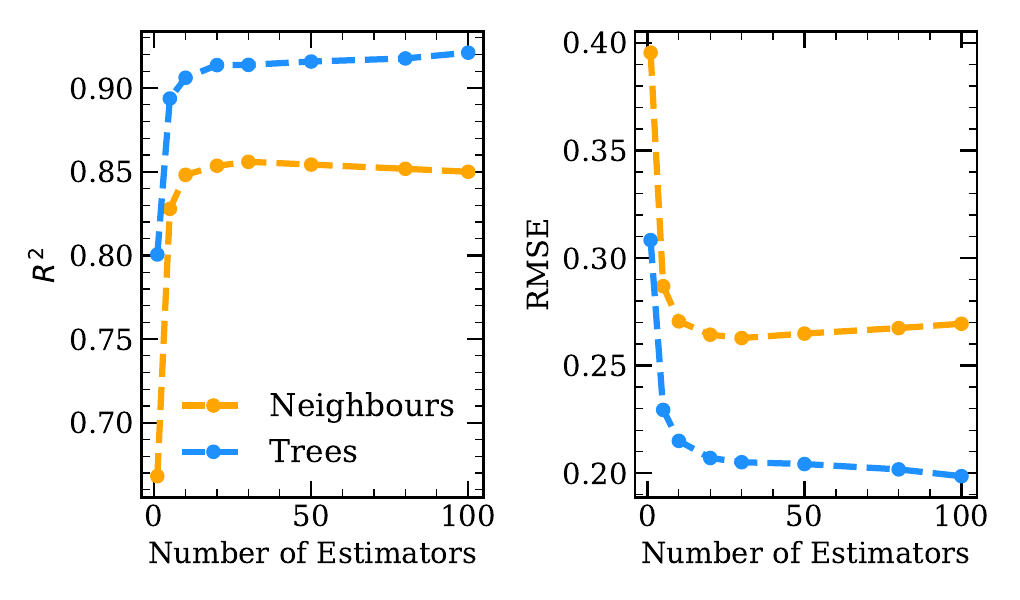}
    \caption{Variation in R$^2$ (left) and RMSE (right) with the change in the number of estimators, neighbours for KNN (dashed orange line), and trees for RF (dashed blue line). These statistics were computed by considering the difference between true and predicted values. These results assume the spectra have S/N=10 and the templates are at $z=0.3$. }
    \label{fig:neighbour_check}
\end{figure}

For the RF, we focus on tuning two hyperparameters: the number of trees and the maximum depth (MD). Figure~\ref{fig:tree_check} shows the changes in $R^2$ and RMSE for different values of these hyperparameters. It is important to note that for our templates, MD$=$Max corresponds to a range of values between $44<$MD$<58$, depending on the particular run, due to the random nature of the RF algorithm. Both $R^2$ and RMSE exhibit higher sensitivity to changes in MD compared to the number of trees. Both $R^2$ and RMSE show similar values for MD$>$20 and number of trees $>25$. 
For the KNN algorithm, the hyperparameter that was optimised was the number of neighbours, K. Figure~\ref{fig:neighbour_check} shows the $R^2$ (left) and RMSE (right) values for different values of K. The performance of KNN is dependent on the value of K, but beyond K $\gtrsim20$ there is minimal change in the $R^2$ and/or RMSE statistics. To mitigate overfitting, a larger K value is chosen than the point at which the statistics flatten. Therefore, 100 neighbours are selected. For comparison, the statistics obtained for the RF algorithm, using MD$=$Max, are also plotted. It is noted that although the plateau of the statistics begins around the same number of estimators for both algorithms, the $R^2$ and RMSE values also indicate a better fit of the model. Finally, we also checked the variance around the mean for $R^2$ and RMSE for both algorithms. We set the number of trees or nearest neighbours ranging from 1 to 100, and for each configuration we carried out 30 independent runs of KNN and RF. For each run, we calculated the $R^2$ and RMSE statistics and computed the variance in these statistics. We find low evidence of overfitting as both KNN and RF exhibit minimal variance with standard deviation $\lesssim 10^{-3}$, in $R^2$ and RMSE, regardless of the number of trees/neighbours.

\end{document}